\documentclass[a4paper,onecolumn,11pt,accepted=2025-12-29]{quantumarticle}
\pdfoutput=1

\usepackage[numbers,sort&compress]{natbib}

\usepackage{amsmath, amssymb, physics, amsthm}
\newtheorem{theorem}{Theorem}
\newtheorem{problem}{Problem}
\newtheorem{lemma}{Lemma}

\newcommand{\mO}{\mathcal{O}}
\newcommand{\poly}{\mathrm{poly}}

\usepackage{adjustbox} 
\usepackage{diagbox} 
\usepackage[table]{xcolor} 
\usepackage{subcaption} 

\definecolor{C0}{HTML}{386BBC}
\definecolor{C1}{HTML}{D81B60}
\definecolor{C2}{HTML}{17C188}
\definecolor{C3}{HTML}{EBDD52}

\usepackage{enumitem}
\usepackage[unicode]{hyperref}
\usepackage[nameinlink,capitalize]{cleveref}

\title{Assessing fault-tolerant quantum advantage for {$k$}-SAT with structure}

\author{Martijn Brehm}
\affiliation{University of Amsterdam, Amsterdam, the Netherlands}
\email{m.a.brehm@uva.nl}
\homepage{https://staff.science.uva.nl/m.a.brehm}
\orcid{0009-0004-5157-5661}
\author{Jordi Weggemans}
\affiliation{QuSoft \& CWI, Amsterdam, the Netherlands}
\homepage{https://jordiweggemans.github.io/research/}
\orcid{0000-0002-8469-6900}

\begin{document}

\begin{abstract}
For many problems, quantum algorithms promise speedups over their classical counterparts. However, these results predominantly rely on asymptotic worst-case analysis, which overlooks significant overheads due to error correction and the fact that real-world instances often contain exploitable structure. In this work, we employ the hybrid benchmarking method to evaluate the potential of quantum Backtracking and Grover's algorithm against the 2023 SAT competition main track winner in solving random $k$-SAT instances with tunable structure, designed to represent industry-like scenarios, using both $T$-depth and $T$-count as cost metrics to estimate quantum run times. Our findings reproduce the results of Campbell, Khurana, and Montanaro (Quantum '19) in the unstructured case using hybrid benchmarking. However, we offer a more sobering perspective in practically relevant regimes: almost all quantum speedups vanish, even asymptotically, when minimal structure is introduced or when $T$-count is considered instead of $T$-depth. Moreover, when the requirement is for the algorithm to \emph{find} a solution within a single day, we find that only Grover's algorithm has the potential to outperform classical algorithms, but only in a very limited regime and only when using $T$-depth. We also discuss how more sophisticated heuristics could restore the asymptotic scaling advantage for quantum backtracking, but our findings suggest that the potential for practical quantum speedups in more structured $k$-SAT solving will remain limited.
\end{abstract}

\section{Introduction}
Quantum computing holds the promise of revolutionizing numerous fields by solving computational problems faster than classical computers. Optimization problems---which have a special status as they arise almost everywhere in practice---are also commonly identified as candidates to benefit from quantum computing~\cite{abbas2023quantum}. 

But how does one show that such a quantum speedup over classical algorithms exists? In theoretical computer science, the de facto way to do this is to compare \emph{asymptotic worst-case complexity}, where the complexity can be measured at different levels of abstraction (like the number of gates, queries to a certain function, etc.). Take, for example, constraint satisfaction problems (CSPs), a broad class of combinatorial problems where one must find an assignment to variables that satisfies a given set of local constraints. Boolean satisfiability $k$-SAT is a canonical CSP: each constraint is a clause containing at most $k$ literals, and the task is to decide whether a single assignment satisfies all clauses simultaneously. For $k$-SAT, the best \emph{known} classical algorithms that succeed on every input yield provable upper bounds (up to polynomial factors) on the number of elementary classical gates of $2^{n(1 - c/k + o(1/k))}$ where $c>0$ is an algorithm-dependent constant~\cite{pudlak2013satisfiability,schoning1999a,paturi2005an}. In contrast, for any fixed $k$, Grover’s algorithm solves $k$-SAT with bounded error using at most $2^{n/2}$ quantum gates (again up to polynomial factors), giving for large constant $k$ an as-good-as quadratic \emph{asymptotic} improvement in the number of ``native operations''.

However, such an asymptotic worst-case analysis has two major shortcomings when it comes to what one wants to know in \emph{practice}:
\begin{enumerate}[label=(\roman*)]
    \item one might not be interested in the worst-case performance---which considers the worst possible performance amongst all problem instances---as the algorithm only has to perform well on a small subset of all possible instances that are encountered in real-life scenarios;
    \item quantum and classical computers use different types of hardware with different `unit costs' (i.e.~implementation resources), so $X$ quantum gates cannot be directly compared with $Y$ classical gates. Moreover, these abstract measures of complexity might not be easily translatable to universal measures like time, memory, energy usage, etc., one actually is interested in.
\end{enumerate}
To cover item (i) classical algorithms are often compared empirically by measuring their performance on benchmark sets. Through decades of progress, we have classical algorithms that can exploit the structure in real-world SAT instances well-enough to solve many large instances efficiently~\cite{biere2009handbook}.

This is not yet feasible for most quantum algorithms, as quantum hardware is still in its infancy. How then can we study the performance of quantum algorithms on instances of real-life interest, while also making assumptions about hardware that does not exist as of yet to take into account item (ii)?

To tackle point (ii), Campbell, Khurana and Montanaro proposed an elegant approach to empirically study the performance of a quantum backtracking algorithm~\cite{montanaro2018quantum} and Grover's algorithm~\cite{GroverQuantum1997} on random instances of CSPs ($k$-SAT and graph colourability), using a cost model that accounts for error-correction~\cite{campbell2019applying}. To go beyond asymptotic worst-case analysis, they write out the constants in the query complexity upper bound of their algorithms, which include as parameters the number of variables $n$ and (for backtracking) the size of the classical backtracking tree $\mathcal{T}$. To estimate the scaling of $\mathcal{T}$ for a class of random problem instances, they sample random instances and compute a fit of the median value of $\mathcal{T}$, giving them query complexity upper bound in terms of just $n$ for both algorithms. They then argue that (if one does not account for the classical processing cost required to decode the surface code) there is a regime such that the estimated number of queries translates into a quantum runtime of $\sim 1$ day, while classical would require anywhere between $15$ and $500$ years~\cite{campbell2019applying}.

While this seems to show the possibility of significant quantum speedups on these problems, there are several objections one could make regarding Ref.~\cite{campbell2019applying}. First, the work only considers random instances at the threshold value $a_k$ of the clauses-to-variable ratio (defined below): SAT instances at the threshold are believed (and experimentally verified) to be amongst the hardest instances~\cite{cheeseman1991really}, implying that they effectively exhibit no structure. Indeed, most classical SAT solvers are empirically tuned to exploit structure, meaning that the algorithms are not optimized to solve these types of problems. 
Second, Ref.~\cite{campbell2019applying} employs a median fit to estimate a parameter involved in the query complexity upper bound: this approach \textit{underestimates} this ``upper bound'', due to an analogue of Jensen's inequality for medians~\cite{merkle2005jensen}. Third, the quantum algorithm is highly parallelized in many aspects---most notably  in its $T$-gate implementation, allowing $T$-depth to be considered instead of $T$-count---an advantage not given to the classical solver. It is known that allowing a moderate amount of classical parallelization can make even a fully quadratic speedup fail to result in a quantum advantage when considering a reasonable maximum on the allowed computing time~\cite{babbush2021focus}. Fourth, the quantum backtracking algorithm only \emph{detects} a solution, but does not find it. This makes the algorithm far less useful than its classical counterpart, in particular for practical applications.\footnote{Even in the main track of the SAT competition, whose winner Ref.~\cite{campbell2019applying} compares against, it is required to provide certificates in the satisfiable and even the unsatisfiable cases.} Using binary search this can be turned into a search algorithm, but this would involve an additional $\mO(\log \mathcal{T}) = \mO(n)$ multiplicative overhead. Fifth and final, the detection algorithm requires an upper bound on $\mathcal{T}$ to run correctly, which costs an additional $\mO(n)$ runs of the binary search algorithm to estimate.\footnote{In Ref.~\cite{campbell2019applying} they do estimate a fit of $\mathcal{T}$, which could in principle be used to configure the detection algorithm. There are two potential problems with this: first, being a fit of the median value for $\mathcal{T}$, we should expect this bound to be incorrect half of the time, so performance guarantees will only be valid on half of the random instances. Second, this means the algorithm can only work when applied to instances from the family it was trained on; in particular, this means that we cannot be sure whether the algorithm works on a specific instance encountered in practice, as we do not have an estimate for its value of $\mathcal{T}$.} This additional $\mO(n^2)$ slowdown could offset the observed speedup.

More recently, Cade, Folkertsma, Niesen and Weggemans proposed an approach that overcomes some of the previously mentioned obstacles~\cite{cade2022quantum,cade2023quantifying}. In this approach, coined \emph{hybrid benchmarking} in Ref.~\cite{ammann2023realistic}, one tightens a query complexity upper bound by parametrizing in many more features of the input instance (i.e.~beyond just input size). As long as these features are efficiently \textit{classically} computable one can evaluate the query complexity upper bound exactly (as opposed to estimating these features, yielding a lower bound only applicable to random data \cite{campbell2019applying}). While Ref.~\cite{campbell2019applying} gives a somewhat positive view of the possibilities of achieving a quantum speedup on instances that are solvable within a day, the works of~\cite{cade2022quantum,cade2023quantifying,ammann2023realistic} give a much more sobering picture for the problems they study.   

This work aims to study the performance of quantum backtracking and Grover's algorithm on $k$-SAT by combining the best of both approaches\footnote{The quantum algorithms considered in \cite{cade2022quantum,cade2023quantifying,ammann2023realistic} were essentially classical algorithms with Grover subroutines, meaning that at most quadratic speedups could be achieved only at the subroutine level. Therefore, focusing on quantum backtracking and vanilla Grover as case studies, as done in Ref.~\cite{campbell2019applying}, seems to have more potential since both algorithms are fully quantum.}: with the hybrid benchmarking of~\cite{cade2022quantum,cade2023quantifying,ammann2023realistic} we can evaluate relatively tight query complexity upper bound of any problem instance (solving the first and second objection listed above), which we use to study random $k$-SAT distributions that have a tunable amount of ``structure'', specifically designed to mimic industry-like instances \cite{levy2017locality,alyahya2023structure}. We then turn the query complexity upper bound into an upper bound on the quantum runtime using the setup of~\cite{campbell2019applying}, where we consider $T$-depth as well as $T$-count (solving the third objection). Finally, we use binary search to turn the detection algorithm into a search algorithm, and use the search algorithm to estimate the required tree size (solving the fourth and fifth objections).

While Ref.~\cite{campbell2019applying} shows that on completely \textit{random} instances at the satisfiability threshold (which are believed to have no exploitable structure) Grover's algorithm and a quantum backtracking detection algorithm can beat classical solvers when the maximum allowed computing time is a day, our guiding question will be: \\
\begin{center}
    \textit{Can fault-tolerant quantum algorithms provide relevant speedups in more structured (and thus more practically motivated) instances of $k$-SAT?}
\end{center}

\paragraph{Contributions} Our main contributions are as follows.
\begin{itemize}
    \item We provide explicit query complexity upper bound of the quantum walk detection algorithm by Belovs (i.e.,~including all constants and other relevant factors), lowering the complexity by a factor of approximately 100 over previously known configurations~\cite{montanaro2018quantum, campbell2019applying, belovs2013quantum}. We provide a similar result for a binary search algorithm based on this detection algorithm.
    
    \item These query complexity upper bound have been implemented into code to estimate the performance on arbitrary $k$-SAT instances. This is achieved by running a classical backtracking algorithm to collect all the required parameters involved in the bounds; these are then combined with the surface code resource estimations from Ref.~\cite{campbell2019applying} to estimate a quantum runtime. We applied the same approach to evaluate query complexity upper bound of Grover's algorithm as presented in~\cite{cade2023quantifying}.
    
    \item We present results of our method applied to random SAT instances with tunable structure, reproducing the results of Ref.~\cite{campbell2019applying} when using $T$-depth (which requires high parallelization) and completely unstructured SAT instances. Beyond this regime, however, we find that almost all quantum speedups vanish, even asymptotically, when minimal structure is introduced or when $T$-count is considered instead of $T$-depth.
    
    \item When the objective is to \emph{find} a solution within a single day, we find that only Grover's algorithm has the potential to outperform classical algorithms. However, this occurs only in a very limited regime and only when using $T$-depth. We show that if current quantum algorithms are restricted from using such parallelization, there is no potential for a speedup that is useful within a one-day computing limit. We further argue that even with improved heuristics, the potential for quantum backtracking algorithms remains limited.
\end{itemize}

\paragraph{Related work}
There are several lines of work that aim to assess the performance of quantum algorithms beyond standard worst-case asymptotic analysis. Since we cannot (yet) run or simulate quantum algorithms on large, arbitrary inputs, each of these approaches proposes a different way to work around this limitation.

Above, we already mentioned a number of works, conceptually similar to ours, that fall under the umbrella of \emph{hybrid benchmarking}~\cite{cade2022quantum, cade2023quantifying, campbell2019applying, ammann2023realistic}. These works parametrise complexity upper bounds of quantum algorithms in terms of additional, efficiently computable features of the input, yielding much tighter expressions. Whenever these features can be computed classically, one can evaluate the quantum complexity on arbitrary inputs. However, computing the features typically requires solving the problem classically, so this approach is particularly suitable for problems where the quantum advantage is expected to be small (for instance, polynomial). A related work studies the performance of the Grover Quantum Approximate Optimization Algorithm on random 3-SAT instances, giving numerical evidence for a quadratic speed-up over classical random sampling~\cite{zhang2024grover}. The authors already note, however, that this advantage is expected to diminish when compared to practical classical SAT solvers, except perhaps on highly unstructured instances---an observation consistent with our conclusions.

Another approach is to simulate quantum algorithms directly at the circuit level. For exact simulation (i.e., state-vector simulation), the required memory scales exponentially with the number of qubits, which already limits the size of the classical problem instances one can explore: solving an $n$-bit CSP with a quantum algorithm requires at least $n$ qubits merely to write down the output string, making large-scale simulations infeasible. Approximate simulation methods, such as tensor-network techniques~\cite{markov2008simulating}, can handle somewhat larger instances, but only when the quantum computation generates limited entanglement. Generally, deeper circuits tend to produce higher entanglement, and since the algorithms we study have very large circuit depth, we expect these methods to perform poorly on the instances we consider (although it may still be worthwhile to test this).

\paragraph{Organisation of paper} In the next section, we introduce the studied algorithms and for each provide a query complexity upper bound, we then introduce the distributions of SAT instances that we study, and motivate our cost model for the quantum algorithms (which we alter slightly from Ref.~\cite{campbell2019applying}). In the section after, we present our results: the observed complexity of the various algorithm on SAT instances with increasing structure. In the final section we conclude.

\section{Algorithms, instances and cost model}
\subsection*{Classical backtracking}
Backtracking is a classical technique for solving constraint satisfaction problems (CSPs). 

\begin{problem}[Constraint satisfaction problem]
  Given a predicate $P : \{0,1\}^n \to \{\textup{true, false}\}$, find an $x\in \{0,1\}^n$ such that $P(x)=\textup{true}$, or output \textup{``not found''} if $P(x)=\textup{false}$ for all $x\in\{0,1\}^n$.
\end{problem}
We call the bit strings $x\in \{0,1\}^n$ \textit{assignments}. Solving a CSP naively requires checking all $2^n$ assignments in the worst case. However, for some CSPs you can (sometimes) detect whether a partial assignment can or cannot be extended to a solution. Formally, this extends the predicate to $P:\{0,1,*\}^n\to\{\text{true, false, indeterminate}\}$, where an `$*$' means a variable is unassigned. When you realize an assignment cannot be extended into a solution, you no longer need to check all assignments that extend it. This is the idea underlying \textit{backtracking}.

Specifically, a backtracking algorithm comes equipped with an implementation of the predicate\footnote{Note that the specific implementation matters. A more advanced predicate could recognize that a partial assignment is mapped to true or false earlier than a more basic one, at the cost of additional processing time.} $P:\{0,1,*\}^n \to \{\textup{true, false, indeterminate}\}$ and a function $h:\{0,1,*\}^n \to \{1, \dots, n\}$,
called a \textit{heuristic}, which, given a partial assignment, returns the next variable to check. The backtracking algorithm starts by creating an empty assignment $x = * \dots *.$ It then runs $h(x),$ which outputs a variable (e.g., variable 2), and $x$ is updated so that this variable is set to true (e.g., set $x := *1*\dots *$). It then calls $P(x)$: if $P(x) = \textup{true},$ $x$ is a solution, and we output it; if $P(x) = \textup{indeterminate},$ we extend $x$ using $h$ (i.e., call $h(x)$ to obtain another variable and set it to true). However, if $P(x) = \textup{false},$ the current assignment $x$ cannot be extended to a solution. This means we can invert our previous decision (e.g., set variable 2 to false instead) and continue. Eventually, the algorithm either finds a solution or exhausts all possibilities and ends up with an empty assignment again, concluding ``not found.''

One can view this procedure as exploring a tree, where each node corresponds to a partial assignment $x$. The root node corresponds to the empty assignment $* \dots *$, and each node has two children: one where the next variable has been set to true, and one where it is false. Leaves correspond to assignments where $P(x)$ returned true or false (dead-ends, complete assignments, or solutions). If $\mathcal{T}$ is the size of the tree, then the backtracking algorithm makes $\mathcal{T}$ queries to $P$ and $h$, and the hope is that $\mathcal{T} \ll 2^n$, so that a large improvement is made over brute-force search.\footnote{Note that a satisfiable instance of your CSP might be solved using far fewer than $\mathcal{T}$ queries, as $\mathcal{T}$ is the size of the entire tree, while one might find a particular solution without searching the entire tree.}

\subsection{Quantum backtracking (detection)}\label{sect:detection}
By a result of Montanaro, any such classical backtracking algorithm exploring a tree of size $\mathcal{T}$ can be turned into a quantum algorithm that detects the existence of a solution with only $\mO(\sqrt{\mathcal{T}n})$ queries to $P$ and $h$, which is almost quadratically better~\cite{montanaro2018quantum}. This is achieved by running a quantum walk algorithm starting from the root of the backtracking tree, using a quantum walk algorithm due to Belovs that can start from an arbitrary starting distribution~\cite{belovs2013quantum}. We provide the details in~\cref{appendix:belovs_bounds}; summarizing, Belovs' result states that, given access to an oracle that determines whether a vertex in a graph is a solution ($P$ in this case), the quantum walk algorithm detects the existence of a solution using $\mO(\sqrt{RW})$ queries, where $W$ is the total weight of the graph, and $R$ is the effective resistance from the support of the starting distribution to the set of solutions. This algorithm requires a priori knowledge of upper bounds on $R$ and $W$, which determine the complexity of the algorithm. Since our tree has unit weights, and since the effective resistance from root to leaves in a tree is at most the depth $n$ of the tree, we recover the complexity $\mO(\sqrt{\mathcal{T}n})$.

In~\cref{appendix:belovs_bounds_detection}, we derive an explicit query complexity upper bound of the quantum walk detection algorithm by Belovs (i.e.,~including all constants and other relevant factors). Given a desired error probability $\delta$, we carefully balance the number of parallel repetitions of the algorithm, the precision of the quantum phase estimation subroutine, and constants in the quantum walk unitary, in order to minimize the total number of queries. This already yields an query complexity upper bound that is a factor $\sim 100$ better over previously known configurations~ \cite{montanaro2018quantum, campbell2019applying,belovs2013quantum}.\footnote{This is in terms of total queries. If we allow parallelisation and compare the maximum number of sequential queries, the speedup will be smaller than $100$. Indeed, we could have optimised for less sequential queries, but that would have resulted in far more total queries. As explained in the introduction, we aim to study the total $T$-count, so we optimised for that.}

Recall that we want to evaluate this query complexity upper bound for specific SAT instances, given a backtracking algorithm for SAT (i.e.~given $P$ and $h$), meaning we need to compute $W$ and $R$. We will attempt to compute $W$ and $R$ exactly. Note that this is unrealistic: in practice, one would have to estimate $W$ and $R$, which requires additional queries to do, and may not yield $W$ and $R$ exactly, but a loose upper bound on them. We will account for these overheads in the next section when we turn detection into search. Hence, we think of the complexity of the detection algorithm as an optimal baseline.

We can estimate $W$ and $R$ by running the classical backtracking algorithm in time $\mO(\mathcal{T})$ to construct the backtracking tree. We can then count the nodes to give us $W=\mathcal{T}$ in time $\mO(\mathcal{T})$ and compute the effective resistance $R$ in time $\mO(\mathcal{T}^3)$.\footnote{The effective resistance of a graph can be computed by inverting the Laplacian matrix of the graph (see e.g.~Theorem~2.2 in \cite{ellens2011effective}), which takes $\mO(\mathcal{T}^3)$ using a naive matrix multiplication algorithm.} Unfortunately, this last step quickly becomes infeasible for large instances, so we settle for the upper bound $R\leq n$ in all our experiments. We note that we can reduce the complexity of computing $R$ to just $\mO(t^3)$, where $t$ is the number of solutions, however in practice this was still too slow.\footnote{This relies on two observations. First, we can remove branches leading to non-solution leaves. Second, long straight paths can then be reduced to single edges. This leaves a system of $\mO(t)$ equations, instead of $\mO(\mathcal{T})$ equations.}

Our choice of $P$ and $h$, i.e., our actually used backtracking algorithm, is the same as in Ref.~\cite{campbell2019applying}. The heuristic $h$ checks variables in order of their number of appearances: e.g., if variable $x_i$ occurs most often in our formula (negated or not), followed by $x_j$, and so on, the first variable we consider is $x_i$, the second is $x_j$, and so on. We always first set a variable to false. The predicate $P$ is defined to iterate over all clauses and literals in order. If one finds a literal is indeterminate, we consider $P(x)$ indeterminate (i.e., even if a later literal in the same clause is satisfied). Otherwise, it behaves as expected: if all literals in all clauses are unsatisfied, we output false, and if one literal in each clause is satisfied, we output true. Our motivation for this choice is to keep $P$ and $h$ as simple as possible, which we come back to in the discussion.

\subsection{Quantum backtracking (search)}\label{sect:search}
The above quantum backtracking algorithm only \emph{detects} whether there exists a solution, i.e.~whether a SAT instance is satisfiable or not. For most practical applications it is desirable to also find the solution if the formula is satisfiable. In our backtracking context (where the graph is a tree), we can turn detection into search by a standard binary search algorithm at the cost of a $\mO(\log \mathcal{T}) = \mO(n)$ overhead. However, to guarantee the $n$ consecutive detection runs are still correct with the desired success probability, we also need to amplify the success probability to $\mO(1/n)$, requiring $\mO(\log n)$ parallel repetitions.

Given this search algorithm, we can estimate the upper bound on $\mathcal{T}$ which the detection algorithm needs to run correctly. This estimation relies on the fact that the detection algorithm requires the upper bound on $\mathcal{T}$ only in the negative case, i.e., for unsatisfiable instances. Hence, when the algorithm outputs ``no marked elements exist" we can always trust it (with the given error probability): if we had given the algorithm an unsatisfiable instance then it would be correct, but if we had given the algorithm a satisfiable instance then it wouldn't have output negatively, regardless of $\mathcal{T}$. 

Thus, we should only doubt the correctness of the detection algorithm when it outputs ``marked elements exist". But our binary search outputs a candidate solution. If the detection algorithm tells us ``marked elements exist", we can force it to find a solution, which we can verify using one query to $P$. If this solution is not valid, we can conclude ``no marked elements exist" with small failure probability.

We can now estimate $\mathcal{T}$ as follows. Define $\mathcal{T}'=1$ and run binary search. If it outputs ``no marked elements exist'', we trust it. Otherwise, we check the candidate solution $x$ by running $P(x)$. If it is correct, we output it. Otherwise, we double $\mathcal{T}'$ and repeat. Once $\mathcal{T}' \geq \mathcal{T}$ becomes a correct upper bound, the procedure is guaranteed to work with the configured success probability, so we require $\lceil \log \mathcal{T} \rceil \in \mO(n)$ repetitions. This algorithm is due to Ref.~\cite[p. 9]{montanaro2018quantum}.

This yields a usable search algorithm which makes $\mO(\sqrt{\mathcal{T}n} n^2 \log n)$ queries to $P$ and $h$ and which does not require any prior knowledge to run (unlike the detection algorithm). In~\cref{appendix:belovs_bounds_search}, we derive an explicit expression of this bound keeping track of all relevant factors (i.e., without asymptotic notation). To evaluate this query complexity upper bound for specific SAT instances, recall that we can run the classical backtracking algorithm to determine $\mathcal{T}$. In addition, we determine the depth of the first solution that the backtracking algorithm finds (note that we can compute this easily using $h$). This allows us to bound the number of repetitions of the detection algorithm in the final iteration. This can have a significant effect, as the last run dominates the complexity since earlier runs have complexity $\mO(\sqrt{\mathcal{T}'n})$ with $\mathcal{T}'$ exponentially smaller than $\mathcal{T}$.

\subsection{Grover's algorithm}\label{sect:grover}
Grover's quantum search algorithm allows one to search a list of $N$ elements with $\mO(\sqrt{N})$ queries to the list. Specifically, given oracle access to a function $f : \{0, 1, ..., N-1\} \to \{0,1\}$, with high probability the algorithm outputs some $x$ such that $f(x)=1$ if it exists, or detects that no such $x$ exists, using only $\mO(\sqrt{N})$ queries to $f$. In the case of SAT, we would search over all $N = 2^n$ assignments, yielding a search algorithm which makes $\mO(\sqrt{2^n})$ queries to the predicate $P$. 

In~\cref{appendix:grover}, we state a very optimised explicit version of this query complexity upper bound due to Ref.~\cite{cade2023quantifying}. This bound is parameterised in terms of both the list size $N$ and the number of solutions $t$ in the list. Hence, to evaluate this bound for a given SAT instance, we need to compute the total number of satisfying assignments to a given SAT instance. Since we already constructed the backtracking tree, we can compute $t$ by taking all solution leaves and counting the number of ways in which this assignment can be extended. 

\subsection{Classical SAT solver}\label{sect:classical_solvers}
Historically, many cutting-edge classical SAT solvers have been backtracking algorithms, one example being the famous DPLL algorithm~\cite{davis1960a,davis1962a}. However, since the mid-90s, SAT solvers have evolved beyond this framework to a framework called \textit{conflict-driven clause learning} (CDCL). Upon reaching a contradiction, they extend the formula with a clause expressing what combination of variables led to the contradiction, so that the mistake is prevented later on. Moreover, instead of simply backtracking up one level, they can realize that several previous levels cannot contain a solution, and they ``backjump'' up multiple levels. This more general approach yields better performance in practice~\cite{silva2021conflict}. This means our classical baseline should not be a backtracking algorithm, but rather one of these more general CDCL solvers. Importantly, this implies that the almost quadratic speedup that quantum backtracking yields is relative to a subpar classical algorithm, harming the possibility of a quantum speedup.

As a classical baseline, we chose the SBVA-CaDiCaL solver: the winner of the 2023 SAT competition, which was the most recent SAT competition at the time.\footnote{Results of the 2023 SAT competition can be found at \url{https://satcompetition.github.io/2023/results.html}.} This algorithm first pre-processes a SAT instance using the SBVA algorithm~\cite{haberlandt2023effective}, and then solves the processed instance using the CaDiCaL 1.9.5 solver~\cite{biere2020cadical}. Our experiments involve SAT instances with varying and increasing clause sizes $k$. We noticed that for $k \geq 4$, the pre-processor started to dominate the runtime and slow down the solving process. Hence, we decided to drop the pre-processing and simply run CaDiCaL 1.9.5 by itself. During our work, version 2.0.0 of CaDiCaL was released, so we opted to use this version instead \cite{BiereFallerFazekasFleuryFroleyks-CAV24}.
For this algorithm, the complexity measure will simply be the observed runtime.

\subsection{Instances}\label{sect:instances}
\begin{table*}[t]
\centering
    \begin{tabular}{c|cccccccccc}
        $k$ & 3 & 4 & 5 & 6 & 7 & 8 & 9 & 10 & 11 & 12 \\\hline
        $\alpha_k$ & 4.27 & 9.93 & 21.12 & 43.27 & 87.79 & 176.54 & 354.01 & 708.92 & 1418.71 & 2838.28
    \end{tabular}
    \caption{Satisfiability threshold ratios $\alpha_k=m/n$ of clauses and variables in a random $k$-SAT instance, as taken from Table~10 in Ref.~\cite{campbell2019applying}.}
    \label{tab:ratios}
\end{table*}
We consider random $k$-SAT instances, which we generate using the CNFgen tool.\footnote{See \url{https://massimolauria.net/cnfgen/}.} Specifically, each clause is picked uniformly randomly and independently from all possible clauses, excluding those which contain duplicate variables. In addition, we consider random SAT instances that have an increasing amount of ``structure." There is an increasing body of work substantiating the notion of structure in SAT instances~\cite{alyahya2023structure}. We use a simplified version of the similarity-popularity model from~\cite{levy2017locality}, which are specifically designed to exhibit some characteristics of many real-world SAT instances, which works as follows. Say we wish to sample a SAT formula with $n$ variables and $m$ clauses. Assign each variable $x_i$ and clause $j$ a uniformly random angle $\theta \in [0,2\pi)$. To sample the formula, iterate over each clause $j$ and sample its $k$ variables according to the following distribution:
\[
\mathbb{P}[x_i \text{ in clause } j] = \frac{1}{1+\left(\dfrac{i (\pi - |\pi - |\theta_i - \theta_j||)}{N}\right)^{1/\beta}},
\]
where $N$ is a normalization coefficient. When a variable is added to a clause, it is negated with probability $1/2$. Intuitively, the occurrence of $i$ in the denominator means we expect variable $x_{i}$ to occur much more often than variable $x_{i+1}$ (``popularity"), while $|\theta_i - \theta_j|$ means variables and clauses with similar angles are more likely to end up together (``similarity"). We get to set the parameter $\beta$.\footnote{In~\cite{levy2017locality}, $\beta$ is denoted by $T$ as it can intuitively be understood as some notion of ``temperature,'' but we have switched notation to prevent confusion with the ``$T$-gate.'' Similarly, they denote the normalization coefficient $N$ by $R$, which we have changed to prevent confusion with the effective resistance $R$.} As we increase $\beta$, we decrease the popularity and similarity effects, and hence the exploitable structure in the instances. In particular, as $\beta \to \infty$, we recover the uniformly random model outlined above. 

For both models, we consider $3$, $4$, $\dots$, $12$-SAT, choosing the number of clauses $m$ such that the ratio $m/n$ is the threshold of satisfiability $\alpha_k$. This is defined so that as $n \to \infty$, a random $k$-SAT instance is satisfiable with probability 1 for any $\alpha < \alpha_k$ and with probability 0 for any $\alpha > \alpha_k$. The existence of such thresholds (known as the Satisfiability Conjecture) has been proven for large $k$~\cite{ding2015proof}; however, for small values of $k \geq 3$, only bounds are known. We list the ratios we use in~\cref{tab:ratios}, which we took from Table~10 in Ref.~\cite{campbell2019applying}.

We argue that increasing $k$ is another method of reducing exploitable structure. First, SAT solvers have a harder time making use of larger clauses, as finding contradictions (i.e., clauses that are unsatisfied by a current assignment) or inferring that certain variables must be true (i.e., through unit propagation -- the observation that if a clause contains a single non-unsatisfied literal, that literal must be set to true) will be harder. Additionally, the satisfiability threshold is known to grow exponentially with $k$, implying that (for random $k$-SAT), increasing $k$ allows you to add exponentially more constraints without harming satisfiability. Finally, this is reflected in the so-called \emph{strong exponential time hypothesis} (SETH)~\cite{impagliazzo2001complexity}. Let $s_k = \inf \left\{c : \text{$k$-SAT can be solved in } 2^{c n} \poly(m) \text{ time} \right\}$. The strong exponential time hypothesis asserts that for every $\epsilon >0$, there exists a $k$ such that $s_k \geq 1-\epsilon$, i.e.~, $\lim_{k \to \infty} s_k = 1$. Therefore, as $k$ becomes very large, this hypothesis conjectures that the optimal SAT solver approaches a scaling of $\sim 2^n \poly(m)$, matching brute-force search. 

Hence, varying $k$ and/or $\beta$ allows us to create distributions of SAT instances that are expected to have varying amounts of exploitable structure. 

\subsection{Cost model}
The cost of the classical SAT solver will simply be the measured runtime. For the quantum algorithms, we evaluate a query complexity upper bound for the given SAT instance, which for the considered algorithms can be directly translated to the \emph{$T$-count} or \emph{$T$-depth}, due to the results in Ref.~\cite{campbell2019applying}. While Ref.~\cite{campbell2019applying} takes $T$-depth of the resulting quantum circuit as their measure of complexity, we instead optimize for the both the $T$-count and $T$-depth. Let us summarize briefly what these two measures are. 

We assume the quantum circuit implementing the algorithm is decomposed in Clifford, Toffoli and $T$-gates, and will be encoded using the surface code~\cite{fowler2012surface}. Clifford gates can be implemented using state injection, which can be done parallel to the implementation of a subsequent Tofolli or $T$ gate. Since Toffoli gates can be implemented using a single layer of $T$ gates~\cite{selinger2013quantum} (using a total of 7 $T$-gates), Ref.~\cite{campbell2019applying} argues that the only relevant gate count for this cost model is then the so-called \emph{$T$-depth} of the circuit, which is defined as the total number of groups of $T$-gates which can be performed simultaneously. By using time-optimal methods~\cite{fowler2012time}, $T$-gates can be implemented fault-tolerantly using a state injection technique where a $T$-state is prepared offline. This comes at the cost of a huge overhead in the total number of qubits that are needed, but this means that every single layer of $T$-gates can be implemented at a time roughly on the order of the measurement time. Hence, the total time cost in this model is simply assumed to be the $T$-depth multiplied by the measurement cost. They also assume that repetitions of the quantum algorithms (to amplify success probability) are done in parallel, adding nothing to the total runtime. By considering measurement times of $10^{-7},10^{-8}$ and $10^{-9}$ seconds (which they call plausible, realistic, and optimistic) they give indications of the quantum runtime.

Recall that we use the same $P$ and $h$ oracles as Ref.~\cite{campbell2019applying}. As such, we can use the circuit implementations of Ref.~\cite{campbell2019applying} to translate our  query complexity upper bound to a gate complexity upper bound. However, we believe \emph{$T$-count}, which is simply the total number of $T$ gates, might be a fairer cost model than $T$-depth, as using the latter requires significant parallelization, which is not granted to the classical algorithms. However, to allow for fair comparison with Ref.~\cite{campbell2019applying} and to leave the final judgement of what cost model is most relevant to the reader, we will present results on both cost models.

We will write $c_{\textup{q}}$ for the time needed to implement a single (layer of) $T$-gate(s). Finally, we assume repetitions of the quantum algorithms (which we need to do to guarantee the required success probability, see \cref{appendix:belovs_bounds}) are all done in serial instead of parallel, as done in Ref.~\cite{campbell2019applying}, to further limit the parallelization in the quantum algorithm. 

\section{Benchmarking results}
\begin{figure}[!h]
\hspace{-1.5cm}
\includegraphics[width=18cm]{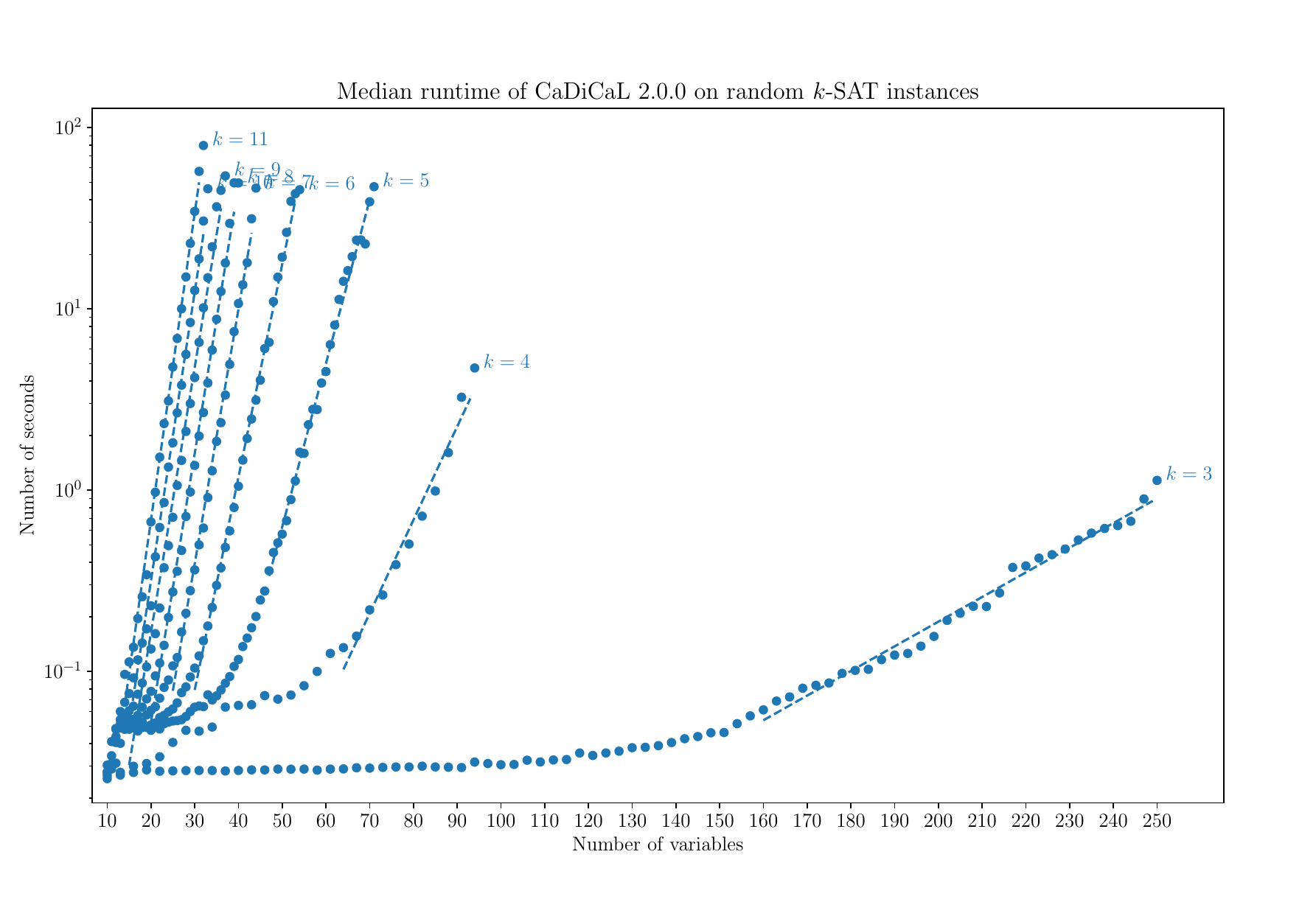}
\caption{Scatter plot of median runtime for the CaDiCaL 2.0.0 solver on random $k$-SAT instances at the satisfiability threshold, see \cref{tab:ratios}. The dotted line indicates the exponential fit of the running time. For small $k$ you can see that we only start fitting the runtime for larger $n$, as the runtime seems to behave differently for small $n$.}
\label{fig:clas_fits}
\end{figure}

We sampled SAT instances for the values of $n$, $\beta$, $k$ listed in \cref{tab:n_ranges}, all at the satisfiability threshold ratio between the number of variables and the number of clauses, as given in~\cref{tab:ratios}.\footnote{We note that the lower bound on $n$ for fitting the classical runtime is often quite high. For small $k$, the runtime initially grows quite fast, and only then stabilises into a pattern of the form $2^{a\cdot n +b}$. We attempted to lower bound $n$ to exclude the small $n$ which behave erratically. This is visualised in \cref{fig:clas_fits}.} For an equal number of satisfiable and unsatisfiable instances (30 each for backtracking, 50 each for CaDiCaL), we compute the median runtime (CaDiCaL) or median $T$-depth and $T$-count (quantum algorithms) for each $n$, and then take a linear least squares fit of the logarithm of these medians. We configure the quantum backtracking algorithms to run with error probability at most $\delta=1/1000$.\footnote{E.g.~Ref.~\cite{campbell2019applying} uses $\delta=1/10$, providing a bigger advantage to the quantum algorithms. We recover their results despite this extra disadvantage.}

This ends up giving us a grid over $k$ and $\beta$. For each entry $(k,\beta)$ we have a fit of the classical runtime, denoted as $2^{s_\textup{c} \cdot n + i_\textup{c}}$, and a fit of the quantum algorithm's $T$-depth or $T$-count denoted $2^{s_\textup{q} \cdot n + i_q}$ (where we use this notation for both $T$-count and $T$-depth, though from context it will be clear which we mean), where $s_\textup{c}$, $i_\textup{c}$, $s_\textup{q}$, $i_\textup{q} \in \mathbb{R}$. The $T$-depth or $T$-count can then directly be converted to a run-time estimate by $c_\textup{q} 2^{s_\textup{q} \cdot n + i_q}$ for some constant $c_\textup{q} > 0$.

Recall that we think of increasing $k$ and $\beta$ as decreasing the exploitable structure in the instances. Hence, we think of the top-left portion of this grid as being very structured, and the bottom-right portion as highly unstructured. It is exactly this bottom-right (i.e.~ random instances for $k=12$) where \cite{campbell2019applying} observe a quantum speedup. The question becomes: what happens as we exit this bottom-right part, and increase the amount of exploitable structure? Studying these results yields three main conclusions.

Our code and data is attached as Ref.~\cite{code}.

\subsection{The four regimes}
\begin{table*}[!h]
    \centering
    $T$\textbf{-depth}\\[5pt] 
    \hspace{-1.3cm}
    \begin{subtable}[t]{0.363\textwidth}
        \begin{tabular}{p{0.5 cm}|p{0.22 cm}p{0.22 cm}p{0.22 cm}p{0.22 cm}p{0.22 cm}p{0.22 cm}p{0.22 cm}p{0.22 cm}}
            \resizebox{0.5cm}{!}{\backslashbox{$k$}{$\beta$}} & $\frac 12$ & 1 & $\frac 32$ & 2 & 3 & 5 & 10 & $\infty$\\\hline
            3& \cellcolor{C0}& \cellcolor{C0}& \cellcolor{C0}& \cellcolor{C0}& \cellcolor{C0} & \cellcolor{C0} & \cellcolor{C0} & \cellcolor{C0}\\
            4& \cellcolor{C0}& \cellcolor{C0}& \cellcolor{C0}& \cellcolor{C0} & \cellcolor{C0} & \cellcolor{C0} & \cellcolor{C0} & \cellcolor{C0}\\
            5& \cellcolor{C0}& \cellcolor{C0}& \cellcolor{C0} & \cellcolor{C0} & \cellcolor{C0} & \cellcolor{C0} & \cellcolor{C0} & \cellcolor{C0}\\
            6& \cellcolor{C0}& \cellcolor{C0} & \cellcolor{C0} & \cellcolor{C0} & \cellcolor{C0} & \cellcolor{C1} & \cellcolor{C1} & \cellcolor{C1}\\
            7 & \cellcolor{C0} & \cellcolor{C0} & \cellcolor{C0} & \cellcolor{C0} & \cellcolor{C1} & \cellcolor{C1} & \cellcolor{C2} & \cellcolor{C2}\\
            8 & \cellcolor{C0} & \cellcolor{C0} & \cellcolor{C0} & \cellcolor{C0} & \cellcolor{C1} & \cellcolor{C1} & \cellcolor{C2} & \cellcolor{C2}\\
            9 & \cellcolor{C0} & \cellcolor{C0} & \cellcolor{C0} & \cellcolor{C0} & \cellcolor{C0} & \cellcolor{C1} & \cellcolor{C1} & \cellcolor{C2}\\
            10 & \cellcolor{C0} & \cellcolor{C0} & \cellcolor{C0} & \cellcolor{C1} & \cellcolor{C1} & \cellcolor{C2} & \cellcolor{C2} & \cellcolor{C2}\\
            11 & \cellcolor{C0} & \cellcolor{C0} & \cellcolor{C1} & \cellcolor{C1} & \cellcolor{C2} & \cellcolor{C3} & \cellcolor{C3} & \cellcolor{C3}\\
            12 & \cellcolor{C0} & \cellcolor{C0} & \cellcolor{C0} & \cellcolor{C1} & \cellcolor{C2} & \cellcolor{C3} & \cellcolor{C3} & \cellcolor{C2}\\
        \end{tabular}
        \caption{Scaling}
        \label{tab:colors_scaling_depth}
    \end{subtable}
    \hfill
    \begin{subtable}[t]{0.3\textwidth}
        \begin{tabular}{p{0.22 cm}p{0.22 cm}p{0.22 cm}p{0.22 cm}p{0.22 cm}p{0.22 cm}p{0.22 cm}p{0.22 cm}}
             $\frac 12$ & 1 & $\frac 32$ & 2 & 3 & 5 & 10 & $\infty$\\\hline
            \cellcolor{C0}& \cellcolor{C0}& \cellcolor{C0}& \cellcolor{C0}& \cellcolor{C0} & \cellcolor{C0} & \cellcolor{C0} & \cellcolor{C0}\\
             \cellcolor{C0}& \cellcolor{C0}& \cellcolor{C0}& \cellcolor{C0} & \cellcolor{C0} & \cellcolor{C0} & \cellcolor{C0} & \cellcolor{C0}\\
             \cellcolor{C0}& \cellcolor{C0}& \cellcolor{C0} & \cellcolor{C0} & \cellcolor{C0} & \cellcolor{C0} & \cellcolor{C0} & \cellcolor{C0}\\
             \cellcolor{C0}& \cellcolor{C0} & \cellcolor{C0} & \cellcolor{C0} & \cellcolor{C0} & \cellcolor{C0} & \cellcolor{C0} & \cellcolor{C0}\\
             \cellcolor{C0} & \cellcolor{C0} & \cellcolor{C0} & \cellcolor{C0} & \cellcolor{C0} & \cellcolor{C0} & \cellcolor{C3} & \cellcolor{C3}\\
             \cellcolor{C0} & \cellcolor{C0} & \cellcolor{C0} & \cellcolor{C0} & \cellcolor{C0} & \cellcolor{C3} & \cellcolor{C3} & \cellcolor{C3}\\
             \cellcolor{C0} & \cellcolor{C0} & \cellcolor{C0} & \cellcolor{C0} & \cellcolor{C0} & \cellcolor{C3} & \cellcolor{C3} & \cellcolor{C3}\\
             \cellcolor{C0} & \cellcolor{C0} & \cellcolor{C0} & \cellcolor{C0} & \cellcolor{C3} & \cellcolor{C3} & \cellcolor{C3} & \cellcolor{C3}\\
             \cellcolor{C0} & \cellcolor{C0} & \cellcolor{C3} & \cellcolor{C3} & \cellcolor{C3} & \cellcolor{C3} & \cellcolor{C3} & \cellcolor{C3}\\
             \cellcolor{C0} & \cellcolor{C0} & \cellcolor{C3} & \cellcolor{C3} & \cellcolor{C3} & \cellcolor{C3} & \cellcolor{C3} & \cellcolor{C3}\\
        \end{tabular}
        \caption{One day, $c_\textup{q}=10^{-9}$}
        \label{tab:colors_one_day_9_depth}
    \end{subtable}
    \hfill
    \begin{subtable}[t]{0.3\textwidth}
        \centering
        \begin{tabular}{p{0.22 cm}p{0.22 cm}p{0.22 cm}p{0.22 cm}p{0.22 cm}p{0.22 cm}p{0.22 cm}p{0.22 cm}}
             $\frac 12$ & 1 & $\frac 32$ & 2 & 3 & 5 & 10 & $\infty$\\\hline
            \cellcolor{C0}& \cellcolor{C0}& \cellcolor{C0}& \cellcolor{C0}& \cellcolor{C0} & \cellcolor{C0} & \cellcolor{C0} & \cellcolor{C0}\\
            \cellcolor{C0}& \cellcolor{C0}& \cellcolor{C0}& \cellcolor{C0} & \cellcolor{C0} & \cellcolor{C0} & \cellcolor{C0} & \cellcolor{C0}\\
            \cellcolor{C0}& \cellcolor{C0}& \cellcolor{C0} & \cellcolor{C0} & \cellcolor{C0} & \cellcolor{C0} & \cellcolor{C0} & \cellcolor{C0}\\
            \cellcolor{C0}& \cellcolor{C0} & \cellcolor{C0} & \cellcolor{C0} & \cellcolor{C0} & \cellcolor{C0} & \cellcolor{C0} & \cellcolor{C0}\\
            \cellcolor{C0} & \cellcolor{C0} & \cellcolor{C0} & \cellcolor{C0} & \cellcolor{C0} & \cellcolor{C0} & \cellcolor{C0} & \cellcolor{C0}\\
            \cellcolor{C0} & \cellcolor{C0} & \cellcolor{C0} & \cellcolor{C0} & \cellcolor{C0} & \cellcolor{C0} & \cellcolor{C0} & \cellcolor{C0}\\
             \cellcolor{C0} & \cellcolor{C0} & \cellcolor{C0} & \cellcolor{C0} & \cellcolor{C0} & \cellcolor{C0} & \cellcolor{C0} & \cellcolor{C0}\\
             \cellcolor{C0} & \cellcolor{C0} & \cellcolor{C0} & \cellcolor{C0} & \cellcolor{C0} & \cellcolor{C0} & \cellcolor{C0} & \cellcolor{C0}\\
             \cellcolor{C0} & \cellcolor{C0} & \cellcolor{C0} & \cellcolor{C0} & \cellcolor{C0} & \cellcolor{C3} & \cellcolor{C3} & \cellcolor{C3}\\
             \cellcolor{C0} & \cellcolor{C0} & \cellcolor{C0} & \cellcolor{C0} & \cellcolor{C3} & \cellcolor{C3} & \cellcolor{C3} & \cellcolor{C3}\\
        \end{tabular}
        \caption{One day, $c_\textup{q}=10^{-6}$}
        \label{tab:colors_one_day_6_depth}
    \end{subtable}
    \vspace{10pt}
    
    $T$\textbf{-count}\\[5pt]
    \hspace{-1.3cm}
    \begin{subtable}[t]{0.363\textwidth}
        \centering
        \begin{tabular}{p{0.5 cm}|p{0.22 cm}p{0.22 cm}p{0.22 cm}p{0.22 cm}p{0.22 cm}p{0.22 cm}p{0.22 cm}p{0.22 cm}}
        \resizebox{0.5cm}{!}{\backslashbox{$k$}{$\beta$}} & $\frac 12$ & 1 & $\frac 32$ & 2 & 3 & 5 & 10 & $\infty$\\\hline
        3& \cellcolor{C0}& \cellcolor{C0}& \cellcolor{C0}& \cellcolor{C0}& \cellcolor{C0} & \footnotesize \cellcolor{C0}  & \footnotesize \cellcolor{C0}  & \footnotesize \cellcolor{C0} \\
        4& \cellcolor{C0}& \cellcolor{C0}& \cellcolor{C0}& \cellcolor{C0}& \cellcolor{C0} & \footnotesize \cellcolor{C0}  & \footnotesize \cellcolor{C0}  & \footnotesize \cellcolor{C0} \\
        5& \cellcolor{C0}& \cellcolor{C0}& \cellcolor{C0}& \cellcolor{C0}& \cellcolor{C0} & \footnotesize \cellcolor{C0}  & \footnotesize \cellcolor{C0}  & \footnotesize \cellcolor{C0} \\
        6& \cellcolor{C0}& \cellcolor{C0}& \cellcolor{C0}& \cellcolor{C0}& \cellcolor{C0} & \footnotesize \cellcolor{C0}  & \footnotesize \cellcolor{C0}  & \footnotesize \cellcolor{C0} \\
        7& \cellcolor{C0}& \cellcolor{C0}& \cellcolor{C0} & \footnotesize \cellcolor{C0}  & \footnotesize \cellcolor{C0}  & \footnotesize \cellcolor{C0}  & \footnotesize \cellcolor{C0}  & \footnotesize \cellcolor{C0} \\
        8& \cellcolor{C0}& \cellcolor{C0}& \cellcolor{C0} & \footnotesize \cellcolor{C0}  & \footnotesize \cellcolor{C0}  & \footnotesize \cellcolor{C0}  & \footnotesize \cellcolor{C0}  & \footnotesize \cellcolor{C0} \\
        9& \cellcolor{C0}& \cellcolor{C0}& \cellcolor{C0} & \footnotesize \cellcolor{C0}  & \footnotesize \cellcolor{C0}  & \footnotesize \cellcolor{C0}  & \footnotesize \cellcolor{C0}  & \footnotesize \cellcolor{C0} \\
        10& \cellcolor{C0}& \cellcolor{C0}& \cellcolor{C0} & \footnotesize \cellcolor{C0}  & \footnotesize \cellcolor{C0}  & \footnotesize \cellcolor{C1}  & \footnotesize \cellcolor{C1}  & \footnotesize \cellcolor{C1} \\
        11& \cellcolor{C0}& \cellcolor{C0}& \cellcolor{C0} & \footnotesize \cellcolor{C0}  & \footnotesize \cellcolor{C1}  & \footnotesize \cellcolor{C1}  & \footnotesize \cellcolor{C1}  & \footnotesize \cellcolor{C1} \\
        12& \cellcolor{C0}& \cellcolor{C0}& \cellcolor{C0} & \footnotesize \cellcolor{C0}  & \footnotesize \cellcolor{C1}  & \footnotesize \cellcolor{C1}  & \footnotesize \cellcolor{C1}  & \footnotesize \cellcolor{C2} \\
        \end{tabular}
        \caption{Scaling}
        \label{tab:colors_scaling_count}
    \end{subtable}
    \hfill
    \begin{subtable}[t]{0.3\textwidth}
        \centering
        \begin{tabular}{p{0.22 cm}p{0.22 cm}p{0.22 cm}p{0.22 cm}p{0.22 cm}p{0.22 cm}p{0.22 cm}p{0.22 cm}}
        $\frac 12$ & 1 & $\frac 32$ & 2 & 3 & 5 & 10 & $\infty$\\\hline
        \cellcolor{C0}& \cellcolor{C0}& \cellcolor{C0}& \cellcolor{C0}& \cellcolor{C0} & \footnotesize \cellcolor{C0}  & \footnotesize \cellcolor{C0}  & \footnotesize \cellcolor{C0} \\
        \cellcolor{C0}& \cellcolor{C0}& \cellcolor{C0}& \cellcolor{C0}& \cellcolor{C0} & \footnotesize \cellcolor{C0}  & \footnotesize \cellcolor{C0}  & \footnotesize \cellcolor{C0} \\
        \cellcolor{C0}& \cellcolor{C0}& \cellcolor{C0}& \cellcolor{C0}& \cellcolor{C0} & \footnotesize \cellcolor{C0}  & \footnotesize \cellcolor{C0}  & \footnotesize \cellcolor{C0} \\
        \cellcolor{C0}& \cellcolor{C0}& \cellcolor{C0}& \cellcolor{C0}& \cellcolor{C0} & \footnotesize \cellcolor{C0}  & \footnotesize \cellcolor{C0}  & \footnotesize \cellcolor{C0} \\
        \cellcolor{C0}& \cellcolor{C0}& \cellcolor{C0} & \footnotesize \cellcolor{C0}  & \footnotesize \cellcolor{C0}  & \footnotesize \cellcolor{C0}  & \footnotesize \cellcolor{C0}  & \footnotesize \cellcolor{C0} \\
        \cellcolor{C0}& \cellcolor{C0}& \cellcolor{C0} & \footnotesize \cellcolor{C0}  & \footnotesize \cellcolor{C0}  & \footnotesize \cellcolor{C0}  & \footnotesize \cellcolor{C0}  & \footnotesize \cellcolor{C0} \\
        \cellcolor{C0}& \cellcolor{C0}& \cellcolor{C0} & \footnotesize \cellcolor{C0}  & \footnotesize \cellcolor{C0}  & \footnotesize \cellcolor{C0}  & \footnotesize \cellcolor{C0}  & \footnotesize \cellcolor{C0} \\
        \cellcolor{C0}& \cellcolor{C0}& \cellcolor{C0} & \footnotesize \cellcolor{C0}  & \footnotesize \cellcolor{C0}  & \footnotesize \cellcolor{C0}  & \footnotesize \cellcolor{C0}  & \footnotesize \cellcolor{C0} \\
        \cellcolor{C0}& \cellcolor{C0}& \cellcolor{C0} & \footnotesize \cellcolor{C0}  & \footnotesize \cellcolor{C0}  & \footnotesize \cellcolor{C0}  & \footnotesize \cellcolor{C0}  & \footnotesize \cellcolor{C0} \\
        \cellcolor{C0}& \cellcolor{C0}& \cellcolor{C0} & \footnotesize \cellcolor{C0}  & \footnotesize \cellcolor{C0}  & \footnotesize \cellcolor{C0}  & \footnotesize \cellcolor{C0}  & \footnotesize \cellcolor{C0} \\
        \end{tabular}
        \caption{One day, $c_\textup{q}=10^{-9}$}
        \label{tab:colors_one_day_9_count}
    \end{subtable}
    \hfill
    \begin{subtable}[t]{0.3\textwidth}
        \centering
        \begin{tabular}{p{0.22 cm}p{0.22 cm}p{0.22 cm}p{0.22 cm}p{0.22 cm}p{0.22 cm}p{0.22 cm}p{0.22 cm}}
        $\frac 12$ & 1 & $\frac 32$ & 2 & 3 & 5 & 10 & $\infty$\\\hline
        \cellcolor{C0}& \cellcolor{C0}& \cellcolor{C0}& \cellcolor{C0}& \cellcolor{C0} & \footnotesize \cellcolor{C0}  & \footnotesize \cellcolor{C0}  & \footnotesize \cellcolor{C0} \\
        \cellcolor{C0}& \cellcolor{C0}& \cellcolor{C0}& \cellcolor{C0}& \cellcolor{C0} & \footnotesize \cellcolor{C0}  & \footnotesize \cellcolor{C0}  & \footnotesize \cellcolor{C0} \\
        \cellcolor{C0}& \cellcolor{C0}& \cellcolor{C0}& \cellcolor{C0}& \cellcolor{C0} & \footnotesize \cellcolor{C0}  & \footnotesize \cellcolor{C0}  & \footnotesize \cellcolor{C0} \\
        \cellcolor{C0}& \cellcolor{C0}& \cellcolor{C0}& \cellcolor{C0}& \cellcolor{C0} & \footnotesize \cellcolor{C0}  & \footnotesize \cellcolor{C0}  & \footnotesize \cellcolor{C0} \\
        \cellcolor{C0}& \cellcolor{C0}& \cellcolor{C0}& \cellcolor{C0}& \cellcolor{C0} & \footnotesize \cellcolor{C0}  & \footnotesize \cellcolor{C0}  & \footnotesize \cellcolor{C0} \\
        \cellcolor{C0}& \cellcolor{C0}& \cellcolor{C0} & \footnotesize \cellcolor{C0}  & \footnotesize \cellcolor{C0}  & \footnotesize \cellcolor{C0}  & \footnotesize \cellcolor{C0}  & \footnotesize \cellcolor{C0} \\
        \cellcolor{C0}& \cellcolor{C0}& \cellcolor{C0} & \footnotesize \cellcolor{C0}  & \footnotesize \cellcolor{C0}  & \footnotesize \cellcolor{C0}  & \footnotesize \cellcolor{C0}  & \footnotesize \cellcolor{C0} \\
        \cellcolor{C0}& \cellcolor{C0}& \cellcolor{C0} & \footnotesize \cellcolor{C0}  & \footnotesize \cellcolor{C0}  & \footnotesize \cellcolor{C0}  & \footnotesize \cellcolor{C0}  & \footnotesize \cellcolor{C0} \\
        \cellcolor{C0}& \cellcolor{C0}& \cellcolor{C0} & \footnotesize \cellcolor{C0}  & \footnotesize \cellcolor{C0}  & \footnotesize \cellcolor{C0}  & \footnotesize \cellcolor{C0}  & \footnotesize \cellcolor{C0} \\
        \cellcolor{C0}& \cellcolor{C0}& \cellcolor{C0} & \footnotesize \cellcolor{C0}  & \footnotesize \cellcolor{C0}  & \footnotesize \cellcolor{C0}  & \footnotesize \cellcolor{C0}  & \footnotesize \cellcolor{C0} \\
        \end{tabular}
        \caption{One day, $c_\textup{q}=10^{-6}$}
        \label{tab:colors_one_day_6_count}
    \end{subtable}

    \caption{Each grid in these tables represents a distribution of $k$-SAT instances with differing amounts of structure $\beta$, fixed to contain an equal number of satisfiable and unsatisfiable instances (where lower $\beta$ means more structure). Cells are coloured based on the best algorithm for the given performance measure. In \cref{tab:colors_scaling_count,tab:colors_scaling_depth} we colour a cell \textcolor{C0}{\textbf{blue}} if the classical algorithm \emph{scales} best (i.e. $s_\textup{c}<s_\textup{q}$), \textcolor{C1}{\textbf{red}} if the quantum backtracking \emph{detection} algorithm scales best, \textcolor{C3}{\textbf{yellow}} if Grover scales best, and \textcolor{C2}{\textbf{green}} if quantum backtracking \emph{search} scales better than classical and Grover. In the other tables, we translate $T$-depth and $T$-count to runtime using two measurement times $c_\textup{q}=10^{-9}$ and $c_\textup{q}=10^{-6}$ and colour based on which algorithm can solve the largest instances in one day given their fits (using the same colour codes).}
    \label{tab:results_colors}
\end{table*}

Consider \cref{tab:colors_scaling_depth}, which displays which algorithm achieves the best scaling in terms of runtime (classical) and $T$-depth (quantum): the lowest constant $s$ in $2^{s\cdot n + i}$. Across the grid, we observe 4 somewhat distinct regimes of SAT instances, reminiscent of a \textit{phase diagram} as encountered in physics. For highly structured instances (top-left) the classical SAT solver easily scales better than all quantum algorithms (\textcolor{C0}{\textbf{blue}}). Moving to the bottom-right, removing some structure, we enter a regime where the quantum backtracking detection algorithm (studied in \cite{campbell2019applying}) starts to scale better than classical (\textcolor{C1}{\textbf{red}}). However, recall that this algorithm does not find a solution unlike all the other considered algorithms, and requires a priori knowledge of the tree size $\mathcal{T}$ to run. To fix this, we needed to turn it into a binary search algorithm, allowing us to then estimate $\mathcal{T}$, suffering an $\mO(n^2\log n)$ overhead. As we remove even more structure from the instances there comes a regime where this binary search algorithm also starts to scale better than the classical SAT solver (\textcolor{C2}{\textbf{green}}). Finally, as we go towards instances that contain barely any exploitable structure, we reach a point where Grover's algorithm starts to scale better than any of the other algorithms (\textcolor{C3}{\textbf{yellow}}).\footnote{Except for the one outlier exactly at the right-bottom corner.} 

These four regimes are in accordance with the SETH: as $k$ grows we expect the classical scaling should approach $\sim 2^n \poly(m)$ for both the modern SAT solver as well as any backtracking algorithm. Quantum backtracking detection will gain a less-than quadratic speedup over this, while the search algorithm will scale even slightly worse. Grover will then eventually scale best, as it achieves a full quadratic speedup over classical. We can think of increasing $\beta$ as having a similar effect: removing structure harms the modern classical solver most, then the detection algorithm, and then the search algorithm, while Grover is agnostic to all of this.

Some representative examples of these regimes are given in~\cref{fig:snippet}. This table shows the estimated \emph{scaling} of the four different algorithms (in terms of runtime for classical and in terms of $T$-depth for quantum) for several selected values of $\beta$.  These four regimes indicate that the classical algorithm outperforms the quantum algorithms on the more structured instances. Though the quantum backtracking algorithm is also able to exploit structure, much of this gain is lost when one is also interested in finding the solution. Only when barely any structure is left does Grover start beating the other algorithms. This means there might be a place for quantum backtracking to solve semi-structured practical SAT instances that are too large for classical solvers. 

However, so far we only discussed asymptotic scaling. Converting the $T$-depth estimates to actual estimated runtimes, the potential for quantum backtracking quickly diminishes. Consider three different $T$-gate implementation times $c_{\textup{q}}$: $10^{-6},10^{-7}$ and $10^{-8}$ seconds.\footnote{We generally omit the unit `seconds'' when talking about values of $c_{\textup{q}}$ from now on.} In~\cref{tab:constants} we use these three values to compute the crossover time for the representative quantum algorithms in these four regimes. The \textit{crossover time} is the time at which the classical runtime matches the quantum runtime (quantum suffers more constant overhead than classical but scales better, so after enough time has passed quantum should obtain an advantage). Formally, we solve $2^{s_\textup{c} \cdot n + i_\textup{c}} =c_\textup{q} \cdot 2^{s_\textup{q} \cdot n + i_q}$ for $n$ and then evaluate either side for that $n$.

\begin{table*}[t]
\begin{tabular}{c|c|c|c|c|}
& \color{white}{High $(\beta=0.5)$} \cellcolor{C0} & \color{white}{Middle $(\beta=2)$} \cellcolor{C1} & Low $(\beta=3)$  \cellcolor{C2} & None ($\beta=\infty$)  \cellcolor{C3} \\\hline
Classical &  $\mathbf{0.0804}$  & $0.432$  & $0.507$  & $0.563$ \\
Detection & $0.282$  & $\mathbf{0.368}$ & $0.394$  & $0.424$ \\
Search & $0.366$  & $0.461$  & $\mathbf{0.46}$  & $0.514$ \\
Grover & $0.509$  & $0.511$  & $0.513$ & $\mathbf{0.509}$ \\\hline
\end{tabular}
    \centering
    \caption{Exponent $s$ of the estimated runtime $2^{sn+i}$ for classical algorithm and $T$-depth for quantum algorithms on $11$-SAT with an equal mix of satisfiable and unsatisfiable instances. For the constant overhead $i$ suffered by the algorithms, see \cref{tab:colors_both_depth}. The values of $\beta$ refer to the amount of structure in the SAT instances. Recall that in the ``High" structure regime, the classical solver scales best. In the ``Middle" structure regime detection scales best. In the ``Low" structure regime, binary search starts to beat classical and Grover (though it never beats detection, as it runs detection multiple times). Finally, in the "None" structure case Grover starts scaling better than both binary search and classical.}
    \label{fig:snippet}
\end{table*}

\begin{table*}[t]
\begin{tabular}{c|c|c|c|c|}
& \color{white}{High $(\beta=0.5)$} \cellcolor{C0} & \color{white}{Middle $(\beta=2)$} \cellcolor{C1} & Low $(\beta=3)$  \cellcolor{C2} & None ($\beta=\infty$)  \cellcolor{C3} \\\hline
$c_\textup{q} = 10^{-6}$ & n.a.  & $\approx$ universe age  & $\gg$ universe age  &  $7$ hours \\
$c_\textup{q} = 10^{-7}$ &  n.a.  &  3 millennia  &  $\gg $ universe age & always \\
$c_\textup{q} = 10^{-8}$ &  n.a.  & 5 hours  &  $\gg$ universe age & always \\
\hline
\end{tabular}
    \centering
    \caption{Crossover times as per~\cref{eq:crossover} for different values of $c_\textup{q}$. In every column, we compare the classical algorithm to the quantum algorithm representative of the given regime, i.e. whose scaling is bolded in~\cref{fig:snippet}. We translate the estimates of the $T$-depth of the quantum algorithms to runtime by multiplying with $c_\textup{q}$.  The values of $\beta$ refer to the amount of structure in the SAT instances.}
    \label{tab:constants}
\end{table*}

Although with a very optimistic $c_{\textup{q}} = 10^{-8}$ the detection algorithm can achieve a crossover time of a few hours, for larger $c_{\textup{q}}$ it quickly grows to be infeasible. It may seem surprising that the crossover time can change so drastically when $c_{\textup{q}}$ drops by one order of magnitude. However, as we show in~\cref{appendix:crossover_time}, solving the above for the crossover time gives:
\begin{align}
t_\textup{crossover}
&= c_\textup{q}^{ \frac{s_\textup{q}}{s_\textup{c}-s_\textup{q}}+1} \left(2^{\frac{s_\textup{q}(i_q - i_c)}{s_\textup{c}-s_\textup{q}}+i_q}\right).
\label{eq:crossover}
\end{align}
This shows that when the scaling of the quantum and classical algorithm are close, small changes in $c_\textup{q}$ can drastically change~\eqref{eq:crossover}. Far more sobering, the more practically usable backtracking search algorithm has a crossover time far exceeding the age of the universe for all constants. This was not just the case in this one example of the regimes where binary search scales better than classical: in all such regimes in~\cref{tab:colors_both_depth} we observe a crossover time of at least 10000 millennia. Even when going to measurement time $c_q = 10^{-9}$, the crossover time is still at least a year, and often still many, many millennia.

In contrast, Grover's algorithm has a much smaller amount of overhead, leading to an immediate crossover point for the smallest two values of $c_{\textup{q}}$ and a crossover point of a few hours for the larger value of $c_{\textup{q}}$.\footnote{Note that we observe this immediate speedup for Grover for the smallest possible instance, i.e. with $n=14$ for $k=11$. However, since $\alpha_{11}=1418.71$, this means this instance already has over $1400$ clauses.} However, this is only for the most random instances ($\beta=0$), and in the cases where Grover scales best but $\beta$ is larger, the advantage of Grover over classical is much smaller resulting in a crossover time of at least several years. 

The above discussions is only limited to using $T$-depth as the estimated cost metric of the quantum algorithms. Studying the $T$-count yields an even more sobering picture. \cref{tab:colors_scaling_count} shows the best scaling algorithms in terms of $T$-count. Just as before we observe that classical scales best initially, while for less structured SAT instances detection starts scaling better, and for even less structured instances binary search also start scaling better than classical. As argued above, by the SETH, we expect to observe the same four regimes for $T$-count, but we conjecture that the values of $k$ we considered are still too small to fully observe all regimes. That is, we imagine the regimes to be shifted down as compared to the $T$-depth (as $T$-depth is a more quantumly favourable cost model).

Note that since speedups in the $T$-count model only start appearing for much larger $k$---so for much less structured instances---this limits the practical utility of these quantum speedups. Even worse, for \textit{every} regime where a quantum algorithm scaled better than the classical algorithm in terms of $T$-count, a quick calculation shows that the crossover time was larger than the age of the universe! 

\subsection{No quantum advantage potential for one day search with structure}
We saw that with respect to scaling there exist four regimes: one where classical scales best, one where quantum backtracking detection scales best, one where quantum backtracking search beats classical and Grover, and one where Grover scales best.  

In~\cref{tab:constants} we considered the crossover time in terms of $T$-depth. We found that only Grover's algorithm seems to offer practical utility, as the binary search algorithm required millennia to start beating classical. To test this, we used the fits of the median runtime $T$-depth to compute the size of the largest instance that classical and quantum algorithms can solve in one day, for measurement times $10^{-9}$ and $10^{-6}$. We then again coloured the grid based on which algorithm can solve the largest instance in one day. 

In~\cref{tab:colors_one_day_9_depth} and~\cref{tab:colors_one_day_6_depth}, we see the results for $c_\textup{q} = 10^{-9}$ and $c_\textup{q} = 10^{-6}$, respectively. We immediately observe that the quantum backtracking algorithms disappeared, showing that they are not able to beat classical solvers or Grover's algorithm, even on the least structured instances. Instead, we see that for almost all instances, the classical solver is the fastest in one day. Only for the least structured instances does Grover's algorithm outperform the classical solver, and Grover's regime grows in size as $c_\textup{q}$ becomes smaller. However, since $c_\textup{q} = 10^{-9}$ is already extremely close to the clock speed of the laptop these experiments were run on, and the implementation time of a fault-tolerant quantum operation is always limited by the classical clock speed, it feels safe to assume that we are already very close to the best possible scenario here.

Moving to $T$-count, we repeat this test for the same measurement times in~\cref{tab:colors_one_day_9_count,tab:colors_one_day_6_count}. We see that no quantum algorithm is able to beat the classical solver within a day, even on the most unstructured instances.

\subsection{Getting an advantage is more difficult on satisfiable instances}
\begin{table*}[ht]
    \centering
    $T$\textbf{-depth}\\[5pt] 
    \begin{subtable}[t]{0.5\textwidth}
        \centering
        \begin{tabular}{p{0.5 cm}|p{0.22 cm}p{0.22 cm}p{0.22 cm}p{0.22 cm}p{0.22 cm}p{0.22 cm}p{0.22 cm}p{0.22 cm}}
            \resizebox{0.5cm}{!}{\backslashbox{$k$}{$\beta$}} & $\frac 12$ & 1 & $\frac 32$ & 2 & 3 & 5 & 10 & $\infty$\\\hline
            3& \cellcolor{C0}& \cellcolor{C0}& \cellcolor{C0}& \cellcolor{C0}& \cellcolor{C0} & \cellcolor{C0} & \cellcolor{C0} & \cellcolor{C0}\\
            4& \cellcolor{C0}& \cellcolor{C0}& \cellcolor{C0}& \cellcolor{C0} & \cellcolor{C0} & \cellcolor{C0} & \cellcolor{C0} & \cellcolor{C0}\\
            5& \cellcolor{C0}& \cellcolor{C0}& \cellcolor{C0} & \cellcolor{C0} & \cellcolor{C0} & \cellcolor{C0} & \cellcolor{C0} & \cellcolor{C0}\\
            6& \cellcolor{C0}& \cellcolor{C0} & \cellcolor{C0} & \cellcolor{C0} & \cellcolor{C0} & \cellcolor{C0} & \cellcolor{C0} & \cellcolor{C0}\\
            7 & \cellcolor{C0} & \cellcolor{C0} & \cellcolor{C0} & \cellcolor{C0} & \cellcolor{C0} & \cellcolor{C0} & \cellcolor{C0} & \cellcolor{C0}\\
            8 & \cellcolor{C0} & \cellcolor{C0} & \cellcolor{C0} & \cellcolor{C0} & \cellcolor{C0} & \cellcolor{C0} & \cellcolor{C0} & \cellcolor{C0}\\
            9 & \cellcolor{C0} & \cellcolor{C0} & \cellcolor{C0} & \cellcolor{C0} & \cellcolor{C0} & \cellcolor{C0} & \cellcolor{C0} & \cellcolor{C0}\\
            10 & \cellcolor{C0} & \cellcolor{C0} & \cellcolor{C0} & \cellcolor{C0} & \cellcolor{C1} & \cellcolor{C1} & \cellcolor{C1} & \cellcolor{C0}\\
            11 & \cellcolor{C0} & \cellcolor{C0} & \cellcolor{C1} & \cellcolor{C1} & \cellcolor{C2} & \cellcolor{C1} & \cellcolor{C3} & \cellcolor{C1}\\
            12 & \cellcolor{C0} & \cellcolor{C0} & \cellcolor{C1} & \cellcolor{C1} & \cellcolor{C3} & \cellcolor{C3} & \cellcolor{C3} & \cellcolor{C3}\\
        \end{tabular}
        \caption{Scaling, satisfiable}
        \label{tab:colors_sat_depth}
    \end{subtable}
    \hfill
    \begin{subtable}[t]{0.48\textwidth}
        \centering
        \begin{tabular}{p{0.22 cm}p{0.22 cm}p{0.22 cm}p{0.22 cm}p{0.22 cm}p{0.22 cm}p{0.22 cm}p{0.22 cm}}
             $\frac 12$ & 1 & $\frac 32$ & 2 & 3 & 5 & 10 & $\infty$\\\hline
            \cellcolor{C0}& \cellcolor{C0}& \cellcolor{C0}& \cellcolor{C0}& \cellcolor{C0} & \footnotesize \cellcolor{C0}  & \footnotesize \cellcolor{C0}  & \footnotesize \cellcolor{C0} \\
            \cellcolor{C0}& \cellcolor{C0}& \cellcolor{C0}& \cellcolor{C0} & \footnotesize \cellcolor{C0}  & \footnotesize \cellcolor{C0}  & \footnotesize \cellcolor{C0}  & \footnotesize \cellcolor{C0} \\
            \cellcolor{C0}& \cellcolor{C0}& \cellcolor{C0} & \footnotesize \cellcolor{C0}  & \footnotesize \cellcolor{C0}  & \footnotesize \cellcolor{C0}  & \footnotesize \cellcolor{C1}  & \footnotesize \cellcolor{C1} \\
            \cellcolor{C0}& \cellcolor{C0} & \footnotesize \cellcolor{C0}  & \footnotesize \cellcolor{C0}  & \footnotesize \cellcolor{C0}  & \footnotesize \cellcolor{C1}  & \footnotesize \cellcolor{C1}  & \footnotesize \cellcolor{C1} \\
            \footnotesize \cellcolor{C0}  & \footnotesize \cellcolor{C0}  & \footnotesize \cellcolor{C0}  & \footnotesize \cellcolor{C0}  & \footnotesize \cellcolor{C1}  & \footnotesize \cellcolor{C1}  & \footnotesize \cellcolor{C2}  & \footnotesize \cellcolor{C2} \\
            \footnotesize \cellcolor{C0}  & \footnotesize \cellcolor{C0}  & \footnotesize \cellcolor{C0}  & \footnotesize \cellcolor{C0}  & \footnotesize \cellcolor{C1}  & \footnotesize \cellcolor{C1}  & \footnotesize \cellcolor{C2}  & \footnotesize \cellcolor{C2} \\
            \footnotesize \cellcolor{C0}  & \footnotesize \cellcolor{C0}  & \footnotesize \cellcolor{C0}  & \footnotesize \cellcolor{C0}  & \footnotesize \cellcolor{C0}  & \footnotesize \cellcolor{C1}  & \footnotesize \cellcolor{C1}  & \footnotesize \cellcolor{C2} \\
            \footnotesize \cellcolor{C0}  & \footnotesize \cellcolor{C0}  & \footnotesize \cellcolor{C0}  & \footnotesize \cellcolor{C1}  & \footnotesize \cellcolor{C1}  & \footnotesize \cellcolor{C2}  & \footnotesize \cellcolor{C2}  & \footnotesize \cellcolor{C2} \\
            \footnotesize \cellcolor{C0}  & \footnotesize \cellcolor{C0}  & \footnotesize \cellcolor{C1}  & \footnotesize \cellcolor{C1}  & \footnotesize \cellcolor{C2}  & \footnotesize \cellcolor{C3}  & \footnotesize \cellcolor{C3}  & \footnotesize \cellcolor{C3} \\
            \footnotesize \cellcolor{C0}  & \footnotesize \cellcolor{C0}  & \footnotesize \cellcolor{C1}  & \footnotesize \cellcolor{C1}  & \footnotesize \cellcolor{C2}  & \footnotesize \cellcolor{C3}  & \footnotesize \cellcolor{C3}  & \footnotesize \cellcolor{C2} \\
        \end{tabular}
        \caption{Scaling, unsatisfiable}
        \label{tab:colors_unsat_depth}
    \end{subtable}
    \caption{These tables are defined the same way as \cref{tab:colors_scaling_depth}: each grid is coloured based on the scaling of the runtime and $T$-depth of the algorithms. Here (a) is restricted to satisfiable instances and (b) to unsatisfiable.}
    \label{tab:tables}
\end{table*}

Finally, we observe a sharp difference in the potential for quantum speedups between satisfiable and unsatisfiable instances. All the previous results are for an equal mix of satisfiable and unsatisfiable instances. Let us now split these up. In~\cref{tab:colors_unsat_depth} we colour the grid based on which algorithm scales best on unsatisfiable instances for $T$-depth. We see that the results look remarkably similar to the result for a mix of satisfiable and unsatisfiable which we saw in~\cref{tab:colors_scaling_depth}. In contrast,~\cref{tab:colors_sat_depth} shows the best scaling algorithm on satisfiable instances using $T$-depth, and the classical algorithm occupies a much larger area. We do the same for $T$-count in~\cref{tab:colors_sat_count,tab:colors_unsat_count} in the Appendix, which yields a similar result.

We believe there are two possible explanations for this: (i) at least in comparison with Grover, classical algorithms might scale better as the number of solutions grows as it might terminate earlier, for which we give a mathematical argument in~\cref{app:multiple_solutions}; and (ii) the CaDiCaL algorithm might be empirically optimized to quickly work towards solutions on structured instances, which might only help if there is a solution to begin with.

\paragraph{Other data and results}
We listed more detailed results in~\cref{appendix:results}. Specifically, we show the grids over $\beta$ and $k$ containing the actual fits (i.e.~as in \cref{fig:snippet}) one for only satisfiable instances, one for unsatisfiable, one for an equal mix, and all of these for $T$-count, for $T$-depth and for query complexity. 

\section{Conclusion and discussion}
Taking into account an overhead for error correction and a reasonable maximum allowed computing time, we have seen that only Grover's algorithm is able to achieve practical speedups, albeit on highly unstructured SAT instances. However, once instances with even little structure are considered and we do not grant the quantum algorithms parallelization for the $T$-gate implementation, the classical SAT solvers perform much better than Grover's algorithm. The quantum backtracking search algorithm fails to achieve a relevant speedup in any regime of considered SAT instances. This is despite backtracking showing an asymptotic advantage (i.e.~better scaling with $n$) in a slim regime of SAT instances with little structure. In this final section, we discuss three potential objections one might make against our findings.

\subsection{What about better heuristics?}
We saw that there was indeed a small regime of slightly structured SAT instances where quantum backtracking offered an \textit{asymptotic} advantage over both classical SAT solvers and Grover, which shows that at least \textit{in principle} quantum backtracking is able to exploit structure in SAT instances better than our current classical solvers, even when using a rather ``unsophisticated'' heuristic $h$ and predicate $P$. 

The usage of more advanced heuristics and predicates could drastically reduce the tree size $\mathcal{T}$, and hence the scaling of quantum backtracking algorithm. For such a choice of $P$ and $h$, one could then also design quantum circuits to implement these functions similar to what was done in Ref.~\cite{campbell2019applying}, and perform similar experiments to what we have done to if there is now a speedup on more structured SAT instances. However, more complicated heuristics and predicates would likely result in significantly larger gate complexity (and thus $T$-count) than the simple choice of $h$ and $P$ we considered.  Hence, there is be trade-off in achieving a potentially better scaling at the cost of increasing the circuit overhead.

However, based on our results, we expect even for the smartest choices of $P$ and $h$ it might still be difficult to achieve a practical speedup when the maximum allowed computing time is only a day. 

The argument is as follows. First, we will consider the $T$-depth as our cost metric. Consider again~\cref{fig:snippet}, where for $\beta = 3$ we found that quantum backtracking search scaled better than the classical solver. In~\cref{tab:constants} we found that even for the very small value of $c_\textup{q} = 10^{-8}$ the crossover time was larger than the age of the universe. Now suppose that we choose a smarter $P$ and $h$ to improve the scaling of the quantum algorithm. What kind of scaling would we need to achieve to match the size of instances that the classical solver can handle in one day? Assuming that the quantum algorithm achieves a perfect quadratic speedup (recall that we don't have a full quadratic speedup, as we lose a factor $\mO(n^2)$, which in practice consistently gives a speedup of order $1.4-1.45$) over the classical backtracker, and using the overheads from~\cref{appendix:results} (which is an extremely unrealistic assumption), we can compute the required scaling of the classical backtracker to match this performance. For a more reasonable $c_\textup{q} = 10^{-6}$, we find that the classical backtracker would need to have a scaling exponent of $0.4385$, meaning that this ``simple'' classical backtracker would have outperformed last year's SAT competition winner! For $c_\textup{q} = 10^{-7}$, we find an exponent of $0.564$, which is still small but not unreasonable. However, recall that this is all under the very unrealistic assumption that using better heuristics does not lead to increased overhead. Doing the same calculation using $T$-count (again for $\beta =3$),  we find that the overhead for $c_{\textup{q}} = 10^{-6}$ is already so big that a we can never find a speedup (for any type of scaling) and that for $c_{\textup{q}} = 10^{-9}$ we have a scaling exponent of $0.148$, which would give a breakthrough even in $3$-SAT solving (using an $11$-SAT algorithm).

Hence, we conclude that it might be difficult to find an improved heuristic that would allow the quantum backtracker to become useful when the maximum allowed computing time is only one day. 

\subsection{What about improved quantum backtracking algorithms?}
An extension of the quantum backtracking algorithm we use is given by Jarret and Wang~\cite{jarret2018improved}, who propose an efficient method to estimate the effective resistance of the graph, reducing the $\sqrt{n}$ dependence in the complexity and replacing the complexity due to binary search with $\mO(\log (Rt) )$, where $R$ is the actual effective resistance of the graph and $t$ is the number of solutions. We expect that the extra overhead from having to run amplitude estimation will not significantly change the above picture.

Additionally, Ambainis and Kokainis show how to change the dependence on the tree size $\mathcal{T}$ to only the subtree that the backtracking algorithm explored before finding the first solution, offering a speedup for satisfiable instances~\cite{ambainis2017quantum} and addressing the argument that classical algorithms benefit more from multiple solutions by terminating once one has been found. Moreover, since we found that the potential for a quantum speedup was much smaller for satisfiable instances compared to unsatisfiable instances, this modification has the potential to improve the competitiveness of quantum backtracking algorithms. However, since the potential for a speedup even on unsatisfiable instances (the most competitive regime for quantum) is minimal, we believe that this improvement by itself won't lead to different conclusions.

Finally, Piddock expanded Belovs' underlying quantum walk algorithm to handle search in general graphs (recall that the above algorithm only handles search on trees), building heavily on ideas from Jarret and Wang~\cite{jarret2018improved}. This replaces the $n$ overhead for binary search with a $\mO(\log^3(t) )$ overhead, where $t$ is the number of solutions~\cite{piddock2019quantum}. However, a priori $\mO(\log^3(t))$ is not obviously lower than $\mO(n)$, as we have no rigorous relation between the number of variables and the number of solutions in our SAT distributions. Preliminary results from applying our methodology to Piddock's algorithm on random $3$-SAT show slightly better scaling in practice than our binary search approach, but significantly higher constant overhead. The similarity of Piddock's algorithm to Jarret and Wang's further supports our claim that their algorithm most likely won't improve the competitiveness of quantum.


\subsection{How reliable are our results?}
An objection one might make is that our results do not include any quantification of the errors\footnote{This also holds for Campbell, Khurana and Montenaro~\cite{campbell2019applying}, which forms the starting point of this work.}, so how likely is it that a reproduction of our experiments leads to the same conclusions? First, it’s important to note that we employed a large number of different distributions over SAT instances (parameterized by $k$ and $\beta$), sampling multiple instances for each distribution across various instance sizes $n$. For each $n$, we calculated the median and used it in a linear least-squares fit of the logarithm of the median. Consistently, we found that the interquartile range relative to the median was at most a few percent, with some larger outliers. However, these outliers were partially mitigated by their role in determining an exponential fit over multiple medians for different values of $n$. The robustness of our approach is further enhanced by the fact that the results are presented on a grid. Even if individual points contain errors, the overall depiction of the different regimes remains largely unaffected.

More importantly, the aim of this work is not to claim that we can make precise \textit{quantitative} predictions of quantum run times for specific instances, but rather to sufficiently rigorously explore the broader \textit{qualitative} picture. Subjecting our numbers, for example the specific crossover times mentioned in \cref{tab:constants} or the data to used the colouring of the grid in \cref{tab:colors_scaling_depth}, to large perturbations due to errors will still yield the same qualitative conclusions, as it is quantitatively consistent over all grids in~\cref{appendix:results}.

\begin{acknowledgments}
JW was supported by the Dutch Ministry of Economic Affairs and Climate Policy (EZK), as part of the Quantum Delta NL programme. We thank Koen Groenland, Nicolas Resch and the anonymous reviewers for helpful comments.
\end{acknowledgments}

\bibliographystyle{plainnat}
\bibliography{main.bib}
\onecolumn\newpage
\appendix
\section{Explicit query complexity upper-bounds for quantum walk algorithms}\label{appendix:belovs_bounds}
This section provides explicit expressions of the query complexity of the quantum walk algorithms used in this work. We start by providing some background information on Belovs' quantum walk detection algorithm (which we defined in \cref{sect:detection}). We then provide an explicit expression of the exact query complexity of this detection algorithm. Finally, we give an upper-bound on the expected query complexity of the detection-based binary search algorithm defined in the quantum walk search algorithm \cref{sect:search}.

\subsection{Background Belovs' quantum walk}
The two quantum walk algorithms solve the following problem.

\begin{problem}[The graph search problem]
Let $G=(V,E, w)$ be an weighted undirected graph, where $w:E \to \mathbb{R}_+$. Let $M : V \to \{0, 1\}$ be a binary function telling you whether any vertex is ``marked'' or not. The goal is to determine whether there exist marked vertices, i.e. whether $|M^{-1}(1)| > 0$ or not (decision version), or if $M$ is non-empty, to find a marked vertex, i.e. find some $v\in V$ such that $M(v)=1$ (search version).
\end{problem}

A simple classical algorithm for this problem is a random walk/Markov chain. Sample a starting vertex from $\sigma$, check if this vertex is marked, and if not, choose a neighbouring vertex to move to with probability proportional to its weight, and repeat. The expected number of queries to $M$ that such an algorithm requires is called the \textit{hitting time} of the Markov chain from $\sigma$ to $M$, denoted $HT_{\sigma, M}$.\footnote{In this notation the second subscript usually refers to the set of vertices that you are trying to hit with your Markov chain; we allow ourselves to abuse this notation and write instead our oracle $M$ that checks whether a vertex is marked.}

Szegedy showed how to take any Markov chain, and turn it into a quantum algorithm for the same problem that requires quadratically less queries to $M$. Specifically, he showed that given a Markov chain, one can construct a unitary $W$ with the property that a specific quantum state $\ket{\phi}$ is invariant under $W$ if and only if no marked vertices exist. Thus, repeatedly applying $W$ to this state should reveal which case one is in. Szegedy showed that this requires $\mO(\sqrt{HT_{\phi, M}})$ applications of $W$ (i.e.~queries to $M$)~\cite{szegedy2004quantum}.

Two main drawbacks of this algorithm are that 1) it only solves the detection problem (for general graphs at least) and 2) it requires one to construct $\ket{\phi}$, which is a superposition over the entire graph (preventing usage when one only has local access to a graph, i.e., oracle access to neighbouring vertices). 

Using a beautiful connection between classical Markov chains and electrical network theory, Belovs showed that the second drawback can be solved~\cite{belovs2013quantum}. In electrical network theory, one views a graph as an electrical circuit, with edge weight corresponding to resistance at that edge. Imagine that one sends a current into the graph according to the starting distribution, i.e., a current of $\sigma(v)$ enters vertex $v$, and that current has to leave the graph at the marked vertices. If $p_e$ is the current at some edge $e$ and $w_e$ is the weight (resistance) of that edge, recall then that the energy at that edge is $p_e^2/w_e$. The energy of the entire flow of current is the sum of the energy at each edge. The minimal energy of any flow of current from the starting distribution $\sigma$ to the marked vertices $M$ is called the \textit{effective resistance}, denoted $R_{\sigma,M}$. Call the electrical flow the flow through the graph that generates the least energy. Belovs shows that the quantum state $\ket{\phi}$ describing this electrical flow (i.e.~amplitude at an edge is equal to the amount of current at that edge under the electrical flow) is invariant under the quantum walk unitary if and only if marked elements exist.

Using this insight, he then proposes the following alteration: for each vertex $v$ in the support of the starting distribution $\sigma$, add a new vertex $v'$ and a new edge between $v$ and $v'$ with very small positive weight. Note that energy is current squared over resistance: since the weight (i.e. resistance) at these edges is small, that means the energy will be large. Hence, in the positive case, $\ket{\phi}$ will have large overlap with our starting state $\ket{\sigma}$, so that running quantum phase estimation (QPE) will output this phase 0 with probability proportional to the square of the overlap $\braket{\sigma}{\phi}$ (independent of the number of bits of precision). To also make this work in the negative case, we would want to measure something other than phase 0, and Belovs shows that if we use $m\in O(\log \sqrt{HT_{\sigma, M}})$ bits of precision this occurs with high probability. 

Thus, his algorithm is simple: run QPE on $W$ (defined using the altered graph) and $\ket{\sigma}$ with $m\in O(\log \sqrt{HT_{\sigma, M}})$ bits of precision and output ``marked elements exist'' if and only if you measure phase 0. Belovs notes that $HT_{\sigma, M} = 2WR_{\sigma, M}$, where $W$ is the sum of weights of the graph and $R_{\sigma, M}$ is the effective resistance, the sum of energy of the electrical flow. By expressing the query complexity required by the precision $m$ in terms of the number unitary applications in QPE (which is $\mO(2^m)$), the query complexity of Belovs' algorithm becomes $\mO(\sqrt{WR_{\sigma, M}})$. Finally note that this means you need to know (upper-bounds) on $W$ and $R_{\sigma, M}$ to run this algorithm.

\begin{theorem}[Ref.~\cite{belovs2013quantum}, Theorem 4]
    For any starting distribution $\sigma$ Belovs' quantum walk (configured with upper-bounds on $W$ and $R_{\sigma, M}$) solves the Detection problem with probability $>1/2$ with $\mO\left(\sqrt{WR_{\sigma, M}}\right)$ queries to $M$.
\end{theorem}

\subsection{Query complexity and optimization of quantum walk detection}\label{appendix:belovs_bounds_detection}
To move beyond asymptotics, let us consider how to exactly configure Belovs' algorithm, and study the resulting query complexity. There are two questions at play here. Assume some desired success probability $1-\delta$. 
\begin{enumerate}
    \item[(1)] How small do we make the weights at the newly added edges, i.e. how large do we want to set $\braket{\sigma}{\phi}$? This essentially sets the success probability in the positive case, and changes the quantum walk unitary. 
    \item[(2)] What exact number of bits $m\in O(\log \sqrt{WR_{\sigma, M}})$ do we use for QPE? This essentially sets the success probability in the negative case. We will see that the outcome of (1) also influences the number of bits $m$.
\end{enumerate}
Once we answer these questions, the resulting complexity is easy to express: it is exactly $2^m-1$ calls to $W$ and hence queries to $M$. 

For (1), recall that in the positive case QPE outputs phase 0 with probability at least $|\braket{\sigma}{\phi}|^2$ (i.e.~regardless of $m$). In the proof of his Theorem~4, Belovs shows that this probability is at least $\frac{C_1}{1+C_1}$ where $C_1>0$ is a constant which sets the weight of the newly added edges. We therefore set $C_1$ such that $\frac{C_1}{1+C_1} \geq 1-\delta$.

For (2), recall that in the negative case, we want QPE to output anything other than phase 0. Belovs shows that the overlap between the starting state $\ket{\sigma}$ and the eigenvectors of $W$ with phase smaller than some angle $\Theta$ can be upper-bounded by $1/(2C_2)^2$ where $C_2>0$ is another constant. 
This is shown by defining 
\[
    \Theta = \frac{1}{C_2\sqrt{1+C_1RW}} \ ,
\]    
where $R,W$ are upper-bounds on the effective resistance from $\sigma$ to $M$ and sum of weights of the graph, respectively. This means we can lower-bound the probability that QPE outputs (the approximation) of a non-zero phase by $1-(2C_2)^2$. However, unlike the phase 0, these small non-zero phases can't be expressed in any number of bits: this means that we need to set $m$ such that $2^m \leq \Theta$.

However, setting $m$ this large is not necessarily sufficient to guarantee an error probability of at most $1/(2C_2)^2$. The problem is that this only bounds the probability that QPE outputs an approximation of a non-zero phase. This leaves the possibility that such an 
approximation of a non-zero phase could be 0, and we need to bound the probability of this happening. 

We note that Belovs' doesn't provide such an analysis, but one is given in~Ref.~\cite[\textsection~4.1]{campbell2019applying}. However, this analysis omits the above two constants by implicitly setting $C_1=C_2=1$, which yields an algorithm with a relatively small base success probability. This is compensated for by repeating the algorithm many times. Asymptotically this makes sense: it is well known that reducing the error probability of an algorithm to $\delta$ by repeating it requires $\mO(\log 1/\delta)$ repetitions. Conversely, rewriting the above expressions involving $C_1,C_2$ to improve the error probability of the base quantum walk algorithm to $\delta$ requires a $\mO(1/\delta)$ factor slowdown: exponentially worse than repeating. However, we found that the constant in the latter case is much larger than in the former case. As a result, optimizing the number of repetitions as well as the value of the constants $C_1,C_2$ ends up yiedling an algorithm that is in our considered setting roughly a factor 100 more efficient (i.e.~as compared to setting $C_1=C_2$ as done in Ref.~\cite{campbell2019applying}).

Let us analyze the success probability of the algorithm with these constants included. Start by expressing the decomposition of our starting state in the eigenbasis of the quantum walk unitary as 
\(\ket{\sigma} = \sum_j \alpha_j \ket{\phi_j}\), where \(\ket{\phi_j}\) is an eigenvector of \(W\) with eigenphase \(\theta_j\). The state just before applying the inverse QFT during QPE is then given by  
\[
\dfrac{1}{\sqrt{2^m}} \sum_j \alpha_j \left( \sum_{x=0}^{2^m-1} e^{2\pi i \theta_j x} \ket{x} \right) \ket{\phi_j}.
\]
As in Ref.~\cite{campbell2019applying}, we replace the inverse QFT in QPE with a Hadamard transform, relying on the fact that we only need to distinguish a \(0\)-eigenphase from those bounded away from \(0\). To see why this works, note that applying a Hadamard transform to the first register of the state above yields  
\[
\dfrac{1}{\sqrt{2^m}} \sum_j \alpha_j \left( \sum_{x=0}^{2^m-1} e^{2\pi i \theta_j x} H^{\otimes m} \ket{x} \right) \ket{\phi_j} =
\dfrac{1}{2^m} \sum_j \alpha_j \left( \sum_{x=0}^{2^m-1} \sum_{y=0}^{2^m-1} e^{2\pi i \theta_j x} (-1)^{x \cdot y} \ket{y} \right) \ket{\phi_j}.
\]
This is the state we measure, and the algorithm accepts if and only if the first register is measured as \(\ket{0^m}\). By the no-signaling principle, any measurement performed on the second register does not affect the probability distribution over outcomes on the first register. If we perform a measurement on the second register in the eigenbasis \(\{\ketbra{\phi_j}\}\), the second register collapses to \(\ket{\phi_j}\) with probability \(\mathbb{P}[\phi_j] = |\alpha_j|^2\). Therefore, conditioning on the outcome \(\phi_j\), the probability of measuring \(\ket{0^m}\) in the first register is then\footnote{If we directly wrote \(\mathbb{P}[\text{accept}]\) by fixing \(y = 0\), we would obtain \(|\sum_j \alpha_j|^2\) instead of \(\sum_j |\alpha_j|^2\). This form allows us to separate the sum over \(j\), which is necessary.}
\[
\mu_j := \left| \frac{1}{2^m} \sum_{x=0}^{2^m-1} e^{2\pi i \theta_j x} (-1)^{x \cdot 0} \right|^2.
\]
By the total law of probability, the overall acceptance probability is  
\[
\mathbb{P}[\text{accept}] = \sum_j \mathbb{P}[\text{accept} | \phi_j] \mathbb{P}[\phi_j] = \sum_j |\alpha_j|^2 \mu_j.
\]

In the positive case (where a marked element exists), the starting state has a significant overlap with the eigenvector corresponding to eigenphase \(0\), denoted \(\phi_0\). Specifically, we have \(\alpha_0 = \braket{\phi_0}{\sigma} \geq \sqrt{\frac{C_1}{1 + C_1}}\). Thus, with probability \(\frac{C_1}{1 + C_1}\), the measurement yields an approximation of the phase \(\theta_0 = 0\). Since this approximation is always \(0\), we succeed with the same probability, as required. 

The negative case is more complicated, as we don't just want to upper bound the probability of ending up with a non-zero phase, but also that the approximations of these non-zero phases are non-zero. To achieve this, we split up the sum over $j$ based on the size of the corresponding phase $\theta_j$. The point of this is that we can use the effective spectral gap lemma to tightly bound the overlap between our starting state and the eigenvectors with small phases $\leq \Theta$ (i.e.~close to $0$). We can then deal with the large phases with a rougher bound, as these large phases are very unlikely to yield an approximation equal to 0. Specifically, we write:
\[
    \mathbb{P}[\text{accept}] = \frac{1}{2^{2m}} \left( \sum_{j,|\theta_j|\leq \Theta} |\alpha_j|^2 \mu_j + \sum_{j,|\theta_j|> \Theta} |\alpha_j|^2 \mu_j \right) \ . 
\]
The spectral gap lemma \cite[Lemma~2]{belovs2013quantum} tells us that overlap between our starting state $\ket{\sigma}$ and eigenvectors with phases $|\theta_j| \leq \Theta = \frac{1}{C_2\sqrt{1+C_1RW}}$ is at most $\frac{\Theta}{2}\sqrt{1+C_1RW}$. Note also that we always have $\mu_j \leq 1$. This means we can upper bound the first sum as $\left(\frac{1}{C_2\sqrt{1+C_1RW}}\right)^2$. 

We bound the second sum as $ \frac{1}{2^{2m}\Theta^2(1-\Theta_j^2/6)}$ using $\mu_j \leq \frac{1}{2^{2m}\theta_j^2(1-\theta_j^2/6)}$ \cite[\textsection~4.1]{campbell2019applying}, combined with the fact that this upper bound on $\mu$ decreases as $|\theta_j$ grows (i.e. is maximized at $\theta_j=\Theta$) and that $\sum_j|\alpha_j|^2 = 1$. Writing this out gives us:

\[
    \mathbb{P}[\text{accept}] \leq \frac{\Theta^2}{4}(1+C_1RW) + \frac{1}{2^{2m}\Theta^2(1-\Theta^2/6)}
\]
To simplify the above we define
\[
    \Theta^2=\dfrac{4a}{1+C_1RW} \text{ and } M^2=2^{2m}=\dfrac{1+C_1RW}{4b} \ .
\] 
Note that before $\Theta$ depended on the constant $C_2$, but now depends on $a$. As such, from now on we write $C:=C_1$. We can now simplify $p$ as
\[
    \mathbb{P}[\text{accept}] \leq a+\dfrac{b}{a\left(1-\frac{2}{3}\frac{a}{1+CRW}\right)} = a+\dfrac{b}{a\left(1-o(1)\right)} \ .
\]
The above is essentially $a+b/a$, and to minimize this over $a$ we set $1-b/a^2=0$, which gives $b/a^2=1$, so that we set $a=\sqrt{b}$.

Recall that the success probability in the positive case is at least $C/(1+C)$, where we can pick $C>0$. The above expression for the error probability $p$ in the negative case contains the constant $b$. Let us require that the negative success probability $1-p$ is at least $C/(1+C)$, and solve for $b$ in terms of $C$:
\[
    \dfrac{C}{1+C} \leq 1 - \left(\sqrt{b} + \dfrac{\sqrt{b}}{1-\frac{2}{3}\frac{\sqrt{b}}{1+CRW}}\right) \ ,
\]
which is the same as 
\[
\dfrac{1}{1+C} \geq \sqrt{b} + \dfrac{\sqrt{b}}{1-\frac{2}{3}\frac{\sqrt{b}}{1+CRW}}
\geq \sqrt{b} +\sqrt{b}\left(1+\frac{2}{3}\frac{\sqrt{b}}{1+CRW}\right)
\geq 2\sqrt{b},
\]
where the one-to-last inequality follows because  $1+x\leq\dfrac{1}{1-x}$ since $x\in[0,1]$. 

Thus, to guarantee success probability $1-\delta$ in the positive case we pick $C>0$ such that $\dfrac{C}{1+C} = 1-\delta$ (and construct the quantum walk unitary). We can then set $\sqrt{b}=\dfrac{1}{2(1+C)}$ to guarantee the same success probability in the negative case (and determine the number of bits needed for phase estimation). Having guaranteed the right error probability, the query complexity becomes $2^m-1$, where we had 
\[
    M=2^{m}=\sqrt{\dfrac{1+CRW}{4\dfrac{1}{4+4C^2}}}=\sqrt{(1+C^2)(1+CRW)}
\]
so that we need to set $m=\lceil\log\sqrt{(1+C^2)(1+CRW)}\rceil$ (i.e.~we might slightly round up $m$ by 1 so that the query complexity is between $M$ and $2M$).

This is all for a single run of the detection algorithm. We now consider doing $l$ repetitions of the algorithm, which yields query complexity $l(2^m-1)$. We now want to minimize $l(2^m-1)$ subject to the choices of $(C,l)$ guaranteeing error probability $\leq \delta$. We could do so analytically by appealing to, say, a Chernoff bound. For the sake of efficiency, let us work out the optimum numerically. That is, write out the error probability of a majority vote of $l$ runs of the algorithm as 
\[
    \delta \leq (1-p)^l \sum_{i=0}^{\lfloor l/2\rfloor} \dbinom{l}{i} \left(\frac{p}{1-p}\right)^i=\frac{1}{(1+C)^l} \sum_{i=0}^{\lfloor l/2\rfloor} \binom{l}{i} \left(\frac{\frac{C}{1+C}}{\frac{1}{1+C}}\right)^i=\frac{1}{(1+C)^l} \sum_{i=0}^{\lfloor l/2\rfloor} \binom{l}{i} C^i \ .
\]
To optimize, simply increment $l=1,3,5...$, exponentially dropping the error probability. For each $l$, solve the above inequality for $C$, i.e. determine $C$ which gives you exactly the required error probability $1-\delta$. Then compute the number of queries $l(2^m-1)$. Once the number of queries increases from one iteration to the next, output $l$ and $C$ corresponding to the smallest number of queries. 

The reason this works is that we must always have $C>1$, as otherwise the success probability of the base algorithm is $\leq 1/2$. At some point, the number of repetitions alone guarantees the required success probability, so that we can't decrease $C$ further. At this point, the number of queries will start to grow with each iteration, whereas before that the cost of increasing $n$ was offset in part by decreasing $C$.

There is one slight caveat here, which is that the query complexity we are minimizing depends on $R$ and $W$, i.e.~upper-bounds that are dependent on the given problem instance. Although the above finds an optimum relative to some $R$ and $W$, it is not clear that this will also be optimal (or even decent) for other $R$ and $W$. However, we note that $\sqrt{(1+C^2)(1+CRW)} \leq\sqrt{1+C^2}\left(1+\sqrt{CRW}\right)=\sqrt{C(1+C^2)}\sqrt{RW} +\sqrt{1+C^2}$ so that we can rewrite the query complexity as:
\begin{align}
    l\cdot (2^m-1) &= l \cdot\left(2^{\lceil\log \sqrt{(1+C^2)(1+CRW)} \rceil} - 1\right) \label{eq:queries_detection} \\
    &\leq l \cdot\left(2^{\lceil\log \sqrt{C(1+C^2)}\sqrt{RW} +\sqrt{1+C^2} \rceil} - 1\right) \notag\\
    &\leq 2 \cdot l \cdot \sqrt{C(1+C^2)}\sqrt{RW} + 2 \cdot l \cdot \sqrt{1+C^2} - l \notag \ .
\end{align}
We run the above procedure, optimizing just for $2 \cdot l \cdot \sqrt{C(1+C^2)}\sqrt{RW}$, as the two other terms are tiny constants. This means we obtain a globally optimum choice $(C,l)$, with the only exception being tiny instances (which we don't care about regardless). \cref{table:configurations} contains optimal $(C,l)$ for a selection of error probabilities $\delta$. As a comparison, the configuration of the detection algorithm given in Ref.~\cite{campbell2019applying} sets $\delta=1/10$ and ends up requiring $2528\sqrt{RW}$ queries, roughly 100 more than our configuration of $26\sqrt{RW}$.

\begin{table}[!ht]
    \centering
    \begin{tabular}{l|l|l|l|l}
    $\delta$ & $C$ & $l$ & Queries to quantum walk unitary 
    \\\hline
    1/10 & 4.107 & 3 & $25.701 \sqrt{RW+0.243} - 3 $ \\
    1/100 & 3.698 & 13 & $95.751 \sqrt{RW+0.270} - 13 $ \\
    1/1000 & 3.735 & 23 & $171.894 \sqrt{RW+0.268} - 23 $ \\
    1/10000 & 3.778 & 33 & $250.710 \sqrt{RW + 0.265} - 33 $ \\
    1/100000 & 3.813 & 43 & $331.004 \sqrt{RW + 0.262} - 43 $ \\
    1/1000000 & 3.742 & 55 & $412.068 \sqrt{RW + 0.267} - 55 $
    \end{tabular}
    \caption{An overview of configurations of Belovs' quantum walk detection algorithm. Herem $\delta$ is a provable upper-bound on the error probability of the algorithm, $C$ is a constant used in the definition of the quantum walk unitary, $l$ is the number of parallel repetitions. Recall that \cref{eq:queries_detection} gives the number of queries in terms of $C$ and $l$.}
    \label{table:configurations}
\end{table}

\subsection{Query complexity and optimization of quantum walk detection}\label{appendix:belovs_bounds_search}
We now turn the above detection algorithm into a usable search algorithm. As explained, we simply perform binary search on a tree, and since the algorithm requires an upper bound on the tree size (as we use unit weights), we repeat the binary search algorithm to estimate this tree size. See \cref{sect:search} for the details. The overhead is a factor $\mO(n^2\log n)$, where $n$ is the depth of the tree. Our question is now what the exact overhead is, including constants.

If our tree has depth $n$, binary search will make at most $n$ repetitions of the detection algorithm. Each individual run of the detection algorithm is guaranteed to succeed with probability $1-\delta$. To make sure that the entire sequence of $\leq n$ runs of the detection algorithms succeeds with the desired probability $1-\delta$, we would need to reduce the error probability to $\mO(1/n)$, requiring $\mO(\log n)$ repetitions. More precisely, we should repeat $l'$ times, where $l'$ is the solution to 
$$1 - \sqrt[n]{1-\delta} \leq \delta^{l'} \sum_{i=0}^{\lfloor l'/2 \rfloor} \binom{l'}{i} \left(\frac{1-\delta}{\delta}\right)^i \ ,$$
and this is something we can compute in code for our target $\delta$. Suppose we have a satisfiable instance. Then there exists a specific first solution that we find, say at depth $d \leq n$ (this is because we use a fixed heuristic $h$). The overhead caused by binary search is then $d\cdot l'=O(n \log n)$. Instead, if we have an unsatisfiable instance, we would detect this at the root, and only have an overhead of $l' = O(\log n)$.

However, this is assuming we are in the final iteration where we have $\mathcal{T'} \geq \mathcal{T}$. Let us now consider the overhead caused by the previous iterations where we use $\mathcal{T'}$ that is too small. Clearly, we need exactly $\lceil \log \mathcal{T} \rceil$ iterations before we guarantee $\mathcal{T'} \geq \mathcal{T}$. In each such iteration we perform binary search. In principle, such a binary search procedure could terminate at a leaf at any depth (or even at the root). However, since the correctness guarantees of the detection algorithm don't hold when $\mathcal{T'}$ is too small, we can't assume that the number of steps is indeed small \footnote{One might object that the correctness of the detection in the positive case doesn't rely on the upper bound $\mathcal{T}$. Indeed, given a satisfiable instance, the first call of the detection algorithm will detect this correctly. However, as part of binary search, we then run the detection algorithm at one of the children. Although one of the children is guaranteed to contain a solution in this case, the other child needn't. We can therefore always end up running detection on a tree with on solutions, after which our correctness guarantee fades away.}. We have to assume the worst case of $n$ full iterations for each binary search call. Finally, note that the query complexity depends on the given value of $\mathcal{T'}$ and is therefore smaller for these initial iterations. Specifically, recalling the query complexity of the detection algorithm from \cref{eq:queries_detection}, we can write the final upper-bound on the query complexity as:
\[
    l' \sum_{i=1}^{\lceil \log \mathcal{T} \rceil} 
    n \cdot l \cdot\left(2^{\lceil\log \sqrt{(1+C^2)(1+CR\cdot2^i)} \rceil} - 1\right) \ ,
\]
where in the positive case we can replace the factor $n$ in the final iteration with a factor $d$, and in the negative case we can remove the factor $n$ in the final iteration (as we detect unsatisfiability at the root, and therefore don't run binary search). 

\section{Explicit expected query complexity upper-bounds for Grover}\label{appendix:grover}
This section provides explicit expressions for the expected query complexity of Grover's algorithm. We take this expression from Ref.~\cite{cade2023quantifying}. 

\begin{lemma}[Adapted from Lemma 4 in Ref.~\cite{cade2023quantifying}] 
\label{lem:QSearch}
Let $L$ be a list, $g: L \rightarrow \{0,1\}$ a Boolean function, $N_{\text{samples}}$ a non-negative integer and $\delta > 0$, and write $t = |g^{-1}(1)|$ for the (unknown) number of marked items of $L$. Then, $\textbf{QSearch}(L, N_{\text{samples}},\delta)$ finds and returns an item $x \in L$ such that $g(x) = 1$ with probability at least $1-\delta$ if one exists using an expected number of queries to $g$ that is given by
\[
E_{\textbf{QSearch}}(|L|,t,N_{\text{samples}},\delta) = \frac{|L|}{t}\left(1 - \left(1-\frac{t}{|L|}\right)^{N_{\text{samples}}}\right) + \left(1-\frac{t}{|L|}\right)^{N_{\text{samples}}} c_q E_{\text{Grover}}(|L|,t) \, ,
\]
where
\[
E_{\text{Grover}}(|L|,t)
\leq F(|L|,t) \left(1 + \frac{1}{1 - \frac{F(|L|,t)}{\alpha \sqrt{|L|}}} \right) \, ,
\]
with
\[
F(|L|,t) =
\begin{cases}
    \frac{9}{4}\frac{|L|}{\sqrt{(|L| -t )t}} + \left\lceil\log_{\frac{6}{5}}\left(\frac{|L|}{2\sqrt{(|L| -t )t}} \right) \right\rceil - 3 \leq \frac{\alpha \sqrt{L|}}{3 \sqrt{t}}   &\text{for} \quad 1 \leq t < \frac{|L|}{4} \\
    2.0344 &\text{for} \quad \frac{|L|}{4} \leq t \leq |L|. 
\end{cases}
\label{eq:F(L,t)}
\]
If no marked item exists, then the expected number of queries to $g$ equals the number of queries needed in the worst case (denoted by $W_{\textbf{QSearch}}(|L|, N_{\text{samples}},\delta)$), which is given by
\[
E_{\textbf{QSearch}}(|L|,0,N_{\text{samples}},\delta) = W_{\textbf{QSearch}}(|L|,N_{\text{samples}},\delta) \leq N_{\text{samples}} + \alpha c_q \lceil \log_3(1/\delta)\rceil) \sqrt{|L|} . \\
\]
In the formulas above, $c_q$ is the number of queries to $g$ required to implement the oracle $\mO_g \ket{x}\ket{0} = \ket{x}\ket{g(x)}$, and $\alpha = 9.2$.
\end{lemma}

In our setting, we take $N_\textup{samples} = 0 $ and a failure probability of $\delta=1/1000$. The upper bound on the expected number of queries made by Grover's algorithm as given by~\cref{lem:QSearch} then simplifies to
\[
E_{\text{Grover}}(N,t) \leq 
\begin{cases}
    9.2 \lceil \log_3(1/\delta)\rceil \sqrt{N} = 64.4\sqrt{N} &\text{for} \quad t = 0 \\
    \frac{9}{4}\frac{N}{\sqrt{(N - t)t}} + \left\lceil\log_{\frac{6}{5}}\left(\frac{N}{2\sqrt{(N -t )t}} \right) \right\rceil - 3 \leq \frac{3.1 \sqrt{N}}{\sqrt{t}} &\text{for} \quad 1 \leq t < \frac{N}{4} \\
     2.0344 &\text{for} \quad \frac{N}{4} \leq t \leq N.
\end{cases}
\]

\section{The influence of the $T$-gate overhead factor on the cross-over time}\label{appendix:crossover_time}
For our estimates of classical and quantum scaling in solving a specific class of SAT instances, we focus on the \emph{crossover point}. This is defined as the time at which the expected execution times for solving a given instance size with both quantum and classical algorithms are equal, assuming the quantum algorithm offers a scaling advantage over the classical one. However, the relation between the quantum '`clock speed'' for $T$-gates, needed to primarily account for error correction, and this crossover point is not straightforward, as our results in~\cref{appendix:results} show. To explain this, let us do a simple calculation which explains this behaviour.

For a given class of SAT instances, parametrized in the number of variables $n$, recall that we estimate the classical run time $t_c$ as
\[
t_\textup{c} = 2^{s_\textup{c} \cdot n + i_c}, 
\]
where $s_\textup{c}$ and $i_c$ are parameters obtained through our fitting. Similarly, for the estimated $T$-count of the quantum algorithm, defined as $T_\textup{q}$, we have
\[
T_\textup{q} = 2^{s_\textup{q} \cdot n + i_q}, 
\]
where again $s_\textup{q}$ and $i_q$ are obtained parameters. We can now estimate the quantum run time $t_\textup{q}$ as
\[
t_\textup{q} = c_\textup{q} T_\textup{q},
\]
where $c_\textup{q}$ is the constant which characterizes the time needed to implement a single $T$ gate. The crossover point $t_\textup{crossover}$ is then determined as the time such that
\[
t_\textup{q} = t_\textup{c},
\]
which can be solved for $n$ to obtain
\[
n^* = \frac{\log_2 c_{\textup{q}} + i_q - i_c}{s_\textup{c} - s_\textup{q}}.
\]
Plutting this in our expression for $T_\textup{q}$ gives
\begin{align*}
t_\textup{crossover} &= c_{\textup{q}} 2^{s_\textup{q} \left( \frac{\log_2 c_{\textup{q}} + i_q - i_c}{s_\textup{c} - s_\textup{q}}\right) + i_q} \notag\\
&= c_{\textup{q}} \left(2^{\log_2 c_{\textup{q}}}\right)^{\frac{s_\textup{q}}{s_\textup{c}-s_\textup{q}}}\left(2^{\frac{s_\textup{q}(i_q - i_c)}{s_\textup{c}-s_\textup{q}}+i_q}\right) \notag\\
&= c_{\textup{q}}^{ \frac{s_\textup{q}}{s_\textup{c}-s_\textup{q}}+1} \left(2^{\frac{s_\textup{q}(i_q - i_c)}{s_\textup{c}-s_\textup{q}}+i_q}\right),
\end{align*}
which depends non-trivially on $c_q$ and the fitting parameters.

\section{The benefit of stopping earlier with multiple solutions}
\label{app:multiple_solutions}
In this appendix we present an argument as to why classical algorithms seem to perform much better on random satisfiable instances as compared to unsatisfiable instances, comparing both to Grover. Consider a search problem over $N$ elements where $t$ are promised to be marked, where instances are chosen uniformly at random. The query complexity of Grover is the same for all these instances, and scales as $\mO(\sqrt{N/t})$. Hence, the expected query complexity is also $\mO(\sqrt{N/t})$. However, for a deterministic classical solver, it is easy to show that the expected query complexity is $\mO(N/t)$, which means it scales quadratically better in $t$ as compared to Grover. We take as our brute force search algorithm the naive algorithm that simply checks for every element in $[N]$ whether it is marked in their standard ordering, and terminates once it sees a marked item. This means that the query complexity is equal to the index of the smallest marked elements, which is for picking $t$ out of $N$ elements uniformly at random equal to $i$ with probability
\[
   P(i)= \frac{\binom{N-i}{t-1}}{\binom{N}{t}}.
\]
Hence, the expected query complexity is equal to
\[
    \sum_{i=1}^N i P(i) =  \sum_{i=1}^N i \frac{\binom{N-i}{t-1}}{\binom{N}{t}} = \frac{N+1}{t+1} = \mO(N/t),
\]
which scales quadratically better than Grover in terms of $t$.

\newpage
\section{Additional numerical results}\label{appendix:results}
\begin{table*}[h]\label{tab:n_ranges}
\centering
\begin{tabular}{c|c|c|c|c|c|c|c|c}
\backslashbox{$k$}{$\beta$} & $\frac 12$ & 1 & $\frac 32$ & 2 & 3 & 5 & 10 & $\infty$\\\hline
3 & & & & & & 160 - 279 & 160 - 262 & 160 - 250\\
 & & & & & & 10 - 56 & 10 - 55 & 10 - 49\\\hline
4 & & & & & 64 - 130 & 64 - 113 & 64 - 103 & 64 - 94\\
 & & & & & 10 - 45 & 10 - 42 & 10 - 41 & 10 - 40\\\hline
5 & & & & 47 - 119 & 47 - 99 & 47 - 86 & 47 - 78 & 47 - 71\\
 & & & & 10 - 45 & 10 - 40 & 10 - 41 & 10 - 39 & 10 - 34\\\hline
6 & & & 30 - 91 & 30 - 85 & 30 - 71 & 30 - 46 & 30 - 60 & 30 - 54\\
 & & & 10 - 42 & 10 - 42 & 10 - 36 & 10 - 35 & 10 - 34 & 10 - 28\\\hline
7 & 25 - 77 & 25 - 105 & 25 - 83 & 25 - 71 & 25 - 60 & 25 - 53 & 25 - 49 & 25 - 44\\
 & 10 - 47 & 10 - 40 & 10 - 35 & 10 - 35 & 10 - 31 & 10 - 29 & 10 - 29 & 10 - 28\\\hline
8 & 20 - 84 & 20 - 80 & 20 - 63 & 20 - 61 & 20 - 52 & 20 - 47 & 20 - 40 & 20 - 40\\
 & 11 - 40 & 11 - 38 & 11 - 33 & 11 - 30 & 11 - 29 & 11 - 25 & 11 - 26 & 10 - 25\\\hline
9 & 15 - 75 & 15 - 66 & 15 - 54 & 15 - 45 & 15 - 39 & 15 - 36 & 15 - 33 & 15 - 37\\
 & 12 - 27 & 12 - 24 & 12 - 21 & 12 - 21 & 12 - 21 & 12 - 21 & 12 - 21 & 11 - 24\\\hline
10 & 15 - 102 & 15 - 73 & 15 - 54 & 15 - 47 & 15 - 42 & 15 - 38 & 15 - 36 & 15 - 33\\
 & 13 - 31 & 13 - 27 & 13 - 24 & 13 - 23 & 13 - 22 & 13 - 21 & 13 - 21 & 12 - 22\\\hline
11 & 14 - 67 & 14 - 50 & 14 - 40 & 14 - 37 & 14 - 34 & 14 - 33 & 14 - 32 & 14 - 32\\
 & 14 - 21 & 14 - 20 & 14 - 20 & 14 - 20 & 14 - 19 & 14 - 19 & 14 - 19 & 13 - 20\\\hline
12 & 15 - 56 & 15 - 41 & 15 - 35 & 15 - 33 & 15 - 28 & 15 - 29 & 15 - 28 & 15 - 30\\
 & 15 - 19 & 15 - 18 & 15 - 18 & 15 - 18 & 15 - 18 & 15 - 17 & 15 - 17 & 14 - 19\\\hline
\end{tabular}
\caption{Ranges of number of variables $n$ considered for each SAT distribution $(k,\beta)$. The top values in each cell are the range of $n$ for CaDiCaL 2.0.0 and the bottom values the range of $n$ for the backtracking algorithm. We consider every integer in these ranges. For each $n$, for CaDiCaL 2.0.0 we always sample 50 satisfiable and 50 unsatisfiable instance. For small $k$ and small $n$ the runtime initially grows quite fast, and only then stabilises into a pattern of the form $2^{a\cdot n + b}$. The relatively large lower-bounds on $n$ is due to us selecting the point where the runtime stabilises into this pattern. No such phenomenon occurs for backtracking. For backtracking we sample 30 satisfiable and 30 unsatisfiable instances. Where the ranges overlap it therefore possible that we use different instances for the two solvers.}
\end{table*}

\begin{table*}[ht]
    \centering
    $T$\textbf{-count}\\[5pt] 
    \begin{subtable}[t]{0.5\textwidth}
        \centering
        \begin{tabular}{p{0.5 cm}|p{0.22 cm}p{0.22 cm}p{0.22 cm}p{0.22 cm}p{0.22 cm}p{0.22 cm}p{0.22 cm}p{0.22 cm}}
            \resizebox{0.5cm}{!}{\backslashbox{$k$}{$\beta$}} & $\frac 12$ & 1 & $\frac 32$ & 2 & 3 & 5 & 10 & $\infty$\\\hline
            3& \cellcolor{C0}& \cellcolor{C0}& \cellcolor{C0}& \cellcolor{C0}& \cellcolor{C0} & \footnotesize \cellcolor{C0}  & \footnotesize \cellcolor{C0}  & \footnotesize \cellcolor{C0} \\
            4& \cellcolor{C0}& \cellcolor{C0}& \cellcolor{C0}& \cellcolor{C0} & \footnotesize \cellcolor{C0}  & \footnotesize \cellcolor{C0}  & \footnotesize \cellcolor{C0}  & \footnotesize \cellcolor{C0} \\
            5& \cellcolor{C0}& \cellcolor{C0}& \cellcolor{C0} & \footnotesize \cellcolor{C0}  & \footnotesize \cellcolor{C0}  & \footnotesize \cellcolor{C0}  & \footnotesize \cellcolor{C0}  & \footnotesize \cellcolor{C0} \\
            6& \cellcolor{C0}& \cellcolor{C0} & \footnotesize \cellcolor{C0}  & \footnotesize \cellcolor{C0}  & \footnotesize \cellcolor{C0}  & \footnotesize \cellcolor{C0}  & \footnotesize \cellcolor{C0}  & \footnotesize \cellcolor{C0} \\
            7 & \footnotesize \cellcolor{C0}  & \footnotesize \cellcolor{C0}  & \footnotesize \cellcolor{C0}  & \footnotesize \cellcolor{C0}  & \footnotesize \cellcolor{C0}  & \footnotesize \cellcolor{C0}  & \footnotesize \cellcolor{C0}  & \footnotesize \cellcolor{C0} \\
            8 & \footnotesize \cellcolor{C0}  & \footnotesize \cellcolor{C0}  & \footnotesize \cellcolor{C0}  & \footnotesize \cellcolor{C0}  & \footnotesize \cellcolor{C0}  & \footnotesize \cellcolor{C0}  & \footnotesize \cellcolor{C0}  & \footnotesize \cellcolor{C0} \\
            9 & \footnotesize \cellcolor{C0}  & \footnotesize \cellcolor{C0}  & \footnotesize \cellcolor{C0}  & \footnotesize \cellcolor{C0}  & \footnotesize \cellcolor{C0}  & \footnotesize \cellcolor{C0}  & \footnotesize \cellcolor{C0}  & \footnotesize \cellcolor{C0} \\
            10 & \footnotesize \cellcolor{C0}  & \footnotesize \cellcolor{C0}  & \footnotesize \cellcolor{C0}  & \footnotesize \cellcolor{C0}  & \footnotesize \cellcolor{C0}  & \footnotesize \cellcolor{C0}  & \footnotesize \cellcolor{C0}  & \footnotesize \cellcolor{C0} \\
            11 & \footnotesize \cellcolor{C0}  & \footnotesize \cellcolor{C0}  & \footnotesize \cellcolor{C0}  & \footnotesize \cellcolor{C0}  & \footnotesize \cellcolor{C1}  & \footnotesize \cellcolor{C1}  & \footnotesize \cellcolor{C1}  & \footnotesize \cellcolor{C0} \\
            12 & \footnotesize \cellcolor{C0}  & \footnotesize \cellcolor{C0}  & \footnotesize \cellcolor{C0}  & \footnotesize \cellcolor{C0}  & \footnotesize \cellcolor{C3}  & \footnotesize \cellcolor{C3}  & \footnotesize \cellcolor{C1}  & \footnotesize \cellcolor{C1} \\
            \end{tabular}
        \caption{Scaling, satisfiable}
        \label{tab:colors_sat_count}
    \end{subtable}
    \hfill
    \begin{subtable}[t]{0.48\textwidth}
        \centering
        \begin{tabular}{p{0.22 cm}p{0.22 cm}p{0.22 cm}p{0.22 cm}p{0.22 cm}p{0.22 cm}p{0.22 cm}p{0.22 cm}}
            $\frac 12$ & 1 & $\frac 32$ & 2 & 3 & 5 & 10 & $\infty$\\\hline
            \cellcolor{C0}& \cellcolor{C0}& \cellcolor{C0}& \cellcolor{C0}& \cellcolor{C0} & \footnotesize \cellcolor{C0}  & \footnotesize \cellcolor{C0}  & \footnotesize \cellcolor{C0} \\
            \cellcolor{C0}& \cellcolor{C0}& \cellcolor{C0}& \cellcolor{C0} & \footnotesize \cellcolor{C0}  & \footnotesize \cellcolor{C0}  & \footnotesize \cellcolor{C0}  & \footnotesize \cellcolor{C0} \\
            \cellcolor{C0}& \cellcolor{C0}& \cellcolor{C0} & \footnotesize \cellcolor{C0}  & \footnotesize \cellcolor{C0}  & \footnotesize \cellcolor{C0}  & \footnotesize \cellcolor{C0}  & \footnotesize \cellcolor{C0} \\
            \cellcolor{C0}& \cellcolor{C0} & \footnotesize \cellcolor{C0}  & \footnotesize \cellcolor{C0}  & \footnotesize \cellcolor{C0}  & \footnotesize \cellcolor{C0}  & \footnotesize \cellcolor{C0}  & \footnotesize \cellcolor{C0} \\
            \footnotesize \cellcolor{C0}  & \footnotesize \cellcolor{C0}  & \footnotesize \cellcolor{C0}  & \footnotesize \cellcolor{C0}  & \footnotesize \cellcolor{C0}  & \footnotesize \cellcolor{C0}  & \footnotesize \cellcolor{C0}  & \footnotesize \cellcolor{C0} \\
            \footnotesize \cellcolor{C0}  & \footnotesize \cellcolor{C0}  & \footnotesize \cellcolor{C0}  & \footnotesize \cellcolor{C0}  & \footnotesize \cellcolor{C0}  & \footnotesize \cellcolor{C0}  & \footnotesize \cellcolor{C1}  & \footnotesize \cellcolor{C0} \\
            \footnotesize \cellcolor{C0}  & \footnotesize \cellcolor{C0}  & \footnotesize \cellcolor{C0}  & \footnotesize \cellcolor{C0}  & \footnotesize \cellcolor{C0}  & \footnotesize \cellcolor{C0}  & \footnotesize \cellcolor{C0}  & \footnotesize \cellcolor{C0} \\
            \footnotesize \cellcolor{C0}  & \footnotesize \cellcolor{C0}  & \footnotesize \cellcolor{C0}  & \footnotesize \cellcolor{C0}  & \footnotesize \cellcolor{C0}  & \footnotesize \cellcolor{C1}  & \footnotesize \cellcolor{C1}  & \footnotesize \cellcolor{C1} \\
            \footnotesize \cellcolor{C0}  & \footnotesize \cellcolor{C0}  & \footnotesize \cellcolor{C0}  & \footnotesize \cellcolor{C0}  & \footnotesize \cellcolor{C1}  & \footnotesize \cellcolor{C1}  & \footnotesize \cellcolor{C1}  & \footnotesize \cellcolor{C1} \\
            \footnotesize \cellcolor{C0}  & \footnotesize \cellcolor{C0}  & \footnotesize \cellcolor{C0}  & \footnotesize \cellcolor{C0}  & \footnotesize \cellcolor{C1}  & \footnotesize \cellcolor{C1}  & \footnotesize \cellcolor{C1}  & \footnotesize \cellcolor{C2} \\
        \end{tabular}
        \caption{Scaling, unsatisfiable}
        \label{tab:colors_unsat_count}
    \end{subtable}
    \caption{These tables are defined the same way as \cref{tab:colors_scaling_depth}: each grid is coloured based on the scaling of the runtime and $T$-count of the algorithms.  Here (a) is restricted to satisfiable instances and (b) to unsatisfiable.}
\end{table*}

\begin{table} \hspace{-2.2cm} \begin{tabular}{p{0.9cm}|p{1.93cm}|p{1.77cm}|p{1.77cm}|p{1.77cm}|p{1.77cm}|p{1.77cm}|p{1.77cm}|p{1.77cm}} \backslashbox{$k$}{$\beta$} & \centering$\frac 12$ & \centering 1 & \centering $\frac 32$ & \centering 2 & \centering 3 & \centering 5 & \centering 10 & \ \ \ \ \ $\infty$ \\\hline
3& \cellcolor{C0}& \cellcolor{C0}& \cellcolor{C0}& \cellcolor{C0}& \cellcolor{C0} & \footnotesize \cellcolor{C0} $.0369n -11.0$  & \footnotesize \cellcolor{C0} $.0502n -12.6$  & \footnotesize \cellcolor{C0} $.0452n -11.4$ \\& \cellcolor{C0}& \cellcolor{C0}& \cellcolor{C0}& \cellcolor{C0}& \cellcolor{C0} & \footnotesize $.186n + 20.7$ \cellcolor{C0} & \footnotesize $.195n + 20.6$ \cellcolor{C0} & \footnotesize $.203n + 20.4$ \cellcolor{C0}
\\& \cellcolor{C0}& \cellcolor{C0}& \cellcolor{C0}& \cellcolor{C0}& \cellcolor{C0} & \footnotesize $.237n + 25.4$ \cellcolor{C0} & \footnotesize $.247n + 25.3$ \cellcolor{C0} & \footnotesize $.255n + 25.1$ \cellcolor{C0}
\\& \cellcolor{C0}& \cellcolor{C0}& \cellcolor{C0}& \cellcolor{C0}& \cellcolor{C0} & \footnotesize $.508n + 9.73$ \cellcolor{C0} & \footnotesize $.508n + 9.72$ \cellcolor{C0} & \footnotesize $.507n + 9.25$ \cellcolor{C0}
\\\hline
4& \cellcolor{C0}& \cellcolor{C0}& \cellcolor{C0}& \cellcolor{C0} & \footnotesize \cellcolor{C0} $.0724n -9.6$  & \footnotesize \cellcolor{C0} $.13n -12.7$  & \footnotesize \cellcolor{C0} $.165n -14.4$  & \footnotesize \cellcolor{C0} $.171n -14.2$ \\& \cellcolor{C0}& \cellcolor{C0}& \cellcolor{C0}& \cellcolor{C0} & \footnotesize $.226n + 20.5$ \cellcolor{C0} & \footnotesize $.247n + 20.4$ \cellcolor{C0} & \footnotesize $.26n + 20.3$ \cellcolor{C0} & \footnotesize $.263n + 20.3$ \cellcolor{C0}
\\& \cellcolor{C0}& \cellcolor{C0}& \cellcolor{C0}& \cellcolor{C0} & \footnotesize $.287n + 25.0$ \cellcolor{C0} & \footnotesize $.309n + 24.9$ \cellcolor{C0} & \footnotesize $.322n + 24.8$ \cellcolor{C0} & \footnotesize $.323n + 24.8$ \cellcolor{C0}
\\& \cellcolor{C0}& \cellcolor{C0}& \cellcolor{C0}& \cellcolor{C0} & \footnotesize $.511n + 9.79$ \cellcolor{C0} & \footnotesize $.512n + 9.77$ \cellcolor{C0} & \footnotesize $.512n + 9.77$ \cellcolor{C0} & \footnotesize $.509n + 9.36$ \cellcolor{C0}
\\\hline
5& \cellcolor{C0}& \cellcolor{C0}& \cellcolor{C0} & \footnotesize \cellcolor{C0} $.0951n -9.27$  & \footnotesize \cellcolor{C0} $.18n -12.7$  & \footnotesize \cellcolor{C0} $.245n -14.7$  & \footnotesize \cellcolor{C0} $.29n -15.9$  & \footnotesize \cellcolor{C0} $.3n -15.7$ \\& \cellcolor{C0}& \cellcolor{C0}& \cellcolor{C0} & \footnotesize $.235n + 20.7$ \cellcolor{C0} & \footnotesize $.268n + 20.4$ \cellcolor{C0} & \footnotesize $.286n + 20.3$ \cellcolor{C0} & \footnotesize $.299n + 20.2$ \cellcolor{C0} & \footnotesize $.31n + 20.1$ \cellcolor{C0}
\\& \cellcolor{C0}& \cellcolor{C0}& \cellcolor{C0} & \footnotesize $.294n + 25.3$ \cellcolor{C0} & \footnotesize $.333n + 24.9$ \cellcolor{C0} & \footnotesize $.351n + 24.7$ \cellcolor{C0} & \footnotesize $.362n + 24.7$ \cellcolor{C0} & \footnotesize $.385n + 24.3$ \cellcolor{C0}
\\& \cellcolor{C0}& \cellcolor{C0}& \cellcolor{C0} & \footnotesize $.509n + 10.1$ \cellcolor{C0} & \footnotesize $.51n + 10.1$ \cellcolor{C0} & \footnotesize $.51n + 10.1$ \cellcolor{C0} & \footnotesize $.51n + 10.1$ \cellcolor{C0} & \footnotesize $.507n + 9.77$ \cellcolor{C0}
\\\hline
6& \cellcolor{C0}& \cellcolor{C0} & \footnotesize \cellcolor{C0} $.081n -7.4$  & \footnotesize \cellcolor{C0} $.164n -10.1$  & \footnotesize \cellcolor{C0} $.249n -12.2$  & \footnotesize \cellcolor{C1} $.327n -13.8$  & \footnotesize \cellcolor{C1} $.34n -13.5$  & \footnotesize \cellcolor{C1} $.39n -15.4$ \\& \cellcolor{C0}& \cellcolor{C0} & \footnotesize $.24n + 20.9$ \cellcolor{C0} & \footnotesize $.266n + 20.7$ \cellcolor{C0} & \footnotesize $.3n + 20.4$ \cellcolor{C0} & \footnotesize $.319n + 20.2$ \cellcolor{C1} & \footnotesize $.331n + 20.1$ \cellcolor{C1} & \footnotesize $.343n + 20.0$ \cellcolor{C1}
\\& \cellcolor{C0}& \cellcolor{C0} & \footnotesize $.303n + 25.4$ \cellcolor{C0} & \footnotesize $.331n + 25.1$ \cellcolor{C0} & \footnotesize $.372n + 24.6$ \cellcolor{C0} & \footnotesize $.394n + 24.4$ \cellcolor{C1} & \footnotesize $.404n + 24.4$ \cellcolor{C1} & \footnotesize $.425n + 24.2$ \cellcolor{C1}
\\& \cellcolor{C0}& \cellcolor{C0} & \footnotesize $.509n + 10.2$ \cellcolor{C0} & \footnotesize $.509n + 10.2$ \cellcolor{C0} & \footnotesize $.51n + 10.2$ \cellcolor{C0} & \footnotesize $.51n + 10.2$ \cellcolor{C1} & \footnotesize $.51n + 10.2$ \cellcolor{C1} & \footnotesize $.506n + 9.89$ \cellcolor{C1}
\\\hline
7 & \footnotesize \cellcolor{C0} $.00334n -4.19$  & \footnotesize \cellcolor{C0} $.0411n -5.58$  & \footnotesize \cellcolor{C0} $.163n -9.37$  & \footnotesize \cellcolor{C0} $.244n -11.3$  & \footnotesize \cellcolor{C1} $.329n -12.9$  & \footnotesize \cellcolor{C1} $.396n -14.1$  & \footnotesize \cellcolor{C2} $.444n -14.9$  & \footnotesize \cellcolor{C2} $.466n -15.3$ \\ & \footnotesize $.157n + 22.1$ \cellcolor{C0} & \footnotesize $.225n + 21.3$ \cellcolor{C0} & \footnotesize $.274n + 20.8$ \cellcolor{C0} & \footnotesize $.298n + 20.6$ \cellcolor{C0} & \footnotesize $.326n + 20.3$ \cellcolor{C1} & \footnotesize $.347n + 20.1$ \cellcolor{C1} & \footnotesize $.358n + 20.1$ \cellcolor{C2} & \footnotesize $.367n + 20.0$ \cellcolor{C2}
\\ & \footnotesize $.214n + 26.7$ \cellcolor{C0} & \footnotesize $.287n + 25.8$ \cellcolor{C0} & \footnotesize $.349n + 25.0$ \cellcolor{C0} & \footnotesize $.367n + 24.9$ \cellcolor{C0} & \footnotesize $.401n + 24.6$ \cellcolor{C1} & \footnotesize $.426n + 24.3$ \cellcolor{C1} & \footnotesize $.439n + 24.2$ \cellcolor{C2} & \footnotesize $.449n + 24.1$ \cellcolor{C2}
\\ & \footnotesize $.507n + 10.3$ \cellcolor{C0} & \footnotesize $.509n + 10.3$ \cellcolor{C0} & \footnotesize $.509n + 10.3$ \cellcolor{C0} & \footnotesize $.509n + 10.3$ \cellcolor{C0} & \footnotesize $.51n + 10.2$ \cellcolor{C1} & \footnotesize $.51n + 10.2$ \cellcolor{C1} & \footnotesize $.51n + 10.2$ \cellcolor{C2} & \footnotesize $.511n + 10.2$ \cellcolor{C2}
\\\hline
8 & \footnotesize \cellcolor{C0} $.00234n -3.95$  & \footnotesize \cellcolor{C0} $.0877n -6.53$  & \footnotesize \cellcolor{C0} $.212n -9.42$  & \footnotesize \cellcolor{C0} $.288n -10.9$  & \footnotesize \cellcolor{C1} $.368n -12.2$  & \footnotesize \cellcolor{C1} $.433n -13.3$  & \footnotesize \cellcolor{C2} $.474n -13.9$  & \footnotesize \cellcolor{C2} $.492n -14.1$ \\ & \footnotesize $.194n + 21.9$ \cellcolor{C0} & \footnotesize $.248n + 21.4$ \cellcolor{C0} & \footnotesize $.295n + 20.8$ \cellcolor{C0} & \footnotesize $.32n + 20.6$ \cellcolor{C0} & \footnotesize $.344n + 20.3$ \cellcolor{C1} & \footnotesize $.368n + 20.1$ \cellcolor{C1} & \footnotesize $.377n + 20.1$ \cellcolor{C2} & \footnotesize $.389n + 19.9$ \cellcolor{C2}
\\ & \footnotesize $.263n + 26.3$ \cellcolor{C0} & \footnotesize $.311n + 25.9$ \cellcolor{C0} & \footnotesize $.371n + 25.1$ \cellcolor{C0} & \footnotesize $.394n + 24.9$ \cellcolor{C0} & \footnotesize $.421n + 24.6$ \cellcolor{C1} & \footnotesize $.464n + 24.0$ \cellcolor{C1} & \footnotesize $.465n + 24.1$ \cellcolor{C2} & \footnotesize $.473n + 24.0$ \cellcolor{C2}
\\ & \footnotesize $.508n + 10.4$ \cellcolor{C0} & \footnotesize $.508n + 10.3$ \cellcolor{C0} & \footnotesize $.509n + 10.3$ \cellcolor{C0} & \footnotesize $.509n + 10.3$ \cellcolor{C0} & \footnotesize $.51n + 10.3$ \cellcolor{C1} & \footnotesize $.51n + 10.3$ \cellcolor{C1} & \footnotesize $.51n + 10.3$ \cellcolor{C2} & \footnotesize $.511n + 10.3$ \cellcolor{C2}
\\\hline
9 & \footnotesize \cellcolor{C0} $.00513n -3.81$  & \footnotesize \cellcolor{C0} $.115n -6.42$  & \footnotesize \cellcolor{C0} $.225n -8.22$  & \footnotesize \cellcolor{C0} $.281n -8.95$  & \footnotesize \cellcolor{C0} $.345n -9.82$  & \footnotesize \cellcolor{C1} $.401n -10.6$  & \footnotesize \cellcolor{C1} $.443n -11.3$  & \footnotesize \cellcolor{C2} $.486n -12.3$ \\ & \footnotesize $.235n + 21.6$ \cellcolor{C0} & \footnotesize $.291n + 21.0$ \cellcolor{C0} & \footnotesize $.329n + 20.6$ \cellcolor{C0} & \footnotesize $.351n + 20.4$ \cellcolor{C0} & \footnotesize $.371n + 20.2$ \cellcolor{C0} & \footnotesize $.386n + 20.1$ \cellcolor{C1} & \footnotesize $.396n + 20.0$ \cellcolor{C1} & \footnotesize $.401n + 20.0$ \cellcolor{C2}
\\ & \footnotesize $.314n + 25.7$ \cellcolor{C0} & \footnotesize $.383n + 25.0$ \cellcolor{C0} & \footnotesize $.411n + 24.8$ \cellcolor{C0} & \footnotesize $.428n + 24.7$ \cellcolor{C0} & \footnotesize $.478n + 24.0$ \cellcolor{C0} & \footnotesize $.478n + 24.0$ \cellcolor{C1} & \footnotesize $.528n + 23.3$ \cellcolor{C1} & \footnotesize $.484n + 24.1$ \cellcolor{C2}
\\ & \footnotesize $.508n + 10.6$ \cellcolor{C0} & \footnotesize $.509n + 10.5$ \cellcolor{C0} & \footnotesize $.506n + 10.6$ \cellcolor{C0} & \footnotesize $.506n + 10.6$ \cellcolor{C0} & \footnotesize $.506n + 10.6$ \cellcolor{C0} & \footnotesize $.506n + 10.6$ \cellcolor{C1} & \footnotesize $.506n + 10.6$ \cellcolor{C1} & \footnotesize $.508n + 10.5$ \cellcolor{C2}
\\\hline
10 & \footnotesize \cellcolor{C0} $.0196n -3.67$  & \footnotesize \cellcolor{C0} $.182n -6.98$  & \footnotesize \cellcolor{C0} $.299n -8.9$  & \footnotesize \cellcolor{C1} $.366n -9.81$  & \footnotesize \cellcolor{C1} $.439n -10.7$  & \footnotesize \cellcolor{C2} $.499n -11.5$  & \footnotesize \cellcolor{C2} $.543n -12.2$  & \footnotesize \cellcolor{C2} $.529n -12.2$ \\ & \footnotesize $.237n + 22.0$ \cellcolor{C0} & \footnotesize $.297n + 21.3$ \cellcolor{C0} & \footnotesize $.335n + 20.8$ \cellcolor{C0} & \footnotesize $.356n + 20.5$ \cellcolor{C1} & \footnotesize $.378n + 20.3$ \cellcolor{C1} & \footnotesize $.396n + 20.1$ \cellcolor{C2} & \footnotesize $.405n + 20.1$ \cellcolor{C2} & \footnotesize $.413n + 20.0$ \cellcolor{C2}
\\ & \footnotesize $.301n + 26.6$ \cellcolor{C0} & \footnotesize $.368n + 25.6$ \cellcolor{C0} & \footnotesize $.423n + 24.9$ \cellcolor{C0} & \footnotesize $.44n + 24.7$ \cellcolor{C1} & \footnotesize $.462n + 24.4$ \cellcolor{C1} & \footnotesize $.49n + 24.1$ \cellcolor{C2} & \footnotesize $.506n + 23.9$ \cellcolor{C2} & \footnotesize $.494n + 24.2$ \cellcolor{C2}
\\ & \footnotesize $.507n + 10.6$ \cellcolor{C0} & \footnotesize $.508n + 10.6$ \cellcolor{C0} & \footnotesize $.508n + 10.6$ \cellcolor{C0} & \footnotesize $.506n + 10.6$ \cellcolor{C1} & \footnotesize $.507n + 10.6$ \cellcolor{C1} & \footnotesize $.508n + 10.6$ \cellcolor{C2} & \footnotesize $.508n + 10.6$ \cellcolor{C2} & \footnotesize $.506n + 10.6$ \cellcolor{C2}
\\\hline
11 & \footnotesize \cellcolor{C0} $.0804n -4.04$  & \footnotesize \cellcolor{C0} $.245n -6.81$  & \footnotesize \cellcolor{C1} $.366n -8.63$  & \footnotesize \cellcolor{C1} $.432n -9.52$  & \footnotesize \cellcolor{C2} $.507n -10.5$  & \footnotesize \cellcolor{C3} $.543n -10.8$  & \footnotesize \cellcolor{C3} $.576n -11.2$  & \footnotesize \cellcolor{C3} $.563n -11.8$ \\ & \footnotesize $.282n + 21.6$ \cellcolor{C0} & \footnotesize $.327n + 21.0$ \cellcolor{C0} & \footnotesize $.358n + 20.7$ \cellcolor{C1} & \footnotesize $.368n + 20.6$ \cellcolor{C1} & \footnotesize $.394n + 20.3$ \cellcolor{C2} & \footnotesize $.409n + 20.1$ \cellcolor{C3} & \footnotesize $.417n + 20.1$ \cellcolor{C3} & \footnotesize $.424n + 20.0$ \cellcolor{C3}
\\ & \footnotesize $.366n + 25.8$ \cellcolor{C0} & \footnotesize $.389n + 25.5$ \cellcolor{C0} & \footnotesize $.479n + 24.2$ \cellcolor{C1} & \footnotesize $.461n + 24.7$ \cellcolor{C1} & \footnotesize $.46n + 24.7$ \cellcolor{C2} & \footnotesize $.517n + 23.9$ \cellcolor{C3} & \footnotesize $.531n + 23.7$ \cellcolor{C3} & \footnotesize $.514n + 24.1$ \cellcolor{C3}
\\ & \footnotesize $.509n + 10.7$ \cellcolor{C0} & \footnotesize $.511n + 10.6$ \cellcolor{C0} & \footnotesize $.511n + 10.6$ \cellcolor{C1} & \footnotesize $.511n + 10.6$ \cellcolor{C1} & \footnotesize $.513n + 10.6$ \cellcolor{C2} & \footnotesize $.513n + 10.6$ \cellcolor{C3} & \footnotesize $.513n + 10.6$ \cellcolor{C3} & \footnotesize $.509n + 10.7$ \cellcolor{C3}
\\\hline
12 & \footnotesize \cellcolor{C0} $.0888n -2.24$  & \footnotesize \cellcolor{C0} $.249n -5.02$  & \footnotesize \cellcolor{C0} $.355n -6.63$  & \footnotesize \cellcolor{C1} $.417n -7.53$  & \footnotesize \cellcolor{C2} $.512n -9.01$  & \footnotesize \cellcolor{C3} $.555n -9.47$  & \footnotesize \cellcolor{C3} $.587n -9.87$  & \footnotesize \cellcolor{C2} $.593n -11.3$ \\ & \footnotesize $.299n + 21.7$ \cellcolor{C0} & \footnotesize $.343n + 21.1$ \cellcolor{C0} & \footnotesize $.356n + 20.9$ \cellcolor{C0} & \footnotesize $.386n + 20.5$ \cellcolor{C1} & \footnotesize $.402n + 20.3$ \cellcolor{C2} & \footnotesize $.423n + 20.1$ \cellcolor{C3} & \footnotesize $.43n + 20.0$ \cellcolor{C3} & \footnotesize $.431n + 20.1$ \cellcolor{C2}
\\ & \footnotesize $.389n + 25.8$ \cellcolor{C0} & \footnotesize $.497n + 24.1$ \cellcolor{C0} & \footnotesize $.447n + 25.1$ \cellcolor{C0} & \footnotesize $.447n + 25.1$ \cellcolor{C1} & \footnotesize $.447n + 25.1$ \cellcolor{C2} & \footnotesize $.656n + 21.9$ \cellcolor{C3} & \footnotesize $.656n + 21.9$ \cellcolor{C3} & \footnotesize $.502n + 24.4$ \cellcolor{C2}
\\ & \footnotesize $.515n + 10.6$ \cellcolor{C0} & \footnotesize $.566n + 9.66$ \cellcolor{C0} & \footnotesize $.52n + 10.5$ \cellcolor{C0} & \footnotesize $.52n + 10.5$ \cellcolor{C1} & \footnotesize $.52n + 10.5$ \cellcolor{C2} & \footnotesize $.525n + 10.4$ \cellcolor{C3} & \footnotesize $.525n + 10.3$ \cellcolor{C3} & \footnotesize $.512n + 10.7$ \cellcolor{C2}
\\\hline
\end{tabular}
\caption{Each grid in these tables represents a distribution of $k$-SAT instances with differing amounts of structure $\beta$, fixed to contain only an equal mix of \textit{satisfiable} and \textit{unsatisfiable }instances (where lower $\beta$ means more structure). For each grid we sample SAT instances and fit the classical runtime and quantum \textit{$T$-depth} of the form $2^{a\cdot n + b}$. Each cell contains the exponent $a\cdot n+b$ in the order: classical runtime, quantum backtracking detection, quantum backtracking search, Grover. We color a cell blue if the classical algorithm scales best (i.e. has the lowest $a$ value), orange if the detection algorithm scales best, yellow if Grover scales best, and green if search scales better than classical and Grover.}
\label{tab:colors_both_depth}
\end{table}
\begin{table} \hspace{-2.2cm} \begin{tabular}{p{0.9cm}|p{1.93cm}|p{1.77cm}|p{1.77cm}|p{1.77cm}|p{1.77cm}|p{1.77cm}|p{1.77cm}|p{1.77cm}} \backslashbox{$k$}{$\beta$} & \centering$\frac 12$ & \centering 1 & \centering $\frac 32$ & \centering 2 & \centering 3 & \centering 5 & \centering 10 & \ \ \ \ \ $\infty$ \\\hline3& \cellcolor{C0}& \cellcolor{C0}& \cellcolor{C0}& \cellcolor{C0}& \cellcolor{C0} & \footnotesize \cellcolor{C0} $.0299n -9.89$  & \footnotesize \cellcolor{C0} $.0356n -10.5$  & \footnotesize \cellcolor{C0} $.0327n -9.93$ \\& \cellcolor{C0}& \cellcolor{C0}& \cellcolor{C0}& \cellcolor{C0}& \cellcolor{C0} & \footnotesize $.187n + 20.8$ \cellcolor{C0} & \footnotesize $.196n + 20.7$ \cellcolor{C0} & \footnotesize $.205n + 20.5$ \cellcolor{C0}
\\& \cellcolor{C0}& \cellcolor{C0}& \cellcolor{C0}& \cellcolor{C0}& \cellcolor{C0} & \footnotesize $.237n + 25.6$ \cellcolor{C0} & \footnotesize $.248n + 25.4$ \cellcolor{C0} & \footnotesize $.258n + 25.1$ \cellcolor{C0}
\\& \cellcolor{C0}& \cellcolor{C0}& \cellcolor{C0}& \cellcolor{C0}& \cellcolor{C0} & \footnotesize $.44n + 6.82$ \cellcolor{C0} & \footnotesize $.443n + 6.8$ \cellcolor{C0} & \footnotesize $.436n + 6.5$ \cellcolor{C0}
\\\hline
4& \cellcolor{C0}& \cellcolor{C0}& \cellcolor{C0}& \cellcolor{C0} & \footnotesize \cellcolor{C0} $.054n -8.35$  & \footnotesize \cellcolor{C0} $.0582n -7.84$  & \footnotesize \cellcolor{C0} $.0571n -7.32$  & \footnotesize \cellcolor{C0} $.0617n -7.25$ \\& \cellcolor{C0}& \cellcolor{C0}& \cellcolor{C0}& \cellcolor{C0} & \footnotesize $.227n + 20.6$ \cellcolor{C0} & \footnotesize $.247n + 20.4$ \cellcolor{C0} & \footnotesize $.26n + 20.3$ \cellcolor{C0} & \footnotesize $.262n + 20.3$ \cellcolor{C0}
\\& \cellcolor{C0}& \cellcolor{C0}& \cellcolor{C0}& \cellcolor{C0} & \footnotesize $.291n + 25.0$ \cellcolor{C0} & \footnotesize $.309n + 24.9$ \cellcolor{C0} & \footnotesize $.324n + 24.7$ \cellcolor{C0} & \footnotesize $.323n + 24.8$ \cellcolor{C0}
\\& \cellcolor{C0}& \cellcolor{C0}& \cellcolor{C0}& \cellcolor{C0} & \footnotesize $.477n + 6.97$ \cellcolor{C0} & \footnotesize $.472n + 7.08$ \cellcolor{C0} & \footnotesize $.479n + 7.0$ \cellcolor{C0} & \footnotesize $.465n + 6.81$ \cellcolor{C0}
\\\hline
5& \cellcolor{C0}& \cellcolor{C0}& \cellcolor{C0} & \footnotesize \cellcolor{C0} $.0713n -7.98$  & \footnotesize \cellcolor{C0} $.103n -8.92$  & \footnotesize \cellcolor{C0} $.0892n -7.25$  & \footnotesize \cellcolor{C0} $.0885n -6.93$  & \footnotesize \cellcolor{C0} $.0587n -5.12$ \\& \cellcolor{C0}& \cellcolor{C0}& \cellcolor{C0} & \footnotesize $.235n + 20.8$ \cellcolor{C0} & \footnotesize $.268n + 20.4$ \cellcolor{C0} & \footnotesize $.286n + 20.3$ \cellcolor{C0} & \footnotesize $.299n + 20.2$ \cellcolor{C0} & \footnotesize $.309n + 20.1$ \cellcolor{C0}
\\& \cellcolor{C0}& \cellcolor{C0}& \cellcolor{C0} & \footnotesize $.295n + 25.3$ \cellcolor{C0} & \footnotesize $.332n + 24.9$ \cellcolor{C0} & \footnotesize $.351n + 24.7$ \cellcolor{C0} & \footnotesize $.366n + 24.6$ \cellcolor{C0} & \footnotesize $.385n + 24.3$ \cellcolor{C0}
\\& \cellcolor{C0}& \cellcolor{C0}& \cellcolor{C0} & \footnotesize $.483n + 7.46$ \cellcolor{C0} & \footnotesize $.495n + 7.17$ \cellcolor{C0} & \footnotesize $.485n + 7.45$ \cellcolor{C0} & \footnotesize $.485n + 7.44$ \cellcolor{C0} & \footnotesize $.474n + 7.34$ \cellcolor{C0}
\\\hline
6& \cellcolor{C0}& \cellcolor{C0} & \footnotesize \cellcolor{C0} $.065n -6.83$  & \footnotesize \cellcolor{C0} $.126n -8.91$  & \footnotesize \cellcolor{C0} $.174n -10.0$  & \footnotesize \cellcolor{C0} $.187n -9.88$  & \footnotesize \cellcolor{C0} $.187n -9.64$  & \footnotesize \cellcolor{C0} $.167n -8.64$ \\& \cellcolor{C0}& \cellcolor{C0} & \footnotesize $.239n + 20.9$ \cellcolor{C0} & \footnotesize $.267n + 20.7$ \cellcolor{C0} & \footnotesize $.299n + 20.4$ \cellcolor{C0} & \footnotesize $.32n + 20.2$ \cellcolor{C0} & \footnotesize $.331n + 20.1$ \cellcolor{C0} & \footnotesize $.344n + 20.0$ \cellcolor{C0}
\\& \cellcolor{C0}& \cellcolor{C0} & \footnotesize $.3n + 25.4$ \cellcolor{C0} & \footnotesize $.33n + 25.1$ \cellcolor{C0} & \footnotesize $.372n + 24.7$ \cellcolor{C0} & \footnotesize $.393n + 24.5$ \cellcolor{C0} & \footnotesize $.404n + 24.4$ \cellcolor{C0} & \footnotesize $.425n + 24.2$ \cellcolor{C0}
\\& \cellcolor{C0}& \cellcolor{C0} & \footnotesize $.484n + 7.66$ \cellcolor{C0} & \footnotesize $.489n + 7.54$ \cellcolor{C0} & \footnotesize $.494n + 7.41$ \cellcolor{C0} & \footnotesize $.488n + 7.72$ \cellcolor{C0} & \footnotesize $.501n + 7.37$ \cellcolor{C0} & \footnotesize $.47n + 7.53$ \cellcolor{C0}
\\\hline
7 & \footnotesize \cellcolor{C0} $.00228n -4.15$  & \footnotesize \cellcolor{C0} $.0336n -5.35$  & \footnotesize \cellcolor{C0} $.138n -8.77$  & \footnotesize \cellcolor{C0} $.205n -10.5$  & \footnotesize \cellcolor{C0} $.256n -11.4$  & \footnotesize \cellcolor{C0} $.264n -11.1$  & \footnotesize \cellcolor{C0} $.295n -11.7$  & \footnotesize \cellcolor{C0} $.255n -10.2$ \\ & \footnotesize $.158n + 22.1$ \cellcolor{C0} & \footnotesize $.224n + 21.3$ \cellcolor{C0} & \footnotesize $.274n + 20.8$ \cellcolor{C0} & \footnotesize $.297n + 20.6$ \cellcolor{C0} & \footnotesize $.326n + 20.3$ \cellcolor{C0} & \footnotesize $.348n + 20.1$ \cellcolor{C0} & \footnotesize $.358n + 20.1$ \cellcolor{C0} & \footnotesize $.368n + 20.0$ \cellcolor{C0}
\\ & \footnotesize $.217n + 26.6$ \cellcolor{C0} & \footnotesize $.289n + 25.8$ \cellcolor{C0} & \footnotesize $.351n + 25.0$ \cellcolor{C0} & \footnotesize $.367n + 24.9$ \cellcolor{C0} & \footnotesize $.401n + 24.6$ \cellcolor{C0} & \footnotesize $.427n + 24.3$ \cellcolor{C0} & \footnotesize $.439n + 24.2$ \cellcolor{C0} & \footnotesize $.449n + 24.1$ \cellcolor{C0}
\\ & \footnotesize $.458n + 8.12$ \cellcolor{C0} & \footnotesize $.487n + 7.62$ \cellcolor{C0} & \footnotesize $.487n + 7.73$ \cellcolor{C0} & \footnotesize $.483n + 7.87$ \cellcolor{C0} & \footnotesize $.496n + 7.55$ \cellcolor{C0} & \footnotesize $.492n + 7.72$ \cellcolor{C0} & \footnotesize $.473n + 8.12$ \cellcolor{C0} & \footnotesize $.5n + 7.71$ \cellcolor{C0}
\\\hline
8 & \footnotesize \cellcolor{C0} $.00223n -3.97$  & \footnotesize \cellcolor{C0} $.073n -6.2$  & \footnotesize \cellcolor{C0} $.183n -8.95$  & \footnotesize \cellcolor{C0} $.244n -10.2$  & \footnotesize \cellcolor{C0} $.288n -10.7$  & \footnotesize \cellcolor{C0} $.328n -11.5$  & \footnotesize \cellcolor{C0} $.349n -11.7$  & \footnotesize \cellcolor{C0} $.3n -10.2$ \\ & \footnotesize $.196n + 21.9$ \cellcolor{C0} & \footnotesize $.249n + 21.4$ \cellcolor{C0} & \footnotesize $.295n + 20.8$ \cellcolor{C0} & \footnotesize $.319n + 20.6$ \cellcolor{C0} & \footnotesize $.344n + 20.3$ \cellcolor{C0} & \footnotesize $.368n + 20.1$ \cellcolor{C0} & \footnotesize $.377n + 20.1$ \cellcolor{C0} & \footnotesize $.389n + 19.9$ \cellcolor{C0}
\\ & \footnotesize $.264n + 26.3$ \cellcolor{C0} & \footnotesize $.311n + 25.9$ \cellcolor{C0} & \footnotesize $.371n + 25.1$ \cellcolor{C0} & \footnotesize $.391n + 24.9$ \cellcolor{C0} & \footnotesize $.421n + 24.6$ \cellcolor{C0} & \footnotesize $.464n + 24.0$ \cellcolor{C0} & \footnotesize $.465n + 24.1$ \cellcolor{C0} & \footnotesize $.473n + 24.0$ \cellcolor{C0}
\\ & \footnotesize $.466n + 8.21$ \cellcolor{C0} & \footnotesize $.488n + 7.86$ \cellcolor{C0} & \footnotesize $.489n + 7.83$ \cellcolor{C0} & \footnotesize $.478n + 8.12$ \cellcolor{C0} & \footnotesize $.496n + 7.76$ \cellcolor{C0} & \footnotesize $.497n + 7.69$ \cellcolor{C0} & \footnotesize $.491n + 7.93$ \cellcolor{C0} & \footnotesize $.481n + 8.11$ \cellcolor{C0}
\\\hline
9 & \footnotesize \cellcolor{C0} $.00368n -3.83$  & \footnotesize \cellcolor{C0} $.1n -6.29$  & \footnotesize \cellcolor{C0} $.196n -7.97$  & \footnotesize \cellcolor{C0} $.226n -8.3$  & \footnotesize \cellcolor{C0} $.277n -8.93$  & \footnotesize \cellcolor{C0} $.292n -9.07$  & \footnotesize \cellcolor{C0} $.332n -9.8$  & \footnotesize \cellcolor{C0} $.343n -10.1$ \\ & \footnotesize $.23n + 21.7$ \cellcolor{C0} & \footnotesize $.295n + 21.0$ \cellcolor{C0} & \footnotesize $.329n + 20.6$ \cellcolor{C0} & \footnotesize $.35n + 20.4$ \cellcolor{C0} & \footnotesize $.37n + 20.2$ \cellcolor{C0} & \footnotesize $.385n + 20.1$ \cellcolor{C0} & \footnotesize $.396n + 20.0$ \cellcolor{C0} & \footnotesize $.401n + 20.0$ \cellcolor{C0}
\\ & \footnotesize $.314n + 25.7$ \cellcolor{C0} & \footnotesize $.383n + 25.0$ \cellcolor{C0} & \footnotesize $.411n + 24.8$ \cellcolor{C0} & \footnotesize $.428n + 24.7$ \cellcolor{C0} & \footnotesize $.478n + 24.0$ \cellcolor{C0} & \footnotesize $.478n + 24.0$ \cellcolor{C0} & \footnotesize $.528n + 23.3$ \cellcolor{C0} & \footnotesize $.484n + 24.1$ \cellcolor{C0}
\\ & \footnotesize $.452n + 8.79$ \cellcolor{C0} & \footnotesize $.511n + 7.73$ \cellcolor{C0} & \footnotesize $.504n + 7.84$ \cellcolor{C0} & \footnotesize $.447n + 8.63$ \cellcolor{C0} & \footnotesize $.499n + 7.79$ \cellcolor{C0} & \footnotesize $.41n + 9.24$ \cellcolor{C0} & \footnotesize $.486n + 8.28$ \cellcolor{C0} & \footnotesize $.494n + 8.09$ \cellcolor{C0}
\\\hline
10 & \footnotesize \cellcolor{C0} $.016n -3.73$  & \footnotesize \cellcolor{C0} $.172n -7.24$  & \footnotesize \cellcolor{C0} $.281n -9.14$  & \footnotesize \cellcolor{C0} $.339n -10.0$  & \footnotesize \cellcolor{C1} $.389n -10.6$  & \footnotesize \cellcolor{C1} $.439n -11.4$  & \footnotesize \cellcolor{C1} $.466n -11.8$  & \footnotesize \cellcolor{C0} $.412n -10.8$ \\ & \footnotesize $.24n + 22.0$ \cellcolor{C0} & \footnotesize $.298n + 21.2$ \cellcolor{C0} & \footnotesize $.335n + 20.8$ \cellcolor{C0} & \footnotesize $.357n + 20.5$ \cellcolor{C0} & \footnotesize $.378n + 20.3$ \cellcolor{C1} & \footnotesize $.396n + 20.1$ \cellcolor{C1} & \footnotesize $.405n + 20.1$ \cellcolor{C1} & \footnotesize $.412n + 20.0$ \cellcolor{C0}
\\ & \footnotesize $.308n + 26.5$ \cellcolor{C0} & \footnotesize $.368n + 25.6$ \cellcolor{C0} & \footnotesize $.423n + 24.9$ \cellcolor{C0} & \footnotesize $.44n + 24.7$ \cellcolor{C0} & \footnotesize $.462n + 24.4$ \cellcolor{C1} & \footnotesize $.49n + 24.1$ \cellcolor{C1} & \footnotesize $.506n + 23.9$ \cellcolor{C1} & \footnotesize $.494n + 24.2$ \cellcolor{C0}
\\ & \footnotesize $.484n + 8.24$ \cellcolor{C0} & \footnotesize $.492n + 8.0$ \cellcolor{C0} & \footnotesize $.499n + 7.89$ \cellcolor{C0} & \footnotesize $.516n + 7.73$ \cellcolor{C0} & \footnotesize $.463n + 8.64$ \cellcolor{C1} & \footnotesize $.449n + 8.73$ \cellcolor{C1} & \footnotesize $.509n + 7.87$ \cellcolor{C1} & \footnotesize $.488n + 8.29$ \cellcolor{C0}
\\\hline
11 & \footnotesize \cellcolor{C0} $.0838n -4.77$  & \footnotesize \cellcolor{C0} $.243n -7.56$  & \footnotesize \cellcolor{C1} $.359n -9.47$  & \footnotesize \cellcolor{C1} $.414n -10.1$  & \footnotesize \cellcolor{C2} $.485n -11.2$  & \footnotesize \cellcolor{C1} $.508n -11.3$  & \footnotesize \cellcolor{C3} $.534n -11.7$  & \footnotesize \cellcolor{C1} $.464n -10.9$ \\ & \footnotesize $.278n + 21.7$ \cellcolor{C0} & \footnotesize $.321n + 21.2$ \cellcolor{C0} & \footnotesize $.356n + 20.7$ \cellcolor{C1} & \footnotesize $.366n + 20.6$ \cellcolor{C1} & \footnotesize $.393n + 20.3$ \cellcolor{C2} & \footnotesize $.409n + 20.1$ \cellcolor{C1} & \footnotesize $.417n + 20.1$ \cellcolor{C3} & \footnotesize $.424n + 20.0$ \cellcolor{C1}
\\ & \footnotesize $.366n + 25.8$ \cellcolor{C0} & \footnotesize $.389n + 25.5$ \cellcolor{C0} & \footnotesize $.479n + 24.2$ \cellcolor{C1} & \footnotesize $.461n + 24.7$ \cellcolor{C1} & \footnotesize $.46n + 24.7$ \cellcolor{C2} & \footnotesize $.517n + 23.9$ \cellcolor{C1} & \footnotesize $.531n + 23.7$ \cellcolor{C3} & \footnotesize $.514n + 24.1$ \cellcolor{C1}
\\ & \footnotesize $.507n + 7.9$ \cellcolor{C0} & \footnotesize $.534n + 7.41$ \cellcolor{C0} & \footnotesize $.482n + 8.25$ \cellcolor{C1} & \footnotesize $.494n + 8.24$ \cellcolor{C1} & \footnotesize $.579n + 6.62$ \cellcolor{C2} & \footnotesize $.612n + 6.06$ \cellcolor{C1} & \footnotesize $.487n + 8.31$ \cellcolor{C3} & \footnotesize $.487n + 8.33$ \cellcolor{C1}
\\\hline
12 & \footnotesize \cellcolor{C0} $.0991n -3.37$  & \footnotesize \cellcolor{C0} $.26n -6.21$  & \footnotesize \cellcolor{C1} $.372n -8.12$  & \footnotesize \cellcolor{C1} $.447n -9.39$  & \footnotesize \cellcolor{C3} $.538n -10.8$  & \footnotesize \cellcolor{C3} $.58n -11.4$  & \footnotesize \cellcolor{C3} $.556n -10.8$  & \footnotesize \cellcolor{C3} $.539n -11.4$ \\ & \footnotesize $.315n + 21.4$ \cellcolor{C0} & \footnotesize $.348n + 21.0$ \cellcolor{C0} & \footnotesize $.353n + 21.0$ \cellcolor{C1} & \footnotesize $.377n + 20.7$ \cellcolor{C1} & \footnotesize $.402n + 20.3$ \cellcolor{C3} & \footnotesize $.422n + 20.1$ \cellcolor{C3} & \footnotesize $.431n + 20.0$ \cellcolor{C3} & \footnotesize $.431n + 20.1$ \cellcolor{C3}
\\ & \footnotesize $.389n + 25.8$ \cellcolor{C0} & \footnotesize $.497n + 24.1$ \cellcolor{C0} & \footnotesize $.447n + 25.1$ \cellcolor{C1} & \footnotesize $.447n + 25.1$ \cellcolor{C1} & \footnotesize $.447n + 25.1$ \cellcolor{C3} & \footnotesize $.656n + 21.9$ \cellcolor{C3} & \footnotesize $.656n + 21.9$ \cellcolor{C3} & \footnotesize $.502n + 24.4$ \cellcolor{C3}
\\ & \footnotesize $.664n + 5.18$ \cellcolor{C0} & \footnotesize $.502n + 8.16$ \cellcolor{C0} & \footnotesize $.477n + 8.51$ \cellcolor{C1} & \footnotesize $.667n + 5.3$ \cellcolor{C1} & \footnotesize $.426n + 9.35$ \cellcolor{C3} & \footnotesize $.505n + 8.11$ \cellcolor{C3} & \footnotesize $.505n + 8.11$ \cellcolor{C3} & \footnotesize $.478n + 8.47$ \cellcolor{C3}
\\\hline
\end{tabular}
\caption{Each grid in these tables represents a distribution of $k$-SAT instances with differing amounts of structure $\beta$, fixed to contain only \textit{satisfiable} instances (where lower $\beta$ means more structure). For each grid we sample SAT instances and fit the classical runtime and quantum \textit{$T$-depth} of the form $2^{a\cdot n + b}$. Each cell contains the exponent $a\cdot n+b$ in the order: classical runtime, quantum backtracking detection, quantum backtracking search, Grover. We color a cell blue if the classical algorithm scales best (i.e. has the lowest $a$ value), orange if the detection algorithm scales best, yellow if Grover scales best, and green if search scales better than classical and Grover.}
\end{table}
\begin{table} \hspace{-2.2cm}
\begin{tabular}{p{0.9cm}|p{1.93cm}|p{1.77cm}|p{1.77cm}|p{1.77cm}|p{1.77cm}|p{1.77cm}|p{1.77cm}|p{1.77cm}}
\backslashbox{$k$}{$\beta$} & \centering$\frac 12$ & \centering 1 & \centering $\frac 32$ & \centering 2 & \centering 3 & \centering 5 & \centering 10 & \ \ \ \ \ $\infty$ \\\hline
3& \cellcolor{C0}& \cellcolor{C0}& \cellcolor{C0}& \cellcolor{C0}& \cellcolor{C0} & \footnotesize \cellcolor{C0} $.0412n -11.6$  & \footnotesize \cellcolor{C0} $.0586n -13.8$  & \footnotesize \cellcolor{C0} $.0529n -12.5$ \\& \cellcolor{C0}& \cellcolor{C0}& \cellcolor{C0}& \cellcolor{C0}& \cellcolor{C0} & \footnotesize $.186n + 20.6$ \cellcolor{C0} & \footnotesize $.195n + 20.5$ \cellcolor{C0} & \footnotesize $.201n + 20.4$ \cellcolor{C0}
\\& \cellcolor{C0}& \cellcolor{C0}& \cellcolor{C0}& \cellcolor{C0}& \cellcolor{C0} & \footnotesize $.236n + 25.3$ \cellcolor{C0} & \footnotesize $.247n + 25.2$ \cellcolor{C0} & \footnotesize $.26n + 24.9$ \cellcolor{C0}
\\& \cellcolor{C0}& \cellcolor{C0}& \cellcolor{C0}& \cellcolor{C0}& \cellcolor{C0} & \footnotesize $.509n + 10.6$ \cellcolor{C0} & \footnotesize $.509n + 10.6$ \cellcolor{C0} & \footnotesize $.509n + 10.1$ \cellcolor{C0}
\\\hline
4& \cellcolor{C0}& \cellcolor{C0}& \cellcolor{C0}& \cellcolor{C0} & \footnotesize \cellcolor{C0} $.0843n -10.4$  & \footnotesize \cellcolor{C0} $.146n -13.6$  & \footnotesize \cellcolor{C0} $.188n -15.7$  & \footnotesize \cellcolor{C0} $.19n -15.2$ \\& \cellcolor{C0}& \cellcolor{C0}& \cellcolor{C0}& \cellcolor{C0} & \footnotesize $.225n + 20.5$ \cellcolor{C0} & \footnotesize $.247n + 20.4$ \cellcolor{C0} & \footnotesize $.26n + 20.2$ \cellcolor{C0} & \footnotesize $.264n + 20.2$ \cellcolor{C0}
\\& \cellcolor{C0}& \cellcolor{C0}& \cellcolor{C0}& \cellcolor{C0} & \footnotesize $.281n + 25.1$ \cellcolor{C0} & \footnotesize $.313n + 24.8$ \cellcolor{C0} & \footnotesize $.324n + 24.7$ \cellcolor{C0} & \footnotesize $.327n + 24.6$ \cellcolor{C0}
\\& \cellcolor{C0}& \cellcolor{C0}& \cellcolor{C0}& \cellcolor{C0} & \footnotesize $.511n + 10.7$ \cellcolor{C0} & \footnotesize $.512n + 10.6$ \cellcolor{C0} & \footnotesize $.513n + 10.6$ \cellcolor{C0} & \footnotesize $.51n + 10.2$ \cellcolor{C0}
\\\hline
5& \cellcolor{C0}& \cellcolor{C0}& \cellcolor{C0} & \footnotesize \cellcolor{C0} $.102n -9.57$  & \footnotesize \cellcolor{C0} $.192n -13.1$  & \footnotesize \cellcolor{C0} $.258n -15.1$  & \footnotesize \cellcolor{C1} $.301n -16.0$  & \footnotesize \cellcolor{C1} $.318n -16.2$ \\& \cellcolor{C0}& \cellcolor{C0}& \cellcolor{C0} & \footnotesize $.235n + 20.7$ \cellcolor{C0} & \footnotesize $.267n + 20.4$ \cellcolor{C0} & \footnotesize $.287n + 20.3$ \cellcolor{C0} & \footnotesize $.3n + 20.2$ \cellcolor{C1} & \footnotesize $.31n + 20.0$ \cellcolor{C1}
\\& \cellcolor{C0}& \cellcolor{C0}& \cellcolor{C0} & \footnotesize $.294n + 25.3$ \cellcolor{C0} & \footnotesize $.33n + 24.9$ \cellcolor{C0} & \footnotesize $.351n + 24.7$ \cellcolor{C0} & \footnotesize $.366n + 24.6$ \cellcolor{C1} & \footnotesize $.384n + 24.3$ \cellcolor{C1}
\\& \cellcolor{C0}& \cellcolor{C0}& \cellcolor{C0} & \footnotesize $.509n + 11.0$ \cellcolor{C0} & \footnotesize $.511n + 10.9$ \cellcolor{C0} & \footnotesize $.51n + 10.9$ \cellcolor{C0} & \footnotesize $.511n + 10.9$ \cellcolor{C1} & \footnotesize $.508n + 10.6$ \cellcolor{C1}
\\\hline
6& \cellcolor{C0}& \cellcolor{C0} & \footnotesize \cellcolor{C0} $.0882n -7.62$  & \footnotesize \cellcolor{C0} $.171n -10.1$  & \footnotesize \cellcolor{C0} $.258n -12.4$  & \footnotesize \cellcolor{C1} $.337n -13.9$  & \footnotesize \cellcolor{C1} $.35n -13.5$  & \footnotesize \cellcolor{C1} $.407n -15.8$ \\& \cellcolor{C0}& \cellcolor{C0} & \footnotesize $.241n + 20.9$ \cellcolor{C0} & \footnotesize $.266n + 20.7$ \cellcolor{C0} & \footnotesize $.3n + 20.3$ \cellcolor{C0} & \footnotesize $.319n + 20.2$ \cellcolor{C1} & \footnotesize $.332n + 20.1$ \cellcolor{C1} & \footnotesize $.343n + 20.0$ \cellcolor{C1}
\\& \cellcolor{C0}& \cellcolor{C0} & \footnotesize $.304n + 25.3$ \cellcolor{C0} & \footnotesize $.333n + 25.0$ \cellcolor{C0} & \footnotesize $.372n + 24.6$ \cellcolor{C0} & \footnotesize $.394n + 24.4$ \cellcolor{C1} & \footnotesize $.407n + 24.4$ \cellcolor{C1} & \footnotesize $.427n + 24.1$ \cellcolor{C1}
\\& \cellcolor{C0}& \cellcolor{C0} & \footnotesize $.509n + 11.1$ \cellcolor{C0} & \footnotesize $.509n + 11.1$ \cellcolor{C0} & \footnotesize $.511n + 11.0$ \cellcolor{C0} & \footnotesize $.511n + 11.0$ \cellcolor{C1} & \footnotesize $.511n + 11.0$ \cellcolor{C1} & \footnotesize $.507n + 10.7$ \cellcolor{C1}
\\\hline
7 & \footnotesize \cellcolor{C0} $.00394n -4.21$  & \footnotesize \cellcolor{C0} $.0473n -5.76$  & \footnotesize \cellcolor{C0} $.172n -9.55$  & \footnotesize \cellcolor{C0} $.252n -11.3$  & \footnotesize \cellcolor{C1} $.334n -12.9$  & \footnotesize \cellcolor{C1} $.402n -14.1$  & \footnotesize \cellcolor{C2} $.451n -14.9$  & \footnotesize \cellcolor{C2} $.476n -15.4$ \\ & \footnotesize $.155n + 22.1$ \cellcolor{C0} & \footnotesize $.226n + 21.3$ \cellcolor{C0} & \footnotesize $.273n + 20.8$ \cellcolor{C0} & \footnotesize $.298n + 20.5$ \cellcolor{C0} & \footnotesize $.327n + 20.3$ \cellcolor{C1} & \footnotesize $.347n + 20.1$ \cellcolor{C1} & \footnotesize $.358n + 20.1$ \cellcolor{C2} & \footnotesize $.367n + 20.0$ \cellcolor{C2}
\\ & \footnotesize $.21n + 26.8$ \cellcolor{C0} & \footnotesize $.288n + 25.8$ \cellcolor{C0} & \footnotesize $.346n + 25.0$ \cellcolor{C0} & \footnotesize $.365n + 25.0$ \cellcolor{C0} & \footnotesize $.401n + 24.6$ \cellcolor{C1} & \footnotesize $.427n + 24.3$ \cellcolor{C1} & \footnotesize $.439n + 24.2$ \cellcolor{C2} & \footnotesize $.449n + 24.1$ \cellcolor{C2}
\\ & \footnotesize $.508n + 11.2$ \cellcolor{C0} & \footnotesize $.509n + 11.1$ \cellcolor{C0} & \footnotesize $.51n + 11.1$ \cellcolor{C0} & \footnotesize $.51n + 11.1$ \cellcolor{C0} & \footnotesize $.511n + 11.1$ \cellcolor{C1} & \footnotesize $.512n + 11.1$ \cellcolor{C1} & \footnotesize $.512n + 11.1$ \cellcolor{C2} & \footnotesize $.512n + 11.1$ \cellcolor{C2}
\\\hline
8 & \footnotesize \cellcolor{C0} $.00241n -3.93$  & \footnotesize \cellcolor{C0} $.0994n -6.8$  & \footnotesize \cellcolor{C0} $.223n -9.54$  & \footnotesize \cellcolor{C0} $.297n -10.9$  & \footnotesize \cellcolor{C1} $.374n -12.2$  & \footnotesize \cellcolor{C1} $.437n -13.2$  & \footnotesize \cellcolor{C2} $.487n -14.0$  & \footnotesize \cellcolor{C2} $.495n -14.0$ \\ & \footnotesize $.192n + 22.0$ \cellcolor{C0} & \footnotesize $.247n + 21.4$ \cellcolor{C0} & \footnotesize $.294n + 20.8$ \cellcolor{C0} & \footnotesize $.32n + 20.6$ \cellcolor{C0} & \footnotesize $.345n + 20.3$ \cellcolor{C1} & \footnotesize $.368n + 20.1$ \cellcolor{C1} & \footnotesize $.378n + 20.1$ \cellcolor{C2} & \footnotesize $.389n + 19.9$ \cellcolor{C2}
\\ & \footnotesize $.255n + 26.5$ \cellcolor{C0} & \footnotesize $.311n + 25.8$ \cellcolor{C0} & \footnotesize $.372n + 25.1$ \cellcolor{C0} & \footnotesize $.394n + 24.9$ \cellcolor{C0} & \footnotesize $.421n + 24.6$ \cellcolor{C1} & \footnotesize $.455n + 24.2$ \cellcolor{C1} & \footnotesize $.465n + 24.1$ \cellcolor{C2} & \footnotesize $.481n + 23.9$ \cellcolor{C2}
\\ & \footnotesize $.509n + 11.2$ \cellcolor{C0} & \footnotesize $.509n + 11.2$ \cellcolor{C0} & \footnotesize $.509n + 11.2$ \cellcolor{C0} & \footnotesize $.51n + 11.2$ \cellcolor{C0} & \footnotesize $.511n + 11.2$ \cellcolor{C1} & \footnotesize $.512n + 11.2$ \cellcolor{C1} & \footnotesize $.512n + 11.2$ \cellcolor{C2} & \footnotesize $.512n + 11.2$ \cellcolor{C2}
\\\hline
9 & \footnotesize \cellcolor{C0} $.0057n -3.78$  & \footnotesize \cellcolor{C0} $.126n -6.56$  & \footnotesize \cellcolor{C0} $.238n -8.34$  & \footnotesize \cellcolor{C0} $.29n -8.96$  & \footnotesize \cellcolor{C0} $.352n -9.8$  & \footnotesize \cellcolor{C1} $.404n -10.4$  & \footnotesize \cellcolor{C1} $.45n -11.2$  & \footnotesize \cellcolor{C2} $.49n -12.3$ \\ & \footnotesize $.237n + 21.6$ \cellcolor{C0} & \footnotesize $.289n + 21.0$ \cellcolor{C0} & \footnotesize $.329n + 20.6$ \cellcolor{C0} & \footnotesize $.351n + 20.4$ \cellcolor{C0} & \footnotesize $.371n + 20.2$ \cellcolor{C0} & \footnotesize $.386n + 20.1$ \cellcolor{C1} & \footnotesize $.396n + 20.0$ \cellcolor{C1} & \footnotesize $.401n + 20.0$ \cellcolor{C2}
\\ & \footnotesize $.3n + 26.1$ \cellcolor{C0} & \footnotesize $.383n + 25.0$ \cellcolor{C0} & \footnotesize $.411n + 24.8$ \cellcolor{C0} & \footnotesize $.428n + 24.7$ \cellcolor{C0} & \footnotesize $.478n + 24.0$ \cellcolor{C0} & \footnotesize $.478n + 24.0$ \cellcolor{C1} & \footnotesize $.528n + 23.3$ \cellcolor{C1} & \footnotesize $.487n + 24.1$ \cellcolor{C2}
\\ & \footnotesize $.509n + 11.4$ \cellcolor{C0} & \footnotesize $.51n + 11.4$ \cellcolor{C0} & \footnotesize $.508n + 11.4$ \cellcolor{C0} & \footnotesize $.508n + 11.4$ \cellcolor{C0} & \footnotesize $.508n + 11.4$ \cellcolor{C0} & \footnotesize $.508n + 11.4$ \cellcolor{C1} & \footnotesize $.508n + 11.4$ \cellcolor{C1} & \footnotesize $.51n + 11.4$ \cellcolor{C2}
\\\hline
10 & \footnotesize \cellcolor{C0} $.0229n -3.66$  & \footnotesize \cellcolor{C0} $.19n -6.91$  & \footnotesize \cellcolor{C0} $.308n -8.83$  & \footnotesize \cellcolor{C1} $.375n -9.76$  & \footnotesize \cellcolor{C1} $.441n -10.6$  & \footnotesize \cellcolor{C2} $.494n -11.2$  & \footnotesize \cellcolor{C2} $.541n -11.9$  & \footnotesize \cellcolor{C2} $.53n -12.1$ \\ & \footnotesize $.235n + 22.1$ \cellcolor{C0} & \footnotesize $.296n + 21.3$ \cellcolor{C0} & \footnotesize $.333n + 20.8$ \cellcolor{C0} & \footnotesize $.355n + 20.6$ \cellcolor{C1} & \footnotesize $.378n + 20.3$ \cellcolor{C1} & \footnotesize $.396n + 20.1$ \cellcolor{C2} & \footnotesize $.405n + 20.1$ \cellcolor{C2} & \footnotesize $.413n + 20.0$ \cellcolor{C2}
\\ & \footnotesize $.308n + 26.4$ \cellcolor{C0} & \footnotesize $.368n + 25.6$ \cellcolor{C0} & \footnotesize $.423n + 24.9$ \cellcolor{C0} & \footnotesize $.44n + 24.7$ \cellcolor{C1} & \footnotesize $.462n + 24.4$ \cellcolor{C1} & \footnotesize $.49n + 24.1$ \cellcolor{C2} & \footnotesize $.506n + 23.9$ \cellcolor{C2} & \footnotesize $.494n + 24.2$ \cellcolor{C2}
\\ & \footnotesize $.507n + 11.5$ \cellcolor{C0} & \footnotesize $.509n + 11.5$ \cellcolor{C0} & \footnotesize $.509n + 11.5$ \cellcolor{C0} & \footnotesize $.507n + 11.5$ \cellcolor{C1} & \footnotesize $.509n + 11.5$ \cellcolor{C1} & \footnotesize $.51n + 11.5$ \cellcolor{C2} & \footnotesize $.51n + 11.5$ \cellcolor{C2} & \footnotesize $.508n + 11.5$ \cellcolor{C2}
\\\hline
11 & \footnotesize \cellcolor{C0} $.0822n -3.78$  & \footnotesize \cellcolor{C0} $.247n -6.51$  & \footnotesize \cellcolor{C1} $.368n -8.41$  & \footnotesize \cellcolor{C1} $.432n -9.23$  & \footnotesize \cellcolor{C2} $.504n -10.2$  & \footnotesize \cellcolor{C3} $.535n -10.4$  & \footnotesize \cellcolor{C3} $.578n -11.0$  & \footnotesize \cellcolor{C3} $.563n -11.7$ \\ & \footnotesize $.283n + 21.6$ \cellcolor{C0} & \footnotesize $.329n + 21.0$ \cellcolor{C0} & \footnotesize $.357n + 20.7$ \cellcolor{C1} & \footnotesize $.372n + 20.5$ \cellcolor{C1} & \footnotesize $.395n + 20.2$ \cellcolor{C2} & \footnotesize $.408n + 20.1$ \cellcolor{C3} & \footnotesize $.417n + 20.1$ \cellcolor{C3} & \footnotesize $.424n + 20.0$ \cellcolor{C3}
\\ & \footnotesize $.359n + 26.0$ \cellcolor{C0} & \footnotesize $.389n + 25.5$ \cellcolor{C0} & \footnotesize $.46n + 24.6$ \cellcolor{C1} & \footnotesize $.461n + 24.7$ \cellcolor{C1} & \footnotesize $.46n + 24.7$ \cellcolor{C2} & \footnotesize $.517n + 23.9$ \cellcolor{C3} & \footnotesize $.531n + 23.7$ \cellcolor{C3} & \footnotesize $.514n + 24.1$ \cellcolor{C3}
\\ & \footnotesize $.51n + 11.5$ \cellcolor{C0} & \footnotesize $.512n + 11.5$ \cellcolor{C0} & \footnotesize $.512n + 11.5$ \cellcolor{C1} & \footnotesize $.512n + 11.5$ \cellcolor{C1} & \footnotesize $.515n + 11.4$ \cellcolor{C2} & \footnotesize $.515n + 11.4$ \cellcolor{C3} & \footnotesize $.515n + 11.4$ \cellcolor{C3} & \footnotesize $.511n + 11.5$ \cellcolor{C3}
\\\hline
12 & \footnotesize \cellcolor{C0} $.0974n -2.22$  & \footnotesize \cellcolor{C0} $.252n -4.78$  & \footnotesize \cellcolor{C1} $.357n -6.44$  & \footnotesize \cellcolor{C1} $.419n -7.35$  & \footnotesize \cellcolor{C2} $.508n -8.74$  & \footnotesize \cellcolor{C3} $.546n -9.14$  & \footnotesize \cellcolor{C3} $.579n -9.55$  & \footnotesize \cellcolor{C2} $.592n -11.2$ \\ & \footnotesize $.281n + 22.0$ \cellcolor{C0} & \footnotesize $.335n + 21.2$ \cellcolor{C0} & \footnotesize $.357n + 20.9$ \cellcolor{C1} & \footnotesize $.389n + 20.5$ \cellcolor{C1} & \footnotesize $.401n + 20.3$ \cellcolor{C2} & \footnotesize $.422n + 20.1$ \cellcolor{C3} & \footnotesize $.429n + 20.1$ \cellcolor{C3} & \footnotesize $.43n + 20.1$ \cellcolor{C2}
\\ & \footnotesize $.289n + 27.4$ \cellcolor{C0} & \footnotesize $.497n + 24.1$ \cellcolor{C0} & \footnotesize $.497n + 24.1$ \cellcolor{C1} & \footnotesize $.447n + 25.1$ \cellcolor{C1} & \footnotesize $.447n + 25.1$ \cellcolor{C2} & \footnotesize $.656n + 21.9$ \cellcolor{C3} & \footnotesize $.656n + 21.9$ \cellcolor{C3} & \footnotesize $.502n + 24.4$ \cellcolor{C2}
\\ & \footnotesize $.516n + 11.5$ \cellcolor{C0} & \footnotesize $.522n + 11.4$ \cellcolor{C0} & \footnotesize $.522n + 11.4$ \cellcolor{C1} & \footnotesize $.522n + 11.4$ \cellcolor{C1} & \footnotesize $.522n + 11.4$ \cellcolor{C2} & \footnotesize $.527n + 11.3$ \cellcolor{C3} & \footnotesize $.527n + 11.3$ \cellcolor{C3} & \footnotesize $.514n + 11.5$ \cellcolor{C2}
\\\hline
\end{tabular}
\caption{Each grid in these tables represents a distribution of $k$-SAT instances with differing amounts of structure $\beta$, fixed to contain only \textit{unsatisfiable} instances (where lower $\beta$ means more structure). For each grid we sample SAT instances and fit the classical runtime and quantum \textit{$T$-depth} of the form $2^{a\cdot n + b}$. Each cell contains the exponent $a\cdot n+b$ in the order: classical runtime, quantum backtracking detection, quantum backtracking search, Grover. We color a cell blue if the classical algorithm scales best (i.e. has the lowest $a$ value), orange if the detection algorithm scales best, yellow if Grover scales best, and green if search scales better than classical and Grover.}
\end{table}

\begin{table} \hspace{-2.2cm} \begin{tabular}{p{0.9cm}|p{1.93cm}|p{1.77cm}|p{1.77cm}|p{1.77cm}|p{1.77cm}|p{1.77cm}|p{1.77cm}|p{1.77cm}} \backslashbox{$k$}{$\beta$} & \centering$\frac 12$ & \centering 1 & \centering $\frac 32$ & \centering 2 & \centering 3 & \centering 5 & \centering 10 & \ \ \ \ \ $\infty$ \\\hline3& \cellcolor{C0}& \cellcolor{C0}& \cellcolor{C0}& \cellcolor{C0}& \cellcolor{C0} & \footnotesize \cellcolor{C0} $.0369n -11.0$ & \footnotesize \cellcolor{C0} $.0502n -12.6$ & \footnotesize \cellcolor{C0} $.0452n -11.4$\\& \cellcolor{C0}& \cellcolor{C0}& \cellcolor{C0}& \cellcolor{C0}& \cellcolor{C0} & \footnotesize $.29n + 29.9$ \cellcolor{C0} & \footnotesize $.3n + 29.8$ \cellcolor{C0} & \footnotesize $.32n + 29.3$ \cellcolor{C0}
\\& \cellcolor{C0}& \cellcolor{C0}& \cellcolor{C0}& \cellcolor{C0}& \cellcolor{C0} & \footnotesize $.341n + 34.7$ \cellcolor{C0} & \footnotesize $.352n + 34.5$ \cellcolor{C0} & \footnotesize $.372n + 34.0$ \cellcolor{C0}
\\& \cellcolor{C0}& \cellcolor{C0}& \cellcolor{C0}& \cellcolor{C0}& \cellcolor{C0} & \footnotesize $.548n + 15.7$ \cellcolor{C0} & \footnotesize $.549n + 15.7$ \cellcolor{C0} & \footnotesize $.53n + 12.6$ \cellcolor{C0}
\\\hline
4& \cellcolor{C0}& \cellcolor{C0}& \cellcolor{C0}& \cellcolor{C0} & \footnotesize \cellcolor{C0} $.0724n -9.6$ & \footnotesize \cellcolor{C0} $.13n -12.7$ & \footnotesize \cellcolor{C0} $.165n -14.4$ & \footnotesize \cellcolor{C0} $.171n -14.2$\\& \cellcolor{C0}& \cellcolor{C0}& \cellcolor{C0}& \cellcolor{C0} & \footnotesize $.346n + 29.5$ \cellcolor{C0} & \footnotesize $.373n + 29.2$ \cellcolor{C0} & \footnotesize $.388n + 29.0$ \cellcolor{C0} & \footnotesize $.398n + 28.8$ \cellcolor{C0}
\\& \cellcolor{C0}& \cellcolor{C0}& \cellcolor{C0}& \cellcolor{C0} & \footnotesize $.407n + 33.9$ \cellcolor{C0} & \footnotesize $.436n + 33.7$ \cellcolor{C0} & \footnotesize $.451n + 33.5$ \cellcolor{C0} & \footnotesize $.458n + 33.3$ \cellcolor{C0}
\\& \cellcolor{C0}& \cellcolor{C0}& \cellcolor{C0}& \cellcolor{C0} & \footnotesize $.557n + 17.2$ \cellcolor{C0} & \footnotesize $.56n + 17.1$ \cellcolor{C0} & \footnotesize $.561n + 17.1$ \cellcolor{C0} & \footnotesize $.515n + 14.1$ \cellcolor{C0}
\\\hline
5& \cellcolor{C0}& \cellcolor{C0}& \cellcolor{C0} & \footnotesize \cellcolor{C0} $.0951n -9.27$ & \footnotesize \cellcolor{C0} $.18n -12.7$ & \footnotesize \cellcolor{C0} $.245n -14.7$ & \footnotesize \cellcolor{C0} $.29n -15.9$ & \footnotesize \cellcolor{C0} $.3n -15.7$\\& \cellcolor{C0}& \cellcolor{C0}& \cellcolor{C0} & \footnotesize $.351n + 29.9$ \cellcolor{C0} & \footnotesize $.393n + 29.3$ \cellcolor{C0} & \footnotesize $.41n + 29.3$ \cellcolor{C0} & \footnotesize $.427n + 29.1$ \cellcolor{C0} & \footnotesize $.459n + 28.3$ \cellcolor{C0}
\\& \cellcolor{C0}& \cellcolor{C0}& \cellcolor{C0} & \footnotesize $.41n + 34.4$ \cellcolor{C0} & \footnotesize $.458n + 33.8$ \cellcolor{C0} & \footnotesize $.475n + 33.7$ \cellcolor{C0} & \footnotesize $.49n + 33.6$ \cellcolor{C0} & \footnotesize $.534n + 32.5$ \cellcolor{C0}
\\& \cellcolor{C0}& \cellcolor{C0}& \cellcolor{C0} & \footnotesize $.558n + 18.6$ \cellcolor{C0} & \footnotesize $.562n + 18.5$ \cellcolor{C0} & \footnotesize $.561n + 18.5$ \cellcolor{C0} & \footnotesize $.563n + 18.5$ \cellcolor{C0} & \footnotesize $.506n + 15.5$ \cellcolor{C0}
\\\hline
6& \cellcolor{C0}& \cellcolor{C0} & \footnotesize \cellcolor{C0} $.081n -7.4$ & \footnotesize \cellcolor{C0} $.164n -10.1$ & \footnotesize \cellcolor{C0} $.249n -12.2$ & \footnotesize \cellcolor{C0} $.327n -13.8$ & \footnotesize \cellcolor{C0} $.34n -13.5$ & \footnotesize \cellcolor{C0} $.39n -15.4$\\& \cellcolor{C0}& \cellcolor{C0} & \footnotesize $.353n + 30.3$ \cellcolor{C0} & \footnotesize $.38n + 30.1$ \cellcolor{C0} & \footnotesize $.424n + 29.5$ \cellcolor{C0} & \footnotesize $.446n + 29.3$ \cellcolor{C0} & \footnotesize $.46n + 29.2$ \cellcolor{C0} & \footnotesize $.509n + 28.0$ \cellcolor{C0}
\\& \cellcolor{C0}& \cellcolor{C0} & \footnotesize $.416n + 34.8$ \cellcolor{C0} & \footnotesize $.445n + 34.4$ \cellcolor{C0} & \footnotesize $.497n + 33.8$ \cellcolor{C0} & \footnotesize $.52n + 33.6$ \cellcolor{C0} & \footnotesize $.532n + 33.5$ \cellcolor{C0} & \footnotesize $.59n + 32.2$ \cellcolor{C0}
\\& \cellcolor{C0}& \cellcolor{C0} & \footnotesize $.56n + 19.9$ \cellcolor{C0} & \footnotesize $.56n + 19.9$ \cellcolor{C0} & \footnotesize $.567n + 19.7$ \cellcolor{C0} & \footnotesize $.568n + 19.7$ \cellcolor{C0} & \footnotesize $.57n + 19.7$ \cellcolor{C0} & \footnotesize $.501n + 16.9$ \cellcolor{C0}
\\\hline
7 & \footnotesize \cellcolor{C0} $.00334n -4.19$ & \footnotesize \cellcolor{C0} $.0411n -5.58$ & \footnotesize \cellcolor{C0} $.163n -9.37$ & \footnotesize \cellcolor{C0} $.244n -11.3$ & \footnotesize \cellcolor{C0} $.329n -12.9$ & \footnotesize \cellcolor{C0} $.396n -14.1$ & \footnotesize \cellcolor{C0} $.444n -14.9$ & \footnotesize \cellcolor{C0} $.466n -15.3$\\ & \footnotesize $.251n + 32.2$ \cellcolor{C0} & \footnotesize $.329n + 31.3$ \cellcolor{C0} & \footnotesize $.385n + 30.6$ \cellcolor{C0} & \footnotesize $.409n + 30.4$ \cellcolor{C0} & \footnotesize $.444n + 30.0$ \cellcolor{C0} & \footnotesize $.47n + 29.7$ \cellcolor{C0} & \footnotesize $.481n + 29.7$ \cellcolor{C0} & \footnotesize $.493n + 29.5$ \cellcolor{C0}
\\ & \footnotesize $.308n + 36.9$ \cellcolor{C0} & \footnotesize $.391n + 35.8$ \cellcolor{C0} & \footnotesize $.461n + 34.8$ \cellcolor{C0} & \footnotesize $.479n + 34.7$ \cellcolor{C0} & \footnotesize $.519n + 34.3$ \cellcolor{C0} & \footnotesize $.549n + 33.9$ \cellcolor{C0} & \footnotesize $.562n + 33.8$ \cellcolor{C0} & \footnotesize $.575n + 33.7$ \cellcolor{C0}
\\ & \footnotesize $.555n + 21.2$ \cellcolor{C0} & \footnotesize $.562n + 21.1$ \cellcolor{C0} & \footnotesize $.568n + 21.0$ \cellcolor{C0} & \footnotesize $.568n + 21.0$ \cellcolor{C0} & \footnotesize $.574n + 20.9$ \cellcolor{C0} & \footnotesize $.577n + 20.8$ \cellcolor{C0} & \footnotesize $.577n + 20.8$ \cellcolor{C0} & \footnotesize $.579n + 20.8$ \cellcolor{C0}
\\\hline
8 & \footnotesize \cellcolor{C0} $.00234n -3.95$ & \footnotesize \cellcolor{C0} $.0877n -6.53$ & \footnotesize \cellcolor{C0} $.212n -9.42$ & \footnotesize \cellcolor{C0} $.288n -10.9$ & \footnotesize \cellcolor{C0} $.368n -12.2$ & \footnotesize \cellcolor{C0} $.433n -13.3$ & \footnotesize \cellcolor{C0} $.474n -13.9$ & \footnotesize \cellcolor{C0} $.492n -14.1$\\ & \footnotesize $.281n + 32.9$ \cellcolor{C0} & \footnotesize $.337n + 32.3$ \cellcolor{C0} & \footnotesize $.391n + 31.6$ \cellcolor{C0} & \footnotesize $.42n + 31.2$ \cellcolor{C0} & \footnotesize $.447n + 31.0$ \cellcolor{C0} & \footnotesize $.479n + 30.6$ \cellcolor{C0} & \footnotesize $.486n + 30.6$ \cellcolor{C0} & \footnotesize $.503n + 30.4$ \cellcolor{C0}
\\ & \footnotesize $.35n + 37.2$ \cellcolor{C0} & \footnotesize $.4n + 36.8$ \cellcolor{C0} & \footnotesize $.467n + 35.8$ \cellcolor{C0} & \footnotesize $.494n + 35.5$ \cellcolor{C0} & \footnotesize $.523n + 35.2$ \cellcolor{C0} & \footnotesize $.575n + 34.5$ \cellcolor{C0} & \footnotesize $.573n + 34.7$ \cellcolor{C0} & \footnotesize $.587n + 34.4$ \cellcolor{C0}
\\ & \footnotesize $.561n + 22.3$ \cellcolor{C0} & \footnotesize $.563n + 22.3$ \cellcolor{C0} & \footnotesize $.569n + 22.2$ \cellcolor{C0} & \footnotesize $.573n + 22.1$ \cellcolor{C0} & \footnotesize $.575n + 22.1$ \cellcolor{C0} & \footnotesize $.582n + 22.0$ \cellcolor{C0} & \footnotesize $.58n + 22.0$ \cellcolor{C0} & \footnotesize $.584n + 21.9$ \cellcolor{C0}
\\\hline
9 & \footnotesize \cellcolor{C0} $.00513n -3.81$ & \footnotesize \cellcolor{C0} $.115n -6.42$ & \footnotesize \cellcolor{C0} $.225n -8.22$ & \footnotesize \cellcolor{C0} $.281n -8.95$ & \footnotesize \cellcolor{C0} $.345n -9.82$ & \footnotesize \cellcolor{C0} $.401n -10.6$ & \footnotesize \cellcolor{C0} $.443n -11.3$ & \footnotesize \cellcolor{C0} $.486n -12.3$\\ & \footnotesize $.325n + 33.3$ \cellcolor{C0} & \footnotesize $.387n + 32.6$ \cellcolor{C0} & \footnotesize $.432n + 32.1$ \cellcolor{C0} & \footnotesize $.454n + 31.8$ \cellcolor{C0} & \footnotesize $.473n + 31.7$ \cellcolor{C0} & \footnotesize $.489n + 31.6$ \cellcolor{C0} & \footnotesize $.499n + 31.5$ \cellcolor{C0} & \footnotesize $.499n + 31.5$ \cellcolor{C0}
\\ & \footnotesize $.404n + 37.4$ \cellcolor{C0} & \footnotesize $.479n + 36.6$ \cellcolor{C0} & \footnotesize $.514n + 36.3$ \cellcolor{C0} & \footnotesize $.531n + 36.2$ \cellcolor{C0} & \footnotesize $.581n + 35.5$ \cellcolor{C0} & \footnotesize $.581n + 35.5$ \cellcolor{C0} & \footnotesize $.631n + 34.8$ \cellcolor{C0} & \footnotesize $.582n + 35.7$ \cellcolor{C0}
\\ & \footnotesize $.576n + 23.2$ \cellcolor{C0} & \footnotesize $.581n + 23.2$ \cellcolor{C0} & \footnotesize $.588n + 23.1$ \cellcolor{C0} & \footnotesize $.588n + 23.1$ \cellcolor{C0} & \footnotesize $.588n + 23.1$ \cellcolor{C0} & \footnotesize $.588n + 23.1$ \cellcolor{C0} & \footnotesize $.588n + 23.1$ \cellcolor{C0} & \footnotesize $.584n + 23.1$ \cellcolor{C0}
\\\hline
10 & \footnotesize \cellcolor{C0} $.0196n -3.67$ & \footnotesize \cellcolor{C0} $.182n -6.98$ & \footnotesize \cellcolor{C0} $.299n -8.9$ & \footnotesize \cellcolor{C0} $.366n -9.81$ & \footnotesize \cellcolor{C0} $.439n -10.7$ & \footnotesize \cellcolor{C1} $.499n -11.5$ & \footnotesize \cellcolor{C1} $.543n -12.2$ & \footnotesize \cellcolor{C1} $.529n -12.2$\\ & \footnotesize $.31n + 35.0$ \cellcolor{C0} & \footnotesize $.375n + 34.1$ \cellcolor{C0} & \footnotesize $.418n + 33.6$ \cellcolor{C0} & \footnotesize $.441n + 33.3$ \cellcolor{C0} & \footnotesize $.465n + 33.1$ \cellcolor{C0} & \footnotesize $.484n + 32.9$ \cellcolor{C1} & \footnotesize $.494n + 32.8$ \cellcolor{C1} & \footnotesize $.502n + 32.7$ \cellcolor{C1}
\\ & \footnotesize $.373n + 39.6$ \cellcolor{C0} & \footnotesize $.446n + 38.5$ \cellcolor{C0} & \footnotesize $.506n + 37.7$ \cellcolor{C0} & \footnotesize $.525n + 37.5$ \cellcolor{C0} & \footnotesize $.549n + 37.2$ \cellcolor{C0} & \footnotesize $.578n + 36.9$ \cellcolor{C1} & \footnotesize $.595n + 36.6$ \cellcolor{C1} & \footnotesize $.584n + 36.9$ \cellcolor{C1}
\\ & \footnotesize $.568n + 24.6$ \cellcolor{C0} & \footnotesize $.573n + 24.5$ \cellcolor{C0} & \footnotesize $.579n + 24.4$ \cellcolor{C0} & \footnotesize $.58n + 24.4$ \cellcolor{C0} & \footnotesize $.582n + 24.3$ \cellcolor{C0} & \footnotesize $.585n + 24.3$ \cellcolor{C1} & \footnotesize $.585n + 24.3$ \cellcolor{C1} & \footnotesize $.585n + 24.3$ \cellcolor{C1}
\\\hline
11 & \footnotesize \cellcolor{C0} $.0804n -4.04$ & \footnotesize \cellcolor{C0} $.245n -6.81$ & \footnotesize \cellcolor{C0} $.366n -8.63$ & \footnotesize \cellcolor{C0} $.432n -9.52$ & \footnotesize \cellcolor{C1} $.507n -10.5$ & \footnotesize \cellcolor{C1} $.543n -10.8$ & \footnotesize \cellcolor{C1} $.576n -11.2$ & \footnotesize \cellcolor{C1} $.563n -11.8$\\ & \footnotesize $.363n + 35.5$ \cellcolor{C0} & \footnotesize $.409n + 34.9$ \cellcolor{C0} & \footnotesize $.44n + 34.5$ \cellcolor{C0} & \footnotesize $.45n + 34.5$ \cellcolor{C0} & \footnotesize $.476n + 34.1$ \cellcolor{C1} & \footnotesize $.491n + 34.0$ \cellcolor{C1} & \footnotesize $.499n + 33.9$ \cellcolor{C1} & \footnotesize $.508n + 33.9$ \cellcolor{C1}
\\ & \footnotesize $.447n + 39.7$ \cellcolor{C0} & \footnotesize $.471n + 39.4$ \cellcolor{C0} & \footnotesize $.561n + 38.1$ \cellcolor{C0} & \footnotesize $.543n + 38.5$ \cellcolor{C0} & \footnotesize $.542n + 38.5$ \cellcolor{C1} & \footnotesize $.599n + 37.8$ \cellcolor{C1} & \footnotesize $.614n + 37.6$ \cellcolor{C1} & \footnotesize $.597n + 37.9$ \cellcolor{C1}
\\ & \footnotesize $.582n + 25.5$ \cellcolor{C0} & \footnotesize $.584n + 25.4$ \cellcolor{C0} & \footnotesize $.584n + 25.4$ \cellcolor{C0} & \footnotesize $.584n + 25.4$ \cellcolor{C0} & \footnotesize $.586n + 25.4$ \cellcolor{C1} & \footnotesize $.586n + 25.4$ \cellcolor{C1} & \footnotesize $.586n + 25.4$ \cellcolor{C1} & \footnotesize $.587n + 25.4$ \cellcolor{C1}
\\\hline
12 & \footnotesize \cellcolor{C0} $.0888n -2.24$ & \footnotesize \cellcolor{C0} $.249n -5.02$ & \footnotesize \cellcolor{C0} $.355n -6.63$ & \footnotesize \cellcolor{C0} $.417n -7.53$ & \footnotesize \cellcolor{C1} $.512n -9.01$ & \footnotesize \cellcolor{C1} $.555n -9.47$ & \footnotesize \cellcolor{C1} $.587n -9.87$ & \footnotesize \cellcolor{C2} $.593n -11.3$\\ & \footnotesize $.377n + 36.7$ \cellcolor{C0} & \footnotesize $.42n + 36.1$ \cellcolor{C0} & \footnotesize $.433n + 36.0$ \cellcolor{C0} & \footnotesize $.463n + 35.5$ \cellcolor{C0} & \footnotesize $.479n + 35.4$ \cellcolor{C1} & \footnotesize $.499n + 35.1$ \cellcolor{C1} & \footnotesize $.507n + 35.1$ \cellcolor{C1} & \footnotesize $.51n + 35.1$ \cellcolor{C2}
\\ & \footnotesize $.467n + 40.8$ \cellcolor{C0} & \footnotesize $.574n + 39.1$ \cellcolor{C0} & \footnotesize $.524n + 40.1$ \cellcolor{C0} & \footnotesize $.524n + 40.1$ \cellcolor{C0} & \footnotesize $.524n + 40.1$ \cellcolor{C1} & \footnotesize $.732n + 36.9$ \cellcolor{C1} & \footnotesize $.732n + 36.9$ \cellcolor{C1} & \footnotesize $.581n + 39.4$ \cellcolor{C2}
\\ & \footnotesize $.584n + 26.6$ \cellcolor{C0} & \footnotesize $.64n + 25.5$ \cellcolor{C0} & \footnotesize $.586n + 26.5$ \cellcolor{C0} & \footnotesize $.586n + 26.5$ \cellcolor{C0} & \footnotesize $.586n + 26.5$ \cellcolor{C1} & \footnotesize $.589n + 26.5$ \cellcolor{C1} & \footnotesize $.589n + 26.3$ \cellcolor{C1} & \footnotesize $.586n + 26.5$ \cellcolor{C2}
\\\hline
\end{tabular}
\caption{Each grid in these tables represents a distribution of $k$-SAT instances with differing amounts of structure $\beta$, fixed to contain an equal mix of \textit{satisfiable} and \textit{unsatisfiable} instances (where lower $\beta$ means more structure). For each grid we sample SAT instances and fit the classical runtime and quantum \textit{$T$-count} of the form $2^{a\cdot n + b}$. Each cell contains the exponent $a\cdot n+b$ in the order: classical runtime, quantum backtracking detection, quantum backtracking search, Grover. We color a cell blue if the classical algorithm scales best (i.e. has the lowest $a$ value), orange if the detection algorithm scales best, yellow if Grover scales best, and green if search scales better than classical and Grover.}
\end{table}

\begin{table} \hspace{-2.2cm} \begin{tabular}{p{0.9cm}|p{1.93cm}|p{1.77cm}|p{1.77cm}|p{1.77cm}|p{1.77cm}|p{1.77cm}|p{1.77cm}|p{1.77cm}} \backslashbox{$k$}{$\beta$} & \centering$\frac 12$ & \centering 1 & \centering $\frac 32$ & \centering 2 & \centering 3 & \centering 5 & \centering 10 & \ \ \ \ \ $\infty$ \\\hline3& \cellcolor{C0}& \cellcolor{C0}& \cellcolor{C0}& \cellcolor{C0}& \cellcolor{C0} & \footnotesize \cellcolor{C0} $.0299n -9.89$ & \footnotesize \cellcolor{C0} $.0356n -10.5$ & \footnotesize \cellcolor{C0} $.0327n -9.93$\\& \cellcolor{C0}& \cellcolor{C0}& \cellcolor{C0}& \cellcolor{C0}& \cellcolor{C0} & \footnotesize $.291n + 30.1$ \cellcolor{C0} & \footnotesize $.301n + 29.9$ \cellcolor{C0} & \footnotesize $.322n + 29.4$ \cellcolor{C0}
\\& \cellcolor{C0}& \cellcolor{C0}& \cellcolor{C0}& \cellcolor{C0}& \cellcolor{C0} & \footnotesize $.341n + 34.8$ \cellcolor{C0} & \footnotesize $.354n + 34.6$ \cellcolor{C0} & \footnotesize $.374n + 34.0$ \cellcolor{C0}
\\& \cellcolor{C0}& \cellcolor{C0}& \cellcolor{C0}& \cellcolor{C0}& \cellcolor{C0} & \footnotesize $.48n + 12.8$ \cellcolor{C0} & \footnotesize $.483n + 12.8$ \cellcolor{C0} & \footnotesize $.458n + 9.87$ \cellcolor{C0}
\\\hline
4& \cellcolor{C0}& \cellcolor{C0}& \cellcolor{C0}& \cellcolor{C0} & \footnotesize \cellcolor{C0} $.054n -8.35$ & \footnotesize \cellcolor{C0} $.0582n -7.84$ & \footnotesize \cellcolor{C0} $.0571n -7.32$ & \footnotesize \cellcolor{C0} $.0617n -7.25$\\& \cellcolor{C0}& \cellcolor{C0}& \cellcolor{C0}& \cellcolor{C0} & \footnotesize $.348n + 29.5$ \cellcolor{C0} & \footnotesize $.374n + 29.2$ \cellcolor{C0} & \footnotesize $.389n + 29.0$ \cellcolor{C0} & \footnotesize $.397n + 28.8$ \cellcolor{C0}
\\& \cellcolor{C0}& \cellcolor{C0}& \cellcolor{C0}& \cellcolor{C0} & \footnotesize $.411n + 33.9$ \cellcolor{C0} & \footnotesize $.435n + 33.7$ \cellcolor{C0} & \footnotesize $.453n + 33.5$ \cellcolor{C0} & \footnotesize $.458n + 33.3$ \cellcolor{C0}
\\& \cellcolor{C0}& \cellcolor{C0}& \cellcolor{C0}& \cellcolor{C0} & \footnotesize $.524n + 14.3$ \cellcolor{C0} & \footnotesize $.521n + 14.4$ \cellcolor{C0} & \footnotesize $.528n + 14.3$ \cellcolor{C0} & \footnotesize $.472n + 11.5$ \cellcolor{C0}
\\\hline
5& \cellcolor{C0}& \cellcolor{C0}& \cellcolor{C0} & \footnotesize \cellcolor{C0} $.0713n -7.98$ & \footnotesize \cellcolor{C0} $.103n -8.92$ & \footnotesize \cellcolor{C0} $.0892n -7.25$ & \footnotesize \cellcolor{C0} $.0885n -6.93$ & \footnotesize \cellcolor{C0} $.0587n -5.12$\\& \cellcolor{C0}& \cellcolor{C0}& \cellcolor{C0} & \footnotesize $.351n + 29.9$ \cellcolor{C0} & \footnotesize $.394n + 29.3$ \cellcolor{C0} & \footnotesize $.41n + 29.3$ \cellcolor{C0} & \footnotesize $.427n + 29.1$ \cellcolor{C0} & \footnotesize $.458n + 28.3$ \cellcolor{C0}
\\& \cellcolor{C0}& \cellcolor{C0}& \cellcolor{C0} & \footnotesize $.411n + 34.4$ \cellcolor{C0} & \footnotesize $.458n + 33.8$ \cellcolor{C0} & \footnotesize $.475n + 33.7$ \cellcolor{C0} & \footnotesize $.494n + 33.5$ \cellcolor{C0} & \footnotesize $.534n + 32.5$ \cellcolor{C0}
\\& \cellcolor{C0}& \cellcolor{C0}& \cellcolor{C0} & \footnotesize $.532n + 15.9$ \cellcolor{C0} & \footnotesize $.547n + 15.6$ \cellcolor{C0} & \footnotesize $.537n + 15.9$ \cellcolor{C0} & \footnotesize $.538n + 15.8$ \cellcolor{C0} & \footnotesize $.473n + 13.1$ \cellcolor{C0}
\\\hline
6& \cellcolor{C0}& \cellcolor{C0} & \footnotesize \cellcolor{C0} $.065n -6.83$ & \footnotesize \cellcolor{C0} $.126n -8.91$ & \footnotesize \cellcolor{C0} $.174n -10.0$ & \footnotesize \cellcolor{C0} $.187n -9.88$ & \footnotesize \cellcolor{C0} $.187n -9.64$ & \footnotesize \cellcolor{C0} $.167n -8.64$\\& \cellcolor{C0}& \cellcolor{C0} & \footnotesize $.353n + 30.3$ \cellcolor{C0} & \footnotesize $.38n + 30.1$ \cellcolor{C0} & \footnotesize $.424n + 29.5$ \cellcolor{C0} & \footnotesize $.446n + 29.3$ \cellcolor{C0} & \footnotesize $.459n + 29.2$ \cellcolor{C0} & \footnotesize $.51n + 28.0$ \cellcolor{C0}
\\& \cellcolor{C0}& \cellcolor{C0} & \footnotesize $.414n + 34.8$ \cellcolor{C0} & \footnotesize $.443n + 34.5$ \cellcolor{C0} & \footnotesize $.496n + 33.8$ \cellcolor{C0} & \footnotesize $.519n + 33.6$ \cellcolor{C0} & \footnotesize $.532n + 33.5$ \cellcolor{C0} & \footnotesize $.59n + 32.2$ \cellcolor{C0}
\\& \cellcolor{C0}& \cellcolor{C0} & \footnotesize $.536n + 17.3$ \cellcolor{C0} & \footnotesize $.541n + 17.2$ \cellcolor{C0} & \footnotesize $.551n + 17.0$ \cellcolor{C0} & \footnotesize $.546n + 17.3$ \cellcolor{C0} & \footnotesize $.561n + 16.9$ \cellcolor{C0} & \footnotesize $.466n + 14.5$ \cellcolor{C0}
\\\hline
7 & \footnotesize \cellcolor{C0} $.00228n -4.15$ & \footnotesize \cellcolor{C0} $.0336n -5.35$ & \footnotesize \cellcolor{C0} $.138n -8.77$ & \footnotesize \cellcolor{C0} $.205n -10.5$ & \footnotesize \cellcolor{C0} $.256n -11.4$ & \footnotesize \cellcolor{C0} $.264n -11.1$ & \footnotesize \cellcolor{C0} $.295n -11.7$ & \footnotesize \cellcolor{C0} $.255n -10.2$\\ & \footnotesize $.253n + 32.2$ \cellcolor{C0} & \footnotesize $.328n + 31.3$ \cellcolor{C0} & \footnotesize $.386n + 30.6$ \cellcolor{C0} & \footnotesize $.409n + 30.4$ \cellcolor{C0} & \footnotesize $.444n + 30.0$ \cellcolor{C0} & \footnotesize $.471n + 29.7$ \cellcolor{C0} & \footnotesize $.481n + 29.7$ \cellcolor{C0} & \footnotesize $.493n + 29.5$ \cellcolor{C0}
\\ & \footnotesize $.311n + 36.8$ \cellcolor{C0} & \footnotesize $.392n + 35.8$ \cellcolor{C0} & \footnotesize $.463n + 34.8$ \cellcolor{C0} & \footnotesize $.479n + 34.7$ \cellcolor{C0} & \footnotesize $.519n + 34.3$ \cellcolor{C0} & \footnotesize $.55n + 33.9$ \cellcolor{C0} & \footnotesize $.562n + 33.8$ \cellcolor{C0} & \footnotesize $.575n + 33.7$ \cellcolor{C0}
\\ & \footnotesize $.506n + 19.0$ \cellcolor{C0} & \footnotesize $.541n + 18.4$ \cellcolor{C0} & \footnotesize $.545n + 18.4$ \cellcolor{C0} & \footnotesize $.542n + 18.6$ \cellcolor{C0} & \footnotesize $.561n + 18.2$ \cellcolor{C0} & \footnotesize $.559n + 18.3$ \cellcolor{C0} & \footnotesize $.539n + 18.7$ \cellcolor{C0} & \footnotesize $.568n + 18.3$ \cellcolor{C0}
\\\hline
8 & \footnotesize \cellcolor{C0} $.00223n -3.97$ & \footnotesize \cellcolor{C0} $.073n -6.2$ & \footnotesize \cellcolor{C0} $.183n -8.95$ & \footnotesize \cellcolor{C0} $.244n -10.2$ & \footnotesize \cellcolor{C0} $.288n -10.7$ & \footnotesize \cellcolor{C0} $.328n -11.5$ & \footnotesize \cellcolor{C0} $.349n -11.7$ & \footnotesize \cellcolor{C0} $.3n -10.2$\\ & \footnotesize $.284n + 32.8$ \cellcolor{C0} & \footnotesize $.338n + 32.2$ \cellcolor{C0} & \footnotesize $.391n + 31.6$ \cellcolor{C0} & \footnotesize $.42n + 31.2$ \cellcolor{C0} & \footnotesize $.447n + 31.0$ \cellcolor{C0} & \footnotesize $.479n + 30.6$ \cellcolor{C0} & \footnotesize $.486n + 30.6$ \cellcolor{C0} & \footnotesize $.502n + 30.4$ \cellcolor{C0}
\\ & \footnotesize $.351n + 37.2$ \cellcolor{C0} & \footnotesize $.4n + 36.8$ \cellcolor{C0} & \footnotesize $.467n + 35.8$ \cellcolor{C0} & \footnotesize $.491n + 35.6$ \cellcolor{C0} & \footnotesize $.523n + 35.2$ \cellcolor{C0} & \footnotesize $.575n + 34.5$ \cellcolor{C0} & \footnotesize $.573n + 34.7$ \cellcolor{C0} & \footnotesize $.587n + 34.4$ \cellcolor{C0}
\\ & \footnotesize $.519n + 20.2$ \cellcolor{C0} & \footnotesize $.542n + 19.8$ \cellcolor{C0} & \footnotesize $.549n + 19.7$ \cellcolor{C0} & \footnotesize $.542n + 19.9$ \cellcolor{C0} & \footnotesize $.561n + 19.5$ \cellcolor{C0} & \footnotesize $.569n + 19.3$ \cellcolor{C0} & \footnotesize $.561n + 19.6$ \cellcolor{C0} & \footnotesize $.555n + 19.7$ \cellcolor{C0}
\\\hline
9 & \footnotesize \cellcolor{C0} $.00368n -3.83$ & \footnotesize \cellcolor{C0} $.1n -6.29$ & \footnotesize \cellcolor{C0} $.196n -7.97$ & \footnotesize \cellcolor{C0} $.226n -8.3$ & \footnotesize \cellcolor{C0} $.277n -8.93$ & \footnotesize \cellcolor{C0} $.292n -9.07$ & \footnotesize \cellcolor{C0} $.332n -9.8$ & \footnotesize \cellcolor{C0} $.343n -10.1$\\ & \footnotesize $.321n + 33.4$ \cellcolor{C0} & \footnotesize $.391n + 32.6$ \cellcolor{C0} & \footnotesize $.432n + 32.1$ \cellcolor{C0} & \footnotesize $.453n + 31.9$ \cellcolor{C0} & \footnotesize $.473n + 31.7$ \cellcolor{C0} & \footnotesize $.488n + 31.6$ \cellcolor{C0} & \footnotesize $.499n + 31.5$ \cellcolor{C0} & \footnotesize $.499n + 31.5$ \cellcolor{C0}
\\ & \footnotesize $.404n + 37.4$ \cellcolor{C0} & \footnotesize $.479n + 36.6$ \cellcolor{C0} & \footnotesize $.514n + 36.3$ \cellcolor{C0} & \footnotesize $.531n + 36.2$ \cellcolor{C0} & \footnotesize $.581n + 35.5$ \cellcolor{C0} & \footnotesize $.581n + 35.5$ \cellcolor{C0} & \footnotesize $.631n + 34.8$ \cellcolor{C0} & \footnotesize $.582n + 35.7$ \cellcolor{C0}
\\ & \footnotesize $.52n + 21.5$ \cellcolor{C0} & \footnotesize $.584n + 20.3$ \cellcolor{C0} & \footnotesize $.585n + 20.3$ \cellcolor{C0} & \footnotesize $.528n + 21.1$ \cellcolor{C0} & \footnotesize $.581n + 20.3$ \cellcolor{C0} & \footnotesize $.491n + 21.7$ \cellcolor{C0} & \footnotesize $.567n + 20.8$ \cellcolor{C0} & \footnotesize $.569n + 20.7$ \cellcolor{C0}
\\\hline
10 & \footnotesize \cellcolor{C0} $.016n -3.73$ & \footnotesize \cellcolor{C0} $.172n -7.24$ & \footnotesize \cellcolor{C0} $.281n -9.14$ & \footnotesize \cellcolor{C0} $.339n -10.0$ & \footnotesize \cellcolor{C0} $.389n -10.6$ & \footnotesize \cellcolor{C0} $.439n -11.4$ & \footnotesize \cellcolor{C0} $.466n -11.8$ & \footnotesize \cellcolor{C0} $.412n -10.8$\\ & \footnotesize $.313n + 35.0$ \cellcolor{C0} & \footnotesize $.377n + 34.1$ \cellcolor{C0} & \footnotesize $.418n + 33.6$ \cellcolor{C0} & \footnotesize $.442n + 33.3$ \cellcolor{C0} & \footnotesize $.465n + 33.1$ \cellcolor{C0} & \footnotesize $.484n + 32.9$ \cellcolor{C0} & \footnotesize $.494n + 32.8$ \cellcolor{C0} & \footnotesize $.502n + 32.7$ \cellcolor{C0}
\\ & \footnotesize $.38n + 39.4$ \cellcolor{C0} & \footnotesize $.446n + 38.5$ \cellcolor{C0} & \footnotesize $.506n + 37.7$ \cellcolor{C0} & \footnotesize $.525n + 37.5$ \cellcolor{C0} & \footnotesize $.549n + 37.2$ \cellcolor{C0} & \footnotesize $.578n + 36.9$ \cellcolor{C0} & \footnotesize $.595n + 36.6$ \cellcolor{C0} & \footnotesize $.584n + 36.9$ \cellcolor{C0}
\\ & \footnotesize $.545n + 22.2$ \cellcolor{C0} & \footnotesize $.557n + 21.9$ \cellcolor{C0} & \footnotesize $.57n + 21.7$ \cellcolor{C0} & \footnotesize $.59n + 21.4$ \cellcolor{C0} & \footnotesize $.538n + 22.3$ \cellcolor{C0} & \footnotesize $.525n + 22.4$ \cellcolor{C0} & \footnotesize $.585n + 21.5$ \cellcolor{C0} & \footnotesize $.567n + 21.9$ \cellcolor{C0}
\\\hline
11 & \footnotesize \cellcolor{C0} $.0838n -4.77$ & \footnotesize \cellcolor{C0} $.243n -7.56$ & \footnotesize \cellcolor{C0} $.359n -9.47$ & \footnotesize \cellcolor{C0} $.414n -10.1$ & \footnotesize \cellcolor{C1} $.485n -11.2$ & \footnotesize \cellcolor{C1} $.508n -11.3$ & \footnotesize \cellcolor{C1} $.534n -11.7$ & \footnotesize \cellcolor{C0} $.464n -10.9$\\ & \footnotesize $.359n + 35.6$ \cellcolor{C0} & \footnotesize $.403n + 35.0$ \cellcolor{C0} & \footnotesize $.438n + 34.6$ \cellcolor{C0} & \footnotesize $.448n + 34.5$ \cellcolor{C0} & \footnotesize $.475n + 34.1$ \cellcolor{C1} & \footnotesize $.492n + 34.0$ \cellcolor{C1} & \footnotesize $.499n + 33.9$ \cellcolor{C1} & \footnotesize $.508n + 33.9$ \cellcolor{C0}
\\ & \footnotesize $.447n + 39.7$ \cellcolor{C0} & \footnotesize $.471n + 39.4$ \cellcolor{C0} & \footnotesize $.561n + 38.1$ \cellcolor{C0} & \footnotesize $.543n + 38.5$ \cellcolor{C0} & \footnotesize $.542n + 38.5$ \cellcolor{C1} & \footnotesize $.599n + 37.8$ \cellcolor{C1} & \footnotesize $.614n + 37.6$ \cellcolor{C1} & \footnotesize $.597n + 37.9$ \cellcolor{C0}
\\ & \footnotesize $.581n + 22.7$ \cellcolor{C0} & \footnotesize $.607n + 22.2$ \cellcolor{C0} & \footnotesize $.556n + 23.1$ \cellcolor{C0} & \footnotesize $.567n + 23.1$ \cellcolor{C0} & \footnotesize $.652n + 21.4$ \cellcolor{C1} & \footnotesize $.685n + 20.9$ \cellcolor{C1} & \footnotesize $.56n + 23.1$ \cellcolor{C1} & \footnotesize $.564n + 23.1$ \cellcolor{C0}
\\\hline
12 & \footnotesize \cellcolor{C0} $.0991n -3.37$ & \footnotesize \cellcolor{C0} $.26n -6.21$ & \footnotesize \cellcolor{C0} $.372n -8.12$ & \footnotesize \cellcolor{C0} $.447n -9.39$ & \footnotesize \cellcolor{C3} $.538n -10.8$ & \footnotesize \cellcolor{C3} $.58n -11.4$ & \footnotesize \cellcolor{C1} $.556n -10.8$ & \footnotesize \cellcolor{C1} $.539n -11.4$\\ & \footnotesize $.393n + 36.4$ \cellcolor{C0} & \footnotesize $.425n + 36.0$ \cellcolor{C0} & \footnotesize $.43n + 36.0$ \cellcolor{C0} & \footnotesize $.455n + 35.7$ \cellcolor{C0} & \footnotesize $.479n + 35.4$ \cellcolor{C3} & \footnotesize $.498n + 35.1$ \cellcolor{C3} & \footnotesize $.507n + 35.1$ \cellcolor{C1} & \footnotesize $.51n + 35.1$ \cellcolor{C1}
\\ & \footnotesize $.467n + 40.8$ \cellcolor{C0} & \footnotesize $.574n + 39.1$ \cellcolor{C0} & \footnotesize $.524n + 40.1$ \cellcolor{C0} & \footnotesize $.524n + 40.1$ \cellcolor{C0} & \footnotesize $.524n + 40.1$ \cellcolor{C3} & \footnotesize $.732n + 36.9$ \cellcolor{C3} & \footnotesize $.732n + 36.9$ \cellcolor{C1} & \footnotesize $.581n + 39.4$ \cellcolor{C1}
\\ & \footnotesize $.733n + 21.2$ \cellcolor{C0} & \footnotesize $.568n + 24.2$ \cellcolor{C0} & \footnotesize $.543n + 24.5$ \cellcolor{C0} & \footnotesize $.733n + 21.3$ \cellcolor{C0} & \footnotesize $.492n + 25.4$ \cellcolor{C3} & \footnotesize $.568n + 24.2$ \cellcolor{C3} & \footnotesize $.568n + 24.2$ \cellcolor{C1} & \footnotesize $.552n + 24.4$ \cellcolor{C1}
\\\hline
\end{tabular}
\caption{Each grid in these tables represents a distribution of $k$-SAT instances with differing amounts of structure $\beta$, fixed to contain only \textit{satisfiable} instances (where lower $\beta$ means more structure). For each grid we sample SAT instances and fit the classical runtime and quantum \textit{$T$-count} of the form $2^{a\cdot n + b}$. Each cell contains the exponent $a\cdot n+b$ in the order: classical runtime, quantum backtracking detection, quantum backtracking search, Grover. We color a cell blue if the classical algorithm scales best (i.e. has the lowest $a$ value), orange if the detection algorithm scales best, yellow if Grover scales best, and green if search scales better than classical and Grover.}
\end{table}

\begin{table} \hspace{-2.2cm} \begin{tabular}{p{0.9cm}|p{1.93cm}|p{1.77cm}|p{1.77cm}|p{1.77cm}|p{1.77cm}|p{1.77cm}|p{1.77cm}|p{1.77cm}} \backslashbox{$k$}{$\beta$} & \centering$\frac 12$ & \centering 1 & \centering $\frac 32$ & \centering 2 & \centering 3 & \centering 5 & \centering 10 & \ \ \ \ \ $\infty$ \\\hline3& \cellcolor{C0}& \cellcolor{C0}& \cellcolor{C0}& \cellcolor{C0}& \cellcolor{C0} & \footnotesize \cellcolor{C0} $.0412n -11.6$ & \footnotesize \cellcolor{C0} $.0586n -13.8$ & \footnotesize \cellcolor{C0} $.0529n -12.5$\\& \cellcolor{C0}& \cellcolor{C0}& \cellcolor{C0}& \cellcolor{C0}& \cellcolor{C0} & \footnotesize $.289n + 29.8$ \cellcolor{C0} & \footnotesize $.3n + 29.8$ \cellcolor{C0} & \footnotesize $.318n + 29.3$ \cellcolor{C0}
\\& \cellcolor{C0}& \cellcolor{C0}& \cellcolor{C0}& \cellcolor{C0}& \cellcolor{C0} & \footnotesize $.34n + 34.5$ \cellcolor{C0} & \footnotesize $.352n + 34.5$ \cellcolor{C0} & \footnotesize $.377n + 33.8$ \cellcolor{C0}
\\& \cellcolor{C0}& \cellcolor{C0}& \cellcolor{C0}& \cellcolor{C0}& \cellcolor{C0} & \footnotesize $.549n + 16.6$ \cellcolor{C0} & \footnotesize $.55n + 16.5$ \cellcolor{C0} & \footnotesize $.531n + 13.5$ \cellcolor{C0}
\\\hline
4& \cellcolor{C0}& \cellcolor{C0}& \cellcolor{C0}& \cellcolor{C0} & \footnotesize \cellcolor{C0} $.0843n -10.4$ & \footnotesize \cellcolor{C0} $.146n -13.6$ & \footnotesize \cellcolor{C0} $.188n -15.7$ & \footnotesize \cellcolor{C0} $.19n -15.2$\\& \cellcolor{C0}& \cellcolor{C0}& \cellcolor{C0}& \cellcolor{C0} & \footnotesize $.345n + 29.5$ \cellcolor{C0} & \footnotesize $.373n + 29.2$ \cellcolor{C0} & \footnotesize $.389n + 29.0$ \cellcolor{C0} & \footnotesize $.399n + 28.7$ \cellcolor{C0}
\\& \cellcolor{C0}& \cellcolor{C0}& \cellcolor{C0}& \cellcolor{C0} & \footnotesize $.401n + 34.0$ \cellcolor{C0} & \footnotesize $.439n + 33.6$ \cellcolor{C0} & \footnotesize $.453n + 33.5$ \cellcolor{C0} & \footnotesize $.463n + 33.1$ \cellcolor{C0}
\\& \cellcolor{C0}& \cellcolor{C0}& \cellcolor{C0}& \cellcolor{C0} & \footnotesize $.558n + 18.0$ \cellcolor{C0} & \footnotesize $.561n + 18.0$ \cellcolor{C0} & \footnotesize $.562n + 17.9$ \cellcolor{C0} & \footnotesize $.516n + 14.9$ \cellcolor{C0}
\\\hline
5& \cellcolor{C0}& \cellcolor{C0}& \cellcolor{C0} & \footnotesize \cellcolor{C0} $.102n -9.57$ & \footnotesize \cellcolor{C0} $.192n -13.1$ & \footnotesize \cellcolor{C0} $.258n -15.1$ & \footnotesize \cellcolor{C0} $.301n -16.0$ & \footnotesize \cellcolor{C0} $.318n -16.2$\\& \cellcolor{C0}& \cellcolor{C0}& \cellcolor{C0} & \footnotesize $.351n + 29.8$ \cellcolor{C0} & \footnotesize $.393n + 29.3$ \cellcolor{C0} & \footnotesize $.41n + 29.3$ \cellcolor{C0} & \footnotesize $.427n + 29.1$ \cellcolor{C0} & \footnotesize $.459n + 28.3$ \cellcolor{C0}
\\& \cellcolor{C0}& \cellcolor{C0}& \cellcolor{C0} & \footnotesize $.41n + 34.4$ \cellcolor{C0} & \footnotesize $.456n + 33.8$ \cellcolor{C0} & \footnotesize $.475n + 33.7$ \cellcolor{C0} & \footnotesize $.494n + 33.5$ \cellcolor{C0} & \footnotesize $.533n + 32.5$ \cellcolor{C0}
\\& \cellcolor{C0}& \cellcolor{C0}& \cellcolor{C0} & \footnotesize $.558n + 19.5$ \cellcolor{C0} & \footnotesize $.563n + 19.4$ \cellcolor{C0} & \footnotesize $.562n + 19.4$ \cellcolor{C0} & \footnotesize $.564n + 19.3$ \cellcolor{C0} & \footnotesize $.507n + 16.4$ \cellcolor{C0}
\\\hline
6& \cellcolor{C0}& \cellcolor{C0} & \footnotesize \cellcolor{C0} $.0882n -7.62$ & \footnotesize \cellcolor{C0} $.171n -10.1$ & \footnotesize \cellcolor{C0} $.258n -12.4$ & \footnotesize \cellcolor{C0} $.337n -13.9$ & \footnotesize \cellcolor{C0} $.35n -13.5$ & \footnotesize \cellcolor{C0} $.407n -15.8$\\& \cellcolor{C0}& \cellcolor{C0} & \footnotesize $.354n + 30.3$ \cellcolor{C0} & \footnotesize $.38n + 30.1$ \cellcolor{C0} & \footnotesize $.425n + 29.5$ \cellcolor{C0} & \footnotesize $.445n + 29.3$ \cellcolor{C0} & \footnotesize $.461n + 29.2$ \cellcolor{C0} & \footnotesize $.508n + 28.0$ \cellcolor{C0}
\\& \cellcolor{C0}& \cellcolor{C0} & \footnotesize $.418n + 34.7$ \cellcolor{C0} & \footnotesize $.447n + 34.4$ \cellcolor{C0} & \footnotesize $.497n + 33.8$ \cellcolor{C0} & \footnotesize $.52n + 33.6$ \cellcolor{C0} & \footnotesize $.535n + 33.5$ \cellcolor{C0} & \footnotesize $.593n + 32.1$ \cellcolor{C0}
\\& \cellcolor{C0}& \cellcolor{C0} & \footnotesize $.561n + 20.7$ \cellcolor{C0} & \footnotesize $.561n + 20.7$ \cellcolor{C0} & \footnotesize $.568n + 20.6$ \cellcolor{C0} & \footnotesize $.569n + 20.6$ \cellcolor{C0} & \footnotesize $.57n + 20.5$ \cellcolor{C0} & \footnotesize $.503n + 17.7$ \cellcolor{C0}
\\\hline
7 & \footnotesize \cellcolor{C0} $.00394n -4.21$ & \footnotesize \cellcolor{C0} $.0473n -5.76$ & \footnotesize \cellcolor{C0} $.172n -9.55$ & \footnotesize \cellcolor{C0} $.252n -11.3$ & \footnotesize \cellcolor{C0} $.334n -12.9$ & \footnotesize \cellcolor{C0} $.402n -14.1$ & \footnotesize \cellcolor{C0} $.451n -14.9$ & \footnotesize \cellcolor{C0} $.476n -15.4$\\ & \footnotesize $.249n + 32.3$ \cellcolor{C0} & \footnotesize $.329n + 31.3$ \cellcolor{C0} & \footnotesize $.385n + 30.6$ \cellcolor{C0} & \footnotesize $.41n + 30.4$ \cellcolor{C0} & \footnotesize $.444n + 30.0$ \cellcolor{C0} & \footnotesize $.47n + 29.7$ \cellcolor{C0} & \footnotesize $.481n + 29.7$ \cellcolor{C0} & \footnotesize $.493n + 29.5$ \cellcolor{C0}
\\ & \footnotesize $.304n + 36.9$ \cellcolor{C0} & \footnotesize $.392n + 35.8$ \cellcolor{C0} & \footnotesize $.458n + 34.8$ \cellcolor{C0} & \footnotesize $.476n + 34.8$ \cellcolor{C0} & \footnotesize $.519n + 34.3$ \cellcolor{C0} & \footnotesize $.55n + 33.9$ \cellcolor{C0} & \footnotesize $.562n + 33.8$ \cellcolor{C0} & \footnotesize $.575n + 33.7$ \cellcolor{C0}
\\ & \footnotesize $.556n + 22.1$ \cellcolor{C0} & \footnotesize $.563n + 21.9$ \cellcolor{C0} & \footnotesize $.569n + 21.8$ \cellcolor{C0} & \footnotesize $.569n + 21.8$ \cellcolor{C0} & \footnotesize $.575n + 21.7$ \cellcolor{C0} & \footnotesize $.578n + 21.7$ \cellcolor{C0} & \footnotesize $.578n + 21.7$ \cellcolor{C0} & \footnotesize $.58n + 21.6$ \cellcolor{C0}
\\\hline
8 & \footnotesize \cellcolor{C0} $.00241n -3.93$ & \footnotesize \cellcolor{C0} $.0994n -6.8$ & \footnotesize \cellcolor{C0} $.223n -9.54$ & \footnotesize \cellcolor{C0} $.297n -10.9$ & \footnotesize \cellcolor{C0} $.374n -12.2$ & \footnotesize \cellcolor{C0} $.437n -13.2$ & \footnotesize \cellcolor{C1} $.487n -14.0$ & \footnotesize \cellcolor{C0} $.495n -14.0$\\ & \footnotesize $.279n + 32.9$ \cellcolor{C0} & \footnotesize $.337n + 32.3$ \cellcolor{C0} & \footnotesize $.39n + 31.6$ \cellcolor{C0} & \footnotesize $.42n + 31.2$ \cellcolor{C0} & \footnotesize $.447n + 31.0$ \cellcolor{C0} & \footnotesize $.479n + 30.6$ \cellcolor{C0} & \footnotesize $.486n + 30.6$ \cellcolor{C1} & \footnotesize $.503n + 30.4$ \cellcolor{C0}
\\ & \footnotesize $.342n + 37.4$ \cellcolor{C0} & \footnotesize $.4n + 36.7$ \cellcolor{C0} & \footnotesize $.468n + 35.8$ \cellcolor{C0} & \footnotesize $.494n + 35.5$ \cellcolor{C0} & \footnotesize $.523n + 35.2$ \cellcolor{C0} & \footnotesize $.566n + 34.7$ \cellcolor{C0} & \footnotesize $.573n + 34.7$ \cellcolor{C1} & \footnotesize $.595n + 34.3$ \cellcolor{C0}
\\ & \footnotesize $.561n + 23.2$ \cellcolor{C0} & \footnotesize $.563n + 23.2$ \cellcolor{C0} & \footnotesize $.57n + 23.0$ \cellcolor{C0} & \footnotesize $.574n + 23.0$ \cellcolor{C0} & \footnotesize $.576n + 22.9$ \cellcolor{C0} & \footnotesize $.583n + 22.8$ \cellcolor{C0} & \footnotesize $.581n + 22.8$ \cellcolor{C1} & \footnotesize $.586n + 22.8$ \cellcolor{C0}
\\\hline
9 & \footnotesize \cellcolor{C0} $.0057n -3.78$ & \footnotesize \cellcolor{C0} $.126n -6.56$ & \footnotesize \cellcolor{C0} $.238n -8.34$ & \footnotesize \cellcolor{C0} $.29n -8.96$ & \footnotesize \cellcolor{C0} $.352n -9.8$ & \footnotesize \cellcolor{C0} $.404n -10.4$ & \footnotesize \cellcolor{C0} $.45n -11.2$ & \footnotesize \cellcolor{C0} $.49n -12.3$\\ & \footnotesize $.328n + 33.3$ \cellcolor{C0} & \footnotesize $.385n + 32.6$ \cellcolor{C0} & \footnotesize $.432n + 32.1$ \cellcolor{C0} & \footnotesize $.454n + 31.8$ \cellcolor{C0} & \footnotesize $.473n + 31.7$ \cellcolor{C0} & \footnotesize $.489n + 31.6$ \cellcolor{C0} & \footnotesize $.499n + 31.5$ \cellcolor{C0} & \footnotesize $.499n + 31.5$ \cellcolor{C0}
\\ & \footnotesize $.391n + 37.7$ \cellcolor{C0} & \footnotesize $.479n + 36.6$ \cellcolor{C0} & \footnotesize $.514n + 36.3$ \cellcolor{C0} & \footnotesize $.531n + 36.2$ \cellcolor{C0} & \footnotesize $.581n + 35.5$ \cellcolor{C0} & \footnotesize $.581n + 35.5$ \cellcolor{C0} & \footnotesize $.631n + 34.8$ \cellcolor{C0} & \footnotesize $.585n + 35.6$ \cellcolor{C0}
\\ & \footnotesize $.577n + 24.1$ \cellcolor{C0} & \footnotesize $.583n + 24.0$ \cellcolor{C0} & \footnotesize $.589n + 23.9$ \cellcolor{C0} & \footnotesize $.589n + 23.9$ \cellcolor{C0} & \footnotesize $.589n + 23.9$ \cellcolor{C0} & \footnotesize $.589n + 23.9$ \cellcolor{C0} & \footnotesize $.589n + 23.9$ \cellcolor{C0} & \footnotesize $.585n + 24.0$ \cellcolor{C0}
\\\hline
10 & \footnotesize \cellcolor{C0} $.0229n -3.66$ & \footnotesize \cellcolor{C0} $.19n -6.91$ & \footnotesize \cellcolor{C0} $.308n -8.83$ & \footnotesize \cellcolor{C0} $.375n -9.76$ & \footnotesize \cellcolor{C0} $.441n -10.6$ & \footnotesize \cellcolor{C1} $.494n -11.2$ & \footnotesize \cellcolor{C1} $.541n -11.9$ & \footnotesize \cellcolor{C1} $.53n -12.1$\\ & \footnotesize $.308n + 35.0$ \cellcolor{C0} & \footnotesize $.374n + 34.2$ \cellcolor{C0} & \footnotesize $.416n + 33.6$ \cellcolor{C0} & \footnotesize $.44n + 33.3$ \cellcolor{C0} & \footnotesize $.464n + 33.1$ \cellcolor{C0} & \footnotesize $.484n + 32.9$ \cellcolor{C1} & \footnotesize $.494n + 32.8$ \cellcolor{C1} & \footnotesize $.502n + 32.7$ \cellcolor{C1}
\\ & \footnotesize $.38n + 39.4$ \cellcolor{C0} & \footnotesize $.446n + 38.5$ \cellcolor{C0} & \footnotesize $.506n + 37.7$ \cellcolor{C0} & \footnotesize $.525n + 37.5$ \cellcolor{C0} & \footnotesize $.549n + 37.2$ \cellcolor{C0} & \footnotesize $.578n + 36.9$ \cellcolor{C1} & \footnotesize $.595n + 36.6$ \cellcolor{C1} & \footnotesize $.584n + 36.9$ \cellcolor{C1}
\\ & \footnotesize $.568n + 25.4$ \cellcolor{C0} & \footnotesize $.574n + 25.3$ \cellcolor{C0} & \footnotesize $.58n + 25.2$ \cellcolor{C0} & \footnotesize $.582n + 25.2$ \cellcolor{C0} & \footnotesize $.584n + 25.2$ \cellcolor{C0} & \footnotesize $.586n + 25.1$ \cellcolor{C1} & \footnotesize $.586n + 25.1$ \cellcolor{C1} & \footnotesize $.587n + 25.1$ \cellcolor{C1}
\\\hline
11 & \footnotesize \cellcolor{C0} $.0822n -3.78$ & \footnotesize \cellcolor{C0} $.247n -6.51$ & \footnotesize \cellcolor{C0} $.368n -8.41$ & \footnotesize \cellcolor{C0} $.432n -9.23$ & \footnotesize \cellcolor{C1} $.504n -10.2$ & \footnotesize \cellcolor{C1} $.535n -10.4$ & \footnotesize \cellcolor{C1} $.578n -11.0$ & \footnotesize \cellcolor{C1} $.563n -11.7$\\ & \footnotesize $.364n + 35.5$ \cellcolor{C0} & \footnotesize $.411n + 34.9$ \cellcolor{C0} & \footnotesize $.439n + 34.6$ \cellcolor{C0} & \footnotesize $.453n + 34.4$ \cellcolor{C0} & \footnotesize $.477n + 34.1$ \cellcolor{C1} & \footnotesize $.491n + 34.0$ \cellcolor{C1} & \footnotesize $.499n + 33.9$ \cellcolor{C1} & \footnotesize $.508n + 33.9$ \cellcolor{C1}
\\ & \footnotesize $.439n + 39.9$ \cellcolor{C0} & \footnotesize $.471n + 39.4$ \cellcolor{C0} & \footnotesize $.541n + 38.5$ \cellcolor{C0} & \footnotesize $.543n + 38.5$ \cellcolor{C0} & \footnotesize $.542n + 38.5$ \cellcolor{C1} & \footnotesize $.599n + 37.8$ \cellcolor{C1} & \footnotesize $.614n + 37.6$ \cellcolor{C1} & \footnotesize $.597n + 37.9$ \cellcolor{C1}
\\ & \footnotesize $.583n + 26.3$ \cellcolor{C0} & \footnotesize $.586n + 26.3$ \cellcolor{C0} & \footnotesize $.586n + 26.3$ \cellcolor{C0} & \footnotesize $.586n + 26.3$ \cellcolor{C0} & \footnotesize $.588n + 26.3$ \cellcolor{C1} & \footnotesize $.588n + 26.3$ \cellcolor{C1} & \footnotesize $.588n + 26.3$ \cellcolor{C1} & \footnotesize $.588n + 26.2$ \cellcolor{C1}
\\\hline
12 & \footnotesize \cellcolor{C0} $.0974n -2.22$ & \footnotesize \cellcolor{C0} $.252n -4.78$ & \footnotesize \cellcolor{C0} $.357n -6.44$ & \footnotesize \cellcolor{C0} $.419n -7.35$ & \footnotesize \cellcolor{C1} $.508n -8.74$ & \footnotesize \cellcolor{C1} $.546n -9.14$ & \footnotesize \cellcolor{C1} $.579n -9.55$ & \footnotesize \cellcolor{C2} $.592n -11.2$\\ & \footnotesize $.359n + 37.0$ \cellcolor{C0} & \footnotesize $.412n + 36.2$ \cellcolor{C0} & \footnotesize $.434n + 35.9$ \cellcolor{C0} & \footnotesize $.466n + 35.5$ \cellcolor{C0} & \footnotesize $.477n + 35.4$ \cellcolor{C1} & \footnotesize $.498n + 35.1$ \cellcolor{C1} & \footnotesize $.506n + 35.1$ \cellcolor{C1} & \footnotesize $.51n + 35.1$ \cellcolor{C2}
\\ & \footnotesize $.367n + 42.4$ \cellcolor{C0} & \footnotesize $.574n + 39.1$ \cellcolor{C0} & \footnotesize $.574n + 39.1$ \cellcolor{C0} & \footnotesize $.524n + 40.1$ \cellcolor{C0} & \footnotesize $.524n + 40.1$ \cellcolor{C1} & \footnotesize $.732n + 36.9$ \cellcolor{C1} & \footnotesize $.732n + 36.9$ \cellcolor{C1} & \footnotesize $.581n + 39.4$ \cellcolor{C2}
\\ & \footnotesize $.585n + 27.4$ \cellcolor{C0} & \footnotesize $.588n + 27.4$ \cellcolor{C0} & \footnotesize $.588n + 27.4$ \cellcolor{C0} & \footnotesize $.588n + 27.4$ \cellcolor{C0} & \footnotesize $.588n + 27.4$ \cellcolor{C1} & \footnotesize $.59n + 27.4$ \cellcolor{C1} & \footnotesize $.59n + 27.4$ \cellcolor{C1} & \footnotesize $.588n + 27.4$ \cellcolor{C2}
\\\hline
\end{tabular}
\caption{Each grid in these tables represents a distribution of $k$-SAT instances with differing amounts of structure $\beta$, fixed to contain only \textit{unsatisfiable} instances (where lower $\beta$ means more structure). For each grid we sample SAT instances and fit the classical runtime and quantum \textit{$T$-count} of the form $2^{a\cdot n + b}$. Each cell contains the exponent $a\cdot n+b$ in the order: classical runtime, quantum backtracking detection, quantum backtracking search, Grover. We color a cell blue if the classical algorithm scales best (i.e. has the lowest $a$ value), orange if the detection algorithm scales best, yellow if Grover scales best, and green if search scales better than classical and Grover.}
\end{table}

\begin{table} \hspace{-2.2cm} \begin{tabular}{p{0.9cm}|p{1.93cm}|p{1.77cm}|p{1.77cm}|p{1.77cm}|p{1.77cm}|p{1.77cm}|p{1.77cm}|p{1.77cm}} \backslashbox{$k$}{$\beta$} & \centering$\frac 12$ & \centering 1 & \centering $\frac 32$ & \centering 2 & \centering 3 & \centering 5 & \centering 10 & \ \ \ \ \ $\infty$ \\\hline3& \cellcolor{C0}& \cellcolor{C0}& \cellcolor{C0}& \cellcolor{C0}& \cellcolor{C0} & \footnotesize \cellcolor{C0} $.0369n -11.0$  & \footnotesize \cellcolor{C0} $.0502n -12.6$  & \footnotesize \cellcolor{C0} $.0452n -11.4$ \\& \cellcolor{C0}& \cellcolor{C0}& \cellcolor{C0}& \cellcolor{C0}& \cellcolor{C0} & \footnotesize $.183n + 11.2$ \cellcolor{C0} & \footnotesize $.192n + 11.1$ \cellcolor{C0} & \footnotesize $.2n + 10.9$ \cellcolor{C0}
\\& \cellcolor{C0}& \cellcolor{C0}& \cellcolor{C0}& \cellcolor{C0}& \cellcolor{C0} & \footnotesize $.234n + 15.9$ \cellcolor{C0} & \footnotesize $.243n + 15.8$ \cellcolor{C0} & \footnotesize $.252n + 15.6$ \cellcolor{C0}
\\& \cellcolor{C0}& \cellcolor{C0}& \cellcolor{C0}& \cellcolor{C0}& \cellcolor{C0} & \footnotesize $.499n + 5.15$ \cellcolor{C0} & \footnotesize $.499n + 5.15$ \cellcolor{C0} & \footnotesize $.499n + 5.15$ \cellcolor{C0}
\\\hline
4& \cellcolor{C0}& \cellcolor{C0}& \cellcolor{C0}& \cellcolor{C0} & \footnotesize \cellcolor{C0} $.0724n -9.6$  & \footnotesize \cellcolor{C0} $.13n -12.7$  & \footnotesize \cellcolor{C0} $.165n -14.4$  & \footnotesize \cellcolor{C0} $.171n -14.2$ \\& \cellcolor{C0}& \cellcolor{C0}& \cellcolor{C0}& \cellcolor{C0} & \footnotesize $.221n + 11.1$ \cellcolor{C0} & \footnotesize $.242n + 10.9$ \cellcolor{C0} & \footnotesize $.255n + 10.8$ \cellcolor{C0} & \footnotesize $.259n + 10.8$ \cellcolor{C0}
\\& \cellcolor{C0}& \cellcolor{C0}& \cellcolor{C0}& \cellcolor{C0} & \footnotesize $.282n + 15.5$ \cellcolor{C0} & \footnotesize $.304n + 15.4$ \cellcolor{C0} & \footnotesize $.317n + 15.3$ \cellcolor{C0} & \footnotesize $.319n + 15.3$ \cellcolor{C0}
\\& \cellcolor{C0}& \cellcolor{C0}& \cellcolor{C0}& \cellcolor{C0} & \footnotesize $.5n + 5.14$ \cellcolor{C0} & \footnotesize $.499n + 5.14$ \cellcolor{C0} & \footnotesize $.499n + 5.14$ \cellcolor{C0} & \footnotesize $.499n + 5.14$ \cellcolor{C0}
\\\hline
5& \cellcolor{C0}& \cellcolor{C0}& \cellcolor{C0} & \footnotesize \cellcolor{C0} $.0951n -9.27$  & \footnotesize \cellcolor{C0} $.18n -12.7$  & \footnotesize \cellcolor{C0} $.245n -14.7$  & \footnotesize \cellcolor{C0} $.29n -15.9$  & \footnotesize \cellcolor{C0} $.3n -15.7$ \\& \cellcolor{C0}& \cellcolor{C0}& \cellcolor{C0} & \footnotesize $.229n + 11.3$ \cellcolor{C0} & \footnotesize $.262n + 11.0$ \cellcolor{C0} & \footnotesize $.281n + 10.8$ \cellcolor{C0} & \footnotesize $.294n + 10.7$ \cellcolor{C0} & \footnotesize $.306n + 10.5$ \cellcolor{C0}
\\& \cellcolor{C0}& \cellcolor{C0}& \cellcolor{C0} & \footnotesize $.289n + 15.8$ \cellcolor{C0} & \footnotesize $.327n + 15.4$ \cellcolor{C0} & \footnotesize $.345n + 15.2$ \cellcolor{C0} & \footnotesize $.356n + 15.2$ \cellcolor{C0} & \footnotesize $.382n + 14.8$ \cellcolor{C0}
\\& \cellcolor{C0}& \cellcolor{C0}& \cellcolor{C0} & \footnotesize $.5n + 5.14$ \cellcolor{C0} & \footnotesize $.499n + 5.14$ \cellcolor{C0} & \footnotesize $.499n + 5.14$ \cellcolor{C0} & \footnotesize $.499n + 5.14$ \cellcolor{C0} & \footnotesize $.499n + 5.15$ \cellcolor{C0}
\\\hline
6& \cellcolor{C0}& \cellcolor{C0} & \footnotesize \cellcolor{C0} $.081n -7.4$  & \footnotesize \cellcolor{C0} $.164n -10.1$  & \footnotesize \cellcolor{C0} $.249n -12.2$  & \footnotesize \cellcolor{C1} $.327n -13.8$  & \footnotesize \cellcolor{C1} $.34n -13.5$  & \footnotesize \cellcolor{C1} $.39n -15.4$ \\& \cellcolor{C0}& \cellcolor{C0} & \footnotesize $.234n + 11.4$ \cellcolor{C0} & \footnotesize $.261n + 11.2$ \cellcolor{C0} & \footnotesize $.293n + 10.9$ \cellcolor{C0} & \footnotesize $.313n + 10.7$ \cellcolor{C1} & \footnotesize $.326n + 10.6$ \cellcolor{C1} & \footnotesize $.34n + 10.5$ \cellcolor{C1}
\\& \cellcolor{C0}& \cellcolor{C0} & \footnotesize $.297n + 15.9$ \cellcolor{C0} & \footnotesize $.325n + 15.6$ \cellcolor{C0} & \footnotesize $.366n + 15.2$ \cellcolor{C0} & \footnotesize $.388n + 15.0$ \cellcolor{C1} & \footnotesize $.398n + 14.9$ \cellcolor{C1} & \footnotesize $.422n + 14.6$ \cellcolor{C1}
\\& \cellcolor{C0}& \cellcolor{C0} & \footnotesize $.499n + 5.14$ \cellcolor{C0} & \footnotesize $.499n + 5.14$ \cellcolor{C0} & \footnotesize $.499n + 5.14$ \cellcolor{C0} & \footnotesize $.499n + 5.15$ \cellcolor{C1} & \footnotesize $.499n + 5.15$ \cellcolor{C1} & \footnotesize $.498n + 5.16$ \cellcolor{C1}
\\\hline
7 & \footnotesize \cellcolor{C0} $.00334n -4.19$  & \footnotesize \cellcolor{C0} $.0411n -5.58$  & \footnotesize \cellcolor{C0} $.163n -9.37$  & \footnotesize \cellcolor{C0} $.244n -11.3$  & \footnotesize \cellcolor{C1} $.329n -12.9$  & \footnotesize \cellcolor{C1} $.396n -14.1$  & \footnotesize \cellcolor{C2} $.444n -14.9$  & \footnotesize \cellcolor{C2} $.466n -15.3$ \\ & \footnotesize $.152n + 12.6$ \cellcolor{C0} & \footnotesize $.218n + 11.8$ \cellcolor{C0} & \footnotesize $.267n + 11.3$ \cellcolor{C0} & \footnotesize $.291n + 11.1$ \cellcolor{C0} & \footnotesize $.321n + 10.8$ \cellcolor{C1} & \footnotesize $.342n + 10.6$ \cellcolor{C1} & \footnotesize $.353n + 10.5$ \cellcolor{C2} & \footnotesize $.362n + 10.5$ \cellcolor{C2}
\\ & \footnotesize $.208n + 17.2$ \cellcolor{C0} & \footnotesize $.28n + 16.3$ \cellcolor{C0} & \footnotesize $.343n + 15.5$ \cellcolor{C0} & \footnotesize $.36n + 15.4$ \cellcolor{C0} & \footnotesize $.396n + 15.0$ \cellcolor{C1} & \footnotesize $.421n + 14.8$ \cellcolor{C1} & \footnotesize $.434n + 14.7$ \cellcolor{C2} & \footnotesize $.444n + 14.6$ \cellcolor{C2}
\\ & \footnotesize $.499n + 5.14$ \cellcolor{C0} & \footnotesize $.499n + 5.14$ \cellcolor{C0} & \footnotesize $.499n + 5.15$ \cellcolor{C0} & \footnotesize $.499n + 5.15$ \cellcolor{C0} & \footnotesize $.499n + 5.15$ \cellcolor{C1} & \footnotesize $.499n + 5.15$ \cellcolor{C1} & \footnotesize $.499n + 5.15$ \cellcolor{C2} & \footnotesize $.499n + 5.15$ \cellcolor{C2}
\\\hline
8 & \footnotesize \cellcolor{C0} $.00234n -3.95$  & \footnotesize \cellcolor{C0} $.0877n -6.53$  & \footnotesize \cellcolor{C0} $.212n -9.42$  & \footnotesize \cellcolor{C0} $.288n -10.9$  & \footnotesize \cellcolor{C1} $.368n -12.2$  & \footnotesize \cellcolor{C1} $.433n -13.3$  & \footnotesize \cellcolor{C2} $.474n -13.9$  & \footnotesize \cellcolor{C2} $.492n -14.1$ \\ & \footnotesize $.188n + 12.4$ \cellcolor{C0} & \footnotesize $.241n + 11.9$ \cellcolor{C0} & \footnotesize $.289n + 11.3$ \cellcolor{C0} & \footnotesize $.314n + 11.0$ \cellcolor{C0} & \footnotesize $.339n + 10.8$ \cellcolor{C1} & \footnotesize $.361n + 10.6$ \cellcolor{C1} & \footnotesize $.371n + 10.5$ \cellcolor{C2} & \footnotesize $.382n + 10.4$ \cellcolor{C2}
\\ & \footnotesize $.256n + 16.8$ \cellcolor{C0} & \footnotesize $.304n + 16.4$ \cellcolor{C0} & \footnotesize $.365n + 15.6$ \cellcolor{C0} & \footnotesize $.389n + 15.3$ \cellcolor{C0} & \footnotesize $.415n + 15.1$ \cellcolor{C1} & \footnotesize $.457n + 14.5$ \cellcolor{C1} & \footnotesize $.458n + 14.6$ \cellcolor{C2} & \footnotesize $.466n + 14.5$ \cellcolor{C2}
\\ & \footnotesize $.499n + 5.14$ \cellcolor{C0} & \footnotesize $.499n + 5.14$ \cellcolor{C0} & \footnotesize $.499n + 5.14$ \cellcolor{C0} & \footnotesize $.499n + 5.15$ \cellcolor{C0} & \footnotesize $.499n + 5.15$ \cellcolor{C1} & \footnotesize $.499n + 5.15$ \cellcolor{C1} & \footnotesize $.499n + 5.15$ \cellcolor{C2} & \footnotesize $.498n + 5.16$ \cellcolor{C2}
\\\hline
9 & \footnotesize \cellcolor{C0} $.00513n -3.81$  & \footnotesize \cellcolor{C0} $.115n -6.42$  & \footnotesize \cellcolor{C0} $.225n -8.22$  & \footnotesize \cellcolor{C0} $.281n -8.95$  & \footnotesize \cellcolor{C0} $.345n -9.82$  & \footnotesize \cellcolor{C1} $.401n -10.6$  & \footnotesize \cellcolor{C1} $.443n -11.3$  & \footnotesize \cellcolor{C2} $.486n -12.3$ \\ & \footnotesize $.229n + 12.1$ \cellcolor{C0} & \footnotesize $.285n + 11.5$ \cellcolor{C0} & \footnotesize $.321n + 11.1$ \cellcolor{C0} & \footnotesize $.343n + 10.8$ \cellcolor{C0} & \footnotesize $.363n + 10.7$ \cellcolor{C0} & \footnotesize $.378n + 10.6$ \cellcolor{C1} & \footnotesize $.388n + 10.5$ \cellcolor{C1} & \footnotesize $.394n + 10.5$ \cellcolor{C2}
\\ & \footnotesize $.308n + 16.2$ \cellcolor{C0} & \footnotesize $.376n + 15.5$ \cellcolor{C0} & \footnotesize $.403n + 15.3$ \cellcolor{C0} & \footnotesize $.42n + 15.1$ \cellcolor{C0} & \footnotesize $.47n + 14.4$ \cellcolor{C0} & \footnotesize $.47n + 14.4$ \cellcolor{C1} & \footnotesize $.52n + 13.7$ \cellcolor{C1} & \footnotesize $.477n + 14.6$ \cellcolor{C2}
\\ & \footnotesize $.499n + 5.15$ \cellcolor{C0} & \footnotesize $.499n + 5.15$ \cellcolor{C0} & \footnotesize $.498n + 5.16$ \cellcolor{C0} & \footnotesize $.498n + 5.16$ \cellcolor{C0} & \footnotesize $.498n + 5.16$ \cellcolor{C0} & \footnotesize $.498n + 5.16$ \cellcolor{C1} & \footnotesize $.498n + 5.16$ \cellcolor{C1} & \footnotesize $.498n + 5.16$ \cellcolor{C2}
\\\hline
10 & \footnotesize \cellcolor{C0} $.0196n -3.67$  & \footnotesize \cellcolor{C0} $.182n -6.98$  & \footnotesize \cellcolor{C0} $.299n -8.9$  & \footnotesize \cellcolor{C1} $.366n -9.81$  & \footnotesize \cellcolor{C1} $.439n -10.7$  & \footnotesize \cellcolor{C2} $.499n -11.5$  & \footnotesize \cellcolor{C2} $.543n -12.2$  & \footnotesize \cellcolor{C2} $.529n -12.2$ \\ & \footnotesize $.233n + 12.4$ \cellcolor{C0} & \footnotesize $.292n + 11.7$ \cellcolor{C0} & \footnotesize $.328n + 11.2$ \cellcolor{C0} & \footnotesize $.349n + 11.0$ \cellcolor{C1} & \footnotesize $.37n + 10.8$ \cellcolor{C1} & \footnotesize $.387n + 10.6$ \cellcolor{C2} & \footnotesize $.396n + 10.6$ \cellcolor{C2} & \footnotesize $.405n + 10.5$ \cellcolor{C2}
\\ & \footnotesize $.296n + 17.0$ \cellcolor{C0} & \footnotesize $.363n + 16.0$ \cellcolor{C0} & \footnotesize $.417n + 15.3$ \cellcolor{C0} & \footnotesize $.433n + 15.2$ \cellcolor{C1} & \footnotesize $.455n + 14.9$ \cellcolor{C1} & \footnotesize $.481n + 14.6$ \cellcolor{C2} & \footnotesize $.498n + 14.4$ \cellcolor{C2} & \footnotesize $.486n + 14.6$ \cellcolor{C2}
\\ & \footnotesize $.499n + 5.14$ \cellcolor{C0} & \footnotesize $.499n + 5.14$ \cellcolor{C0} & \footnotesize $.499n + 5.15$ \cellcolor{C0} & \footnotesize $.499n + 5.15$ \cellcolor{C1} & \footnotesize $.499n + 5.15$ \cellcolor{C1} & \footnotesize $.498n + 5.15$ \cellcolor{C2} & \footnotesize $.498n + 5.15$ \cellcolor{C2} & \footnotesize $.498n + 5.16$ \cellcolor{C2}
\\\hline
11 & \footnotesize \cellcolor{C0} $.0804n -4.04$  & \footnotesize \cellcolor{C0} $.245n -6.81$  & \footnotesize \cellcolor{C1} $.366n -8.63$  & \footnotesize \cellcolor{C1} $.432n -9.52$  & \footnotesize \cellcolor{C2} $.507n -10.5$  & \footnotesize \cellcolor{C3} $.543n -10.8$  & \footnotesize \cellcolor{C3} $.576n -11.2$  & \footnotesize \cellcolor{C3} $.563n -11.8$ \\ & \footnotesize $.274n + 12.1$ \cellcolor{C0} & \footnotesize $.318n + 11.5$ \cellcolor{C0} & \footnotesize $.348n + 11.1$ \cellcolor{C1} & \footnotesize $.359n + 11.1$ \cellcolor{C1} & \footnotesize $.382n + 10.8$ \cellcolor{C2} & \footnotesize $.397n + 10.6$ \cellcolor{C3} & \footnotesize $.405n + 10.6$ \cellcolor{C3} & \footnotesize $.414n + 10.5$ \cellcolor{C3}
\\ & \footnotesize $.358n + 16.3$ \cellcolor{C0} & \footnotesize $.38n + 16.0$ \cellcolor{C0} & \footnotesize $.469n + 14.7$ \cellcolor{C1} & \footnotesize $.451n + 15.1$ \cellcolor{C1} & \footnotesize $.448n + 15.2$ \cellcolor{C2} & \footnotesize $.505n + 14.4$ \cellcolor{C3} & \footnotesize $.52n + 14.3$ \cellcolor{C3} & \footnotesize $.503n + 14.6$ \cellcolor{C3}
\\ & \footnotesize $.499n + 5.15$ \cellcolor{C0} & \footnotesize $.499n + 5.15$ \cellcolor{C0} & \footnotesize $.499n + 5.15$ \cellcolor{C1} & \footnotesize $.499n + 5.15$ \cellcolor{C1} & \footnotesize $.498n + 5.15$ \cellcolor{C2} & \footnotesize $.498n + 5.15$ \cellcolor{C3} & \footnotesize $.498n + 5.15$ \cellcolor{C3} & \footnotesize $.498n + 5.16$ \cellcolor{C3}
\\\hline
12 & \footnotesize \cellcolor{C0} $.0888n -2.24$  & \footnotesize \cellcolor{C0} $.249n -5.02$  & \footnotesize \cellcolor{C1} $.355n -6.63$  & \footnotesize \cellcolor{C1} $.417n -7.53$  & \footnotesize \cellcolor{C2} $.512n -9.01$  & \footnotesize \cellcolor{C3} $.555n -9.47$  & \footnotesize \cellcolor{C3} $.587n -9.87$  & \footnotesize \cellcolor{C2} $.593n -11.3$ \\ & \footnotesize $.289n + 12.1$ \cellcolor{C0} & \footnotesize $.329n + 11.6$ \cellcolor{C0} & \footnotesize $.342n + 11.5$ \cellcolor{C1} & \footnotesize $.372n + 11.0$ \cellcolor{C1} & \footnotesize $.388n + 10.9$ \cellcolor{C2} & \footnotesize $.404n + 10.7$ \cellcolor{C3} & \footnotesize $.412n + 10.6$ \cellcolor{C3} & \footnotesize $.419n + 10.6$ \cellcolor{C2}
\\ & \footnotesize $.379n + 16.3$ \cellcolor{C0} & \footnotesize $.483n + 14.7$ \cellcolor{C0} & \footnotesize $.433n + 15.6$ \cellcolor{C1} & \footnotesize $.433n + 15.6$ \cellcolor{C1} & \footnotesize $.433n + 15.6$ \cellcolor{C2} & \footnotesize $.638n + 12.5$ \cellcolor{C3} & \footnotesize $.638n + 12.5$ \cellcolor{C3} & \footnotesize $.491n + 14.9$ \cellcolor{C2}
\\ & \footnotesize $.499n + 5.15$ \cellcolor{C0} & \footnotesize $.544n + 4.29$ \cellcolor{C0} & \footnotesize $.498n + 5.15$ \cellcolor{C1} & \footnotesize $.498n + 5.15$ \cellcolor{C1} & \footnotesize $.498n + 5.15$ \cellcolor{C2} & \footnotesize $.498n + 5.16$ \cellcolor{C3} & \footnotesize $.498n + 5.01$ \cellcolor{C3} & \footnotesize $.498n + 5.15$ \cellcolor{C2}
\\\hline
\end{tabular}
\caption{Each grid in these tables represents a distribution of $k$-SAT instances with differing amounts of structure $\beta$, fixed to contain an equal mix of \textit{satisfiable} and \textit{unsatisfiable} instances (where lower $\beta$ means more structure). For each grid we sample SAT instances and fit the classical runtime and quantum \textit{query complexity} of the form $2^{a\cdot n + b}$. Each cell contains the exponent $a\cdot n+b$ in the order: classical runtime, quantum backtracking detection, quantum backtracking search, Grover. We color a cell blue if the classical algorithm scales best (i.e. has the lowest $a$ value), orange if the detection algorithm scales best, yellow if Grover scales best, and green if search scales better than classical and Grover.}
\end{table}

\begin{table} \hspace{-2.2cm} \begin{tabular}{p{0.9cm}|p{1.93cm}|p{1.77cm}|p{1.77cm}|p{1.77cm}|p{1.77cm}|p{1.77cm}|p{1.77cm}|p{1.77cm}} \backslashbox{$k$}{$\beta$} & \centering$\frac 12$ & \centering 1 & \centering $\frac 32$ & \centering 2 & \centering 3 & \centering 5 & \centering 10 & \ \ \ \ \ $\infty$ \\\hline3& \cellcolor{C0}& \cellcolor{C0}& \cellcolor{C0}& \cellcolor{C0}& \cellcolor{C0} & \footnotesize \cellcolor{C0} $.0299n -9.89$  & \footnotesize \cellcolor{C0} $.0356n -10.5$  & \footnotesize \cellcolor{C0} $.0327n -9.93$ \\& \cellcolor{C0}& \cellcolor{C0}& \cellcolor{C0}& \cellcolor{C0}& \cellcolor{C0} & \footnotesize $.184n + 11.3$ \cellcolor{C0} & \footnotesize $.193n + 11.2$ \cellcolor{C0} & \footnotesize $.202n + 11.0$ \cellcolor{C0}
\\& \cellcolor{C0}& \cellcolor{C0}& \cellcolor{C0}& \cellcolor{C0}& \cellcolor{C0} & \footnotesize $.234n + 16.0$ \cellcolor{C0} & \footnotesize $.245n + 15.8$ \cellcolor{C0} & \footnotesize $.254n + 15.6$ \cellcolor{C0}
\\& \cellcolor{C0}& \cellcolor{C0}& \cellcolor{C0}& \cellcolor{C0}& \cellcolor{C0} & \footnotesize $.43n + 2.24$ \cellcolor{C0} & \footnotesize $.433n + 2.22$ \cellcolor{C0} & \footnotesize $.427n + 2.4$ \cellcolor{C0}
\\\hline
4& \cellcolor{C0}& \cellcolor{C0}& \cellcolor{C0}& \cellcolor{C0} & \footnotesize \cellcolor{C0} $.054n -8.35$  & \footnotesize \cellcolor{C0} $.0582n -7.84$  & \footnotesize \cellcolor{C0} $.0571n -7.32$  & \footnotesize \cellcolor{C0} $.0617n -7.25$ \\& \cellcolor{C0}& \cellcolor{C0}& \cellcolor{C0}& \cellcolor{C0} & \footnotesize $.223n + 11.1$ \cellcolor{C0} & \footnotesize $.242n + 10.9$ \cellcolor{C0} & \footnotesize $.255n + 10.8$ \cellcolor{C0} & \footnotesize $.258n + 10.8$ \cellcolor{C0}
\\& \cellcolor{C0}& \cellcolor{C0}& \cellcolor{C0}& \cellcolor{C0} & \footnotesize $.286n + 15.5$ \cellcolor{C0} & \footnotesize $.304n + 15.4$ \cellcolor{C0} & \footnotesize $.319n + 15.3$ \cellcolor{C0} & \footnotesize $.319n + 15.3$ \cellcolor{C0}
\\& \cellcolor{C0}& \cellcolor{C0}& \cellcolor{C0}& \cellcolor{C0} & \footnotesize $.466n + 2.32$ \cellcolor{C0} & \footnotesize $.46n + 2.45$ \cellcolor{C0} & \footnotesize $.466n + 2.37$ \cellcolor{C0} & \footnotesize $.456n + 2.59$ \cellcolor{C0}
\\\hline
5& \cellcolor{C0}& \cellcolor{C0}& \cellcolor{C0} & \footnotesize \cellcolor{C0} $.0713n -7.98$  & \footnotesize \cellcolor{C0} $.103n -8.92$  & \footnotesize \cellcolor{C0} $.0892n -7.25$  & \footnotesize \cellcolor{C0} $.0885n -6.93$  & \footnotesize \cellcolor{C0} $.0587n -5.12$ \\& \cellcolor{C0}& \cellcolor{C0}& \cellcolor{C0} & \footnotesize $.229n + 11.3$ \cellcolor{C0} & \footnotesize $.262n + 11.0$ \cellcolor{C0} & \footnotesize $.28n + 10.8$ \cellcolor{C0} & \footnotesize $.293n + 10.8$ \cellcolor{C0} & \footnotesize $.305n + 10.6$ \cellcolor{C0}
\\& \cellcolor{C0}& \cellcolor{C0}& \cellcolor{C0} & \footnotesize $.29n + 15.8$ \cellcolor{C0} & \footnotesize $.327n + 15.4$ \cellcolor{C0} & \footnotesize $.345n + 15.2$ \cellcolor{C0} & \footnotesize $.36n + 15.1$ \cellcolor{C0} & \footnotesize $.382n + 14.8$ \cellcolor{C0}
\\& \cellcolor{C0}& \cellcolor{C0}& \cellcolor{C0} & \footnotesize $.474n + 2.49$ \cellcolor{C0} & \footnotesize $.484n + 2.24$ \cellcolor{C0} & \footnotesize $.475n + 2.51$ \cellcolor{C0} & \footnotesize $.474n + 2.51$ \cellcolor{C0} & \footnotesize $.467n + 2.73$ \cellcolor{C0}
\\\hline
6& \cellcolor{C0}& \cellcolor{C0} & \footnotesize \cellcolor{C0} $.065n -6.83$  & \footnotesize \cellcolor{C0} $.126n -8.91$  & \footnotesize \cellcolor{C0} $.174n -10.0$  & \footnotesize \cellcolor{C0} $.187n -9.88$  & \footnotesize \cellcolor{C0} $.187n -9.64$  & \footnotesize \cellcolor{C0} $.167n -8.64$ \\& \cellcolor{C0}& \cellcolor{C0} & \footnotesize $.234n + 11.5$ \cellcolor{C0} & \footnotesize $.261n + 11.2$ \cellcolor{C0} & \footnotesize $.293n + 10.9$ \cellcolor{C0} & \footnotesize $.314n + 10.7$ \cellcolor{C0} & \footnotesize $.325n + 10.7$ \cellcolor{C0} & \footnotesize $.341n + 10.4$ \cellcolor{C0}
\\& \cellcolor{C0}& \cellcolor{C0} & \footnotesize $.295n + 16.0$ \cellcolor{C0} & \footnotesize $.324n + 15.6$ \cellcolor{C0} & \footnotesize $.366n + 15.2$ \cellcolor{C0} & \footnotesize $.387n + 15.0$ \cellcolor{C0} & \footnotesize $.398n + 14.9$ \cellcolor{C0} & \footnotesize $.422n + 14.6$ \cellcolor{C0}
\\& \cellcolor{C0}& \cellcolor{C0} & \footnotesize $.475n + 2.61$ \cellcolor{C0} & \footnotesize $.48n + 2.49$ \cellcolor{C0} & \footnotesize $.484n + 2.39$ \cellcolor{C0} & \footnotesize $.477n + 2.7$ \cellcolor{C0} & \footnotesize $.49n + 2.35$ \cellcolor{C0} & \footnotesize $.463n + 2.8$ \cellcolor{C0}
\\\hline
7 & \footnotesize \cellcolor{C0} $.00228n -4.15$  & \footnotesize \cellcolor{C0} $.0336n -5.35$  & \footnotesize \cellcolor{C0} $.138n -8.77$  & \footnotesize \cellcolor{C0} $.205n -10.5$  & \footnotesize \cellcolor{C0} $.256n -11.4$  & \footnotesize \cellcolor{C0} $.264n -11.1$  & \footnotesize \cellcolor{C0} $.295n -11.7$  & \footnotesize \cellcolor{C0} $.255n -10.2$ \\ & \footnotesize $.153n + 12.6$ \cellcolor{C0} & \footnotesize $.218n + 11.8$ \cellcolor{C0} & \footnotesize $.268n + 11.3$ \cellcolor{C0} & \footnotesize $.291n + 11.1$ \cellcolor{C0} & \footnotesize $.321n + 10.8$ \cellcolor{C0} & \footnotesize $.342n + 10.6$ \cellcolor{C0} & \footnotesize $.352n + 10.5$ \cellcolor{C0} & \footnotesize $.362n + 10.5$ \cellcolor{C0}
\\ & \footnotesize $.211n + 17.1$ \cellcolor{C0} & \footnotesize $.282n + 16.3$ \cellcolor{C0} & \footnotesize $.345n + 15.5$ \cellcolor{C0} & \footnotesize $.36n + 15.4$ \cellcolor{C0} & \footnotesize $.396n + 15.0$ \cellcolor{C0} & \footnotesize $.422n + 14.7$ \cellcolor{C0} & \footnotesize $.434n + 14.7$ \cellcolor{C0} & \footnotesize $.444n + 14.6$ \cellcolor{C0}
\\ & \footnotesize $.45n + 2.96$ \cellcolor{C0} & \footnotesize $.478n + 2.49$ \cellcolor{C0} & \footnotesize $.476n + 2.63$ \cellcolor{C0} & \footnotesize $.473n + 2.76$ \cellcolor{C0} & \footnotesize $.486n + 2.45$ \cellcolor{C0} & \footnotesize $.481n + 2.64$ \cellcolor{C0} & \footnotesize $.461n + 3.03$ \cellcolor{C0} & \footnotesize $.488n + 2.64$ \cellcolor{C0}
\\\hline
8 & \footnotesize \cellcolor{C0} $.00223n -3.97$  & \footnotesize \cellcolor{C0} $.073n -6.2$  & \footnotesize \cellcolor{C0} $.183n -8.95$  & \footnotesize \cellcolor{C0} $.244n -10.2$  & \footnotesize \cellcolor{C0} $.288n -10.7$  & \footnotesize \cellcolor{C0} $.328n -11.5$  & \footnotesize \cellcolor{C0} $.349n -11.7$  & \footnotesize \cellcolor{C0} $.3n -10.2$ \\ & \footnotesize $.19n + 12.4$ \cellcolor{C0} & \footnotesize $.242n + 11.8$ \cellcolor{C0} & \footnotesize $.289n + 11.3$ \cellcolor{C0} & \footnotesize $.314n + 11.0$ \cellcolor{C0} & \footnotesize $.339n + 10.8$ \cellcolor{C0} & \footnotesize $.361n + 10.6$ \cellcolor{C0} & \footnotesize $.371n + 10.5$ \cellcolor{C0} & \footnotesize $.381n + 10.4$ \cellcolor{C0}
\\ & \footnotesize $.258n + 16.7$ \cellcolor{C0} & \footnotesize $.304n + 16.4$ \cellcolor{C0} & \footnotesize $.365n + 15.6$ \cellcolor{C0} & \footnotesize $.385n + 15.4$ \cellcolor{C0} & \footnotesize $.415n + 15.1$ \cellcolor{C0} & \footnotesize $.457n + 14.5$ \cellcolor{C0} & \footnotesize $.458n + 14.6$ \cellcolor{C0} & \footnotesize $.466n + 14.5$ \cellcolor{C0}
\\ & \footnotesize $.457n + 3.0$ \cellcolor{C0} & \footnotesize $.479n + 2.66$ \cellcolor{C0} & \footnotesize $.48n + 2.63$ \cellcolor{C0} & \footnotesize $.468n + 2.94$ \cellcolor{C0} & \footnotesize $.486n + 2.59$ \cellcolor{C0} & \footnotesize $.485n + 2.53$ \cellcolor{C0} & \footnotesize $.48n + 2.77$ \cellcolor{C0} & \footnotesize $.469n + 2.97$ \cellcolor{C0}
\\\hline
9 & \footnotesize \cellcolor{C0} $.00368n -3.83$  & \footnotesize \cellcolor{C0} $.1n -6.29$  & \footnotesize \cellcolor{C0} $.196n -7.97$  & \footnotesize \cellcolor{C0} $.226n -8.3$  & \footnotesize \cellcolor{C0} $.277n -8.93$  & \footnotesize \cellcolor{C0} $.292n -9.07$  & \footnotesize \cellcolor{C0} $.332n -9.8$  & \footnotesize \cellcolor{C0} $.343n -10.1$ \\ & \footnotesize $.225n + 12.1$ \cellcolor{C0} & \footnotesize $.288n + 11.4$ \cellcolor{C0} & \footnotesize $.321n + 11.1$ \cellcolor{C0} & \footnotesize $.342n + 10.9$ \cellcolor{C0} & \footnotesize $.363n + 10.7$ \cellcolor{C0} & \footnotesize $.378n + 10.6$ \cellcolor{C0} & \footnotesize $.388n + 10.5$ \cellcolor{C0} & \footnotesize $.394n + 10.5$ \cellcolor{C0}
\\ & \footnotesize $.308n + 16.2$ \cellcolor{C0} & \footnotesize $.376n + 15.5$ \cellcolor{C0} & \footnotesize $.403n + 15.3$ \cellcolor{C0} & \footnotesize $.42n + 15.1$ \cellcolor{C0} & \footnotesize $.47n + 14.4$ \cellcolor{C0} & \footnotesize $.47n + 14.4$ \cellcolor{C0} & \footnotesize $.52n + 13.7$ \cellcolor{C0} & \footnotesize $.477n + 14.6$ \cellcolor{C0}
\\ & \footnotesize $.443n + 3.38$ \cellcolor{C0} & \footnotesize $.501n + 2.34$ \cellcolor{C0} & \footnotesize $.496n + 2.42$ \cellcolor{C0} & \footnotesize $.439n + 3.21$ \cellcolor{C0} & \footnotesize $.491n + 2.37$ \cellcolor{C0} & \footnotesize $.402n + 3.82$ \cellcolor{C0} & \footnotesize $.477n + 2.86$ \cellcolor{C0} & \footnotesize $.484n + 2.71$ \cellcolor{C0}
\\\hline
10 & \footnotesize \cellcolor{C0} $.016n -3.73$  & \footnotesize \cellcolor{C0} $.172n -7.24$  & \footnotesize \cellcolor{C0} $.281n -9.14$  & \footnotesize \cellcolor{C0} $.339n -10.0$  & \footnotesize \cellcolor{C1} $.389n -10.6$  & \footnotesize \cellcolor{C1} $.439n -11.4$  & \footnotesize \cellcolor{C1} $.466n -11.8$  & \footnotesize \cellcolor{C1} $.412n -10.8$ \\ & \footnotesize $.236n + 12.4$ \cellcolor{C0} & \footnotesize $.293n + 11.7$ \cellcolor{C0} & \footnotesize $.329n + 11.2$ \cellcolor{C0} & \footnotesize $.35n + 11.0$ \cellcolor{C0} & \footnotesize $.37n + 10.8$ \cellcolor{C1} & \footnotesize $.387n + 10.6$ \cellcolor{C1} & \footnotesize $.396n + 10.6$ \cellcolor{C1} & \footnotesize $.405n + 10.5$ \cellcolor{C1}
\\ & \footnotesize $.303n + 16.9$ \cellcolor{C0} & \footnotesize $.363n + 16.0$ \cellcolor{C0} & \footnotesize $.417n + 15.3$ \cellcolor{C0} & \footnotesize $.433n + 15.2$ \cellcolor{C0} & \footnotesize $.455n + 14.9$ \cellcolor{C1} & \footnotesize $.481n + 14.6$ \cellcolor{C1} & \footnotesize $.498n + 14.4$ \cellcolor{C1} & \footnotesize $.486n + 14.6$ \cellcolor{C1}
\\ & \footnotesize $.477n + 2.75$ \cellcolor{C0} & \footnotesize $.483n + 2.54$ \cellcolor{C0} & \footnotesize $.49n + 2.42$ \cellcolor{C0} & \footnotesize $.509n + 2.24$ \cellcolor{C0} & \footnotesize $.455n + 3.17$ \cellcolor{C1} & \footnotesize $.439n + 3.28$ \cellcolor{C1} & \footnotesize $.499n + 2.42$ \cellcolor{C1} & \footnotesize $.48n + 2.81$ \cellcolor{C1}
\\\hline
11 & \footnotesize \cellcolor{C0} $.0838n -4.77$  & \footnotesize \cellcolor{C0} $.243n -7.56$  & \footnotesize \cellcolor{C1} $.359n -9.47$  & \footnotesize \cellcolor{C1} $.414n -10.1$  & \footnotesize \cellcolor{C2} $.485n -11.2$  & \footnotesize \cellcolor{C2} $.508n -11.3$  & \footnotesize \cellcolor{C3} $.534n -11.7$  & \footnotesize \cellcolor{C1} $.464n -10.9$ \\ & \footnotesize $.27n + 12.1$ \cellcolor{C0} & \footnotesize $.312n + 11.6$ \cellcolor{C0} & \footnotesize $.347n + 11.2$ \cellcolor{C1} & \footnotesize $.356n + 11.1$ \cellcolor{C1} & \footnotesize $.381n + 10.8$ \cellcolor{C2} & \footnotesize $.398n + 10.6$ \cellcolor{C2} & \footnotesize $.405n + 10.6$ \cellcolor{C3} & \footnotesize $.414n + 10.5$ \cellcolor{C1}
\\ & \footnotesize $.358n + 16.3$ \cellcolor{C0} & \footnotesize $.38n + 16.0$ \cellcolor{C0} & \footnotesize $.469n + 14.7$ \cellcolor{C1} & \footnotesize $.451n + 15.1$ \cellcolor{C1} & \footnotesize $.448n + 15.2$ \cellcolor{C2} & \footnotesize $.505n + 14.4$ \cellcolor{C2} & \footnotesize $.52n + 14.3$ \cellcolor{C3} & \footnotesize $.503n + 14.6$ \cellcolor{C1}
\\ & \footnotesize $.497n + 2.4$ \cellcolor{C0} & \footnotesize $.522n + 1.94$ \cellcolor{C0} & \footnotesize $.47n + 2.78$ \cellcolor{C1} & \footnotesize $.482n + 2.77$ \cellcolor{C1} & \footnotesize $.564n + 1.19$ \cellcolor{C2} & \footnotesize $.597n + 0.632$ \cellcolor{C2} & \footnotesize $.472n + 2.88$ \cellcolor{C3} & \footnotesize $.476n + 2.84$ \cellcolor{C1}
\\\hline
12 & \footnotesize \cellcolor{C0} $.0991n -3.37$  & \footnotesize \cellcolor{C0} $.26n -6.21$  & \footnotesize \cellcolor{C1} $.372n -8.12$  & \footnotesize \cellcolor{C2} $.447n -9.39$  & \footnotesize \cellcolor{C3} $.538n -10.8$  & \footnotesize \cellcolor{C3} $.58n -11.4$  & \footnotesize \cellcolor{C3} $.556n -10.8$  & \footnotesize \cellcolor{C3} $.539n -11.4$ \\ & \footnotesize $.305n + 11.9$ \cellcolor{C0} & \footnotesize $.334n + 11.5$ \cellcolor{C0} & \footnotesize $.34n + 11.5$ \cellcolor{C1} & \footnotesize $.364n + 11.2$ \cellcolor{C2} & \footnotesize $.388n + 10.9$ \cellcolor{C3} & \footnotesize $.404n + 10.7$ \cellcolor{C3} & \footnotesize $.413n + 10.6$ \cellcolor{C3} & \footnotesize $.419n + 10.6$ \cellcolor{C3}
\\ & \footnotesize $.379n + 16.3$ \cellcolor{C0} & \footnotesize $.483n + 14.7$ \cellcolor{C0} & \footnotesize $.433n + 15.6$ \cellcolor{C1} & \footnotesize $.433n + 15.6$ \cellcolor{C2} & \footnotesize $.433n + 15.6$ \cellcolor{C3} & \footnotesize $.638n + 12.5$ \cellcolor{C3} & \footnotesize $.638n + 12.5$ \cellcolor{C3} & \footnotesize $.491n + 14.9$ \cellcolor{C3}
\\ & \footnotesize $.648n -0.28$ \cellcolor{C0} & \footnotesize $.48n + 2.79$ \cellcolor{C0} & \footnotesize $.455n + 3.14$ \cellcolor{C1} & \footnotesize $.645n -0.0727$ \cellcolor{C2} & \footnotesize $.404n + 3.98$ \cellcolor{C3} & \footnotesize $.478n + 2.83$ \cellcolor{C3} & \footnotesize $.478n + 2.83$ \cellcolor{C3} & \footnotesize $.464n + 2.97$ \cellcolor{C3}
\\\hline
\end{tabular}
\caption{Each grid in these tables represents a distribution of $k$-SAT instances with differing amounts of structure $\beta$, fixed to contain only \textit{satisfiable} instances (where lower $\beta$ means more structure). For each grid we sample SAT instances and fit the classical runtime and quantum \textit{query complexity} of the form $2^{a\cdot n + b}$. Each cell contains the exponent $a\cdot n+b$ in the order: classical runtime, quantum backtracking detection, quantum backtracking search, Grover. We color a cell blue if the classical algorithm scales best (i.e. has the lowest $a$ value), orange if the detection algorithm scales best, yellow if Grover scales best, and green if search scales better than classical and Grover.}
\end{table}

\begin{table} \hspace{-2.2cm} \begin{tabular}{p{0.9cm}|p{1.93cm}|p{1.77cm}|p{1.77cm}|p{1.77cm}|p{1.77cm}|p{1.77cm}|p{1.77cm}|p{1.77cm}} \backslashbox{$k$}{$\beta$} & \centering$\frac 12$ & \centering 1 & \centering $\frac 32$ & \centering 2 & \centering 3 & \centering 5 & \centering 10 & \ \ \ \ \ $\infty$ \\\hline3& \cellcolor{C0}& \cellcolor{C0}& \cellcolor{C0}& \cellcolor{C0}& \cellcolor{C0} & \footnotesize \cellcolor{C0} $.0412n -11.6$  & \footnotesize \cellcolor{C0} $.0586n -13.8$  & \footnotesize \cellcolor{C0} $.0529n -12.5$ \\& \cellcolor{C0}& \cellcolor{C0}& \cellcolor{C0}& \cellcolor{C0}& \cellcolor{C0} & \footnotesize $.183n + 11.1$ \cellcolor{C0} & \footnotesize $.192n + 11.0$ \cellcolor{C0} & \footnotesize $.198n + 10.9$ \cellcolor{C0}
\\& \cellcolor{C0}& \cellcolor{C0}& \cellcolor{C0}& \cellcolor{C0}& \cellcolor{C0} & \footnotesize $.233n + 15.7$ \cellcolor{C0} & \footnotesize $.243n + 15.7$ \cellcolor{C0} & \footnotesize $.257n + 15.4$ \cellcolor{C0}
\\& \cellcolor{C0}& \cellcolor{C0}& \cellcolor{C0}& \cellcolor{C0}& \cellcolor{C0} & \footnotesize $.5n + 6.01$ \cellcolor{C0} & \footnotesize $.5n + 6.01$ \cellcolor{C0} & \footnotesize $.5n + 6.01$ \cellcolor{C0}
\\\hline
4& \cellcolor{C0}& \cellcolor{C0}& \cellcolor{C0}& \cellcolor{C0} & \footnotesize \cellcolor{C0} $.0843n -10.4$  & \footnotesize \cellcolor{C0} $.146n -13.6$  & \footnotesize \cellcolor{C0} $.188n -15.7$  & \footnotesize \cellcolor{C0} $.19n -15.2$ \\& \cellcolor{C0}& \cellcolor{C0}& \cellcolor{C0}& \cellcolor{C0} & \footnotesize $.22n + 11.1$ \cellcolor{C0} & \footnotesize $.242n + 10.9$ \cellcolor{C0} & \footnotesize $.255n + 10.8$ \cellcolor{C0} & \footnotesize $.26n + 10.7$ \cellcolor{C0}
\\& \cellcolor{C0}& \cellcolor{C0}& \cellcolor{C0}& \cellcolor{C0} & \footnotesize $.276n + 15.6$ \cellcolor{C0} & \footnotesize $.308n + 15.3$ \cellcolor{C0} & \footnotesize $.319n + 15.2$ \cellcolor{C0} & \footnotesize $.324n + 15.1$ \cellcolor{C0}
\\& \cellcolor{C0}& \cellcolor{C0}& \cellcolor{C0}& \cellcolor{C0} & \footnotesize $.5n + 6.01$ \cellcolor{C0} & \footnotesize $.5n + 6.01$ \cellcolor{C0} & \footnotesize $.5n + 6.01$ \cellcolor{C0} & \footnotesize $.5n + 6.01$ \cellcolor{C0}
\\\hline
5& \cellcolor{C0}& \cellcolor{C0}& \cellcolor{C0} & \footnotesize \cellcolor{C0} $.102n -9.57$  & \footnotesize \cellcolor{C0} $.192n -13.1$  & \footnotesize \cellcolor{C0} $.258n -15.1$  & \footnotesize \cellcolor{C1} $.301n -16.0$  & \footnotesize \cellcolor{C1} $.318n -16.2$ \\& \cellcolor{C0}& \cellcolor{C0}& \cellcolor{C0} & \footnotesize $.23n + 11.2$ \cellcolor{C0} & \footnotesize $.262n + 11.0$ \cellcolor{C0} & \footnotesize $.281n + 10.8$ \cellcolor{C0} & \footnotesize $.294n + 10.7$ \cellcolor{C1} & \footnotesize $.307n + 10.5$ \cellcolor{C1}
\\& \cellcolor{C0}& \cellcolor{C0}& \cellcolor{C0} & \footnotesize $.288n + 15.8$ \cellcolor{C0} & \footnotesize $.324n + 15.4$ \cellcolor{C0} & \footnotesize $.345n + 15.2$ \cellcolor{C0} & \footnotesize $.36n + 15.1$ \cellcolor{C1} & \footnotesize $.38n + 14.8$ \cellcolor{C1}
\\& \cellcolor{C0}& \cellcolor{C0}& \cellcolor{C0} & \footnotesize $.5n + 6.01$ \cellcolor{C0} & \footnotesize $.5n + 6.01$ \cellcolor{C0} & \footnotesize $.5n + 6.01$ \cellcolor{C0} & \footnotesize $.5n + 6.01$ \cellcolor{C1} & \footnotesize $.5n + 6.01$ \cellcolor{C1}
\\\hline
6& \cellcolor{C0}& \cellcolor{C0} & \footnotesize \cellcolor{C0} $.0882n -7.62$  & \footnotesize \cellcolor{C0} $.171n -10.1$  & \footnotesize \cellcolor{C0} $.258n -12.4$  & \footnotesize \cellcolor{C1} $.337n -13.9$  & \footnotesize \cellcolor{C1} $.35n -13.5$  & \footnotesize \cellcolor{C1} $.407n -15.8$ \\& \cellcolor{C0}& \cellcolor{C0} & \footnotesize $.235n + 11.4$ \cellcolor{C0} & \footnotesize $.26n + 11.2$ \cellcolor{C0} & \footnotesize $.294n + 10.9$ \cellcolor{C0} & \footnotesize $.313n + 10.7$ \cellcolor{C1} & \footnotesize $.327n + 10.6$ \cellcolor{C1} & \footnotesize $.34n + 10.5$ \cellcolor{C1}
\\& \cellcolor{C0}& \cellcolor{C0} & \footnotesize $.298n + 15.9$ \cellcolor{C0} & \footnotesize $.327n + 15.5$ \cellcolor{C0} & \footnotesize $.366n + 15.1$ \cellcolor{C0} & \footnotesize $.388n + 15.0$ \cellcolor{C1} & \footnotesize $.401n + 14.9$ \cellcolor{C1} & \footnotesize $.424n + 14.6$ \cellcolor{C1}
\\& \cellcolor{C0}& \cellcolor{C0} & \footnotesize $.5n + 6.01$ \cellcolor{C0} & \footnotesize $.5n + 6.01$ \cellcolor{C0} & \footnotesize $.5n + 6.01$ \cellcolor{C0} & \footnotesize $.5n + 6.01$ \cellcolor{C1} & \footnotesize $.5n + 6.01$ \cellcolor{C1} & \footnotesize $.5n + 6.01$ \cellcolor{C1}
\\\hline
7 & \footnotesize \cellcolor{C0} $.00394n -4.21$  & \footnotesize \cellcolor{C0} $.0473n -5.76$  & \footnotesize \cellcolor{C0} $.172n -9.55$  & \footnotesize \cellcolor{C0} $.252n -11.3$  & \footnotesize \cellcolor{C1} $.334n -12.9$  & \footnotesize \cellcolor{C1} $.402n -14.1$  & \footnotesize \cellcolor{C2} $.451n -14.9$  & \footnotesize \cellcolor{C2} $.476n -15.4$ \\ & \footnotesize $.149n + 12.6$ \cellcolor{C0} & \footnotesize $.219n + 11.8$ \cellcolor{C0} & \footnotesize $.267n + 11.3$ \cellcolor{C0} & \footnotesize $.292n + 11.1$ \cellcolor{C0} & \footnotesize $.321n + 10.8$ \cellcolor{C1} & \footnotesize $.341n + 10.6$ \cellcolor{C1} & \footnotesize $.353n + 10.5$ \cellcolor{C2} & \footnotesize $.361n + 10.5$ \cellcolor{C2}
\\ & \footnotesize $.205n + 17.2$ \cellcolor{C0} & \footnotesize $.281n + 16.3$ \cellcolor{C0} & \footnotesize $.34n + 15.5$ \cellcolor{C0} & \footnotesize $.358n + 15.5$ \cellcolor{C0} & \footnotesize $.396n + 15.0$ \cellcolor{C1} & \footnotesize $.422n + 14.8$ \cellcolor{C1} & \footnotesize $.434n + 14.7$ \cellcolor{C2} & \footnotesize $.444n + 14.6$ \cellcolor{C2}
\\ & \footnotesize $.5n + 6.01$ \cellcolor{C0} & \footnotesize $.5n + 6.01$ \cellcolor{C0} & \footnotesize $.5n + 6.01$ \cellcolor{C0} & \footnotesize $.5n + 6.01$ \cellcolor{C0} & \footnotesize $.5n + 6.01$ \cellcolor{C1} & \footnotesize $.5n + 6.01$ \cellcolor{C1} & \footnotesize $.5n + 6.01$ \cellcolor{C2} & \footnotesize $.5n + 6.01$ \cellcolor{C2}
\\\hline
8 & \footnotesize \cellcolor{C0} $.00241n -3.93$  & \footnotesize \cellcolor{C0} $.0994n -6.8$  & \footnotesize \cellcolor{C0} $.223n -9.54$  & \footnotesize \cellcolor{C0} $.297n -10.9$  & \footnotesize \cellcolor{C1} $.374n -12.2$  & \footnotesize \cellcolor{C1} $.437n -13.2$  & \footnotesize \cellcolor{C2} $.487n -14.0$  & \footnotesize \cellcolor{C2} $.495n -14.0$ \\ & \footnotesize $.186n + 12.4$ \cellcolor{C0} & \footnotesize $.241n + 11.9$ \cellcolor{C0} & \footnotesize $.289n + 11.3$ \cellcolor{C0} & \footnotesize $.315n + 11.0$ \cellcolor{C0} & \footnotesize $.339n + 10.8$ \cellcolor{C1} & \footnotesize $.361n + 10.6$ \cellcolor{C1} & \footnotesize $.371n + 10.5$ \cellcolor{C2} & \footnotesize $.382n + 10.4$ \cellcolor{C2}
\\ & \footnotesize $.249n + 17.0$ \cellcolor{C0} & \footnotesize $.304n + 16.3$ \cellcolor{C0} & \footnotesize $.366n + 15.5$ \cellcolor{C0} & \footnotesize $.389n + 15.3$ \cellcolor{C0} & \footnotesize $.415n + 15.1$ \cellcolor{C1} & \footnotesize $.448n + 14.7$ \cellcolor{C1} & \footnotesize $.458n + 14.6$ \cellcolor{C2} & \footnotesize $.474n + 14.4$ \cellcolor{C2}
\\ & \footnotesize $.5n + 6.01$ \cellcolor{C0} & \footnotesize $.5n + 6.01$ \cellcolor{C0} & \footnotesize $.5n + 6.01$ \cellcolor{C0} & \footnotesize $.5n + 6.01$ \cellcolor{C0} & \footnotesize $.5n + 6.01$ \cellcolor{C1} & \footnotesize $.5n + 6.01$ \cellcolor{C1} & \footnotesize $.5n + 6.01$ \cellcolor{C2} & \footnotesize $.5n + 6.01$ \cellcolor{C2}
\\\hline
9 & \footnotesize \cellcolor{C0} $.0057n -3.78$  & \footnotesize \cellcolor{C0} $.126n -6.56$  & \footnotesize \cellcolor{C0} $.238n -8.34$  & \footnotesize \cellcolor{C0} $.29n -8.96$  & \footnotesize \cellcolor{C0} $.352n -9.8$  & \footnotesize \cellcolor{C1} $.404n -10.4$  & \footnotesize \cellcolor{C1} $.45n -11.2$  & \footnotesize \cellcolor{C2} $.49n -12.3$ \\ & \footnotesize $.232n + 12.0$ \cellcolor{C0} & \footnotesize $.282n + 11.5$ \cellcolor{C0} & \footnotesize $.322n + 11.1$ \cellcolor{C0} & \footnotesize $.343n + 10.8$ \cellcolor{C0} & \footnotesize $.363n + 10.7$ \cellcolor{C0} & \footnotesize $.378n + 10.5$ \cellcolor{C1} & \footnotesize $.389n + 10.5$ \cellcolor{C1} & \footnotesize $.394n + 10.5$ \cellcolor{C2}
\\ & \footnotesize $.295n + 16.5$ \cellcolor{C0} & \footnotesize $.376n + 15.5$ \cellcolor{C0} & \footnotesize $.403n + 15.3$ \cellcolor{C0} & \footnotesize $.42n + 15.1$ \cellcolor{C0} & \footnotesize $.47n + 14.4$ \cellcolor{C0} & \footnotesize $.47n + 14.4$ \cellcolor{C1} & \footnotesize $.52n + 13.7$ \cellcolor{C1} & \footnotesize $.48n + 14.6$ \cellcolor{C2}
\\ & \footnotesize $.5n + 6.01$ \cellcolor{C0} & \footnotesize $.5n + 6.01$ \cellcolor{C0} & \footnotesize $.5n + 6.01$ \cellcolor{C0} & \footnotesize $.5n + 6.01$ \cellcolor{C0} & \footnotesize $.5n + 6.01$ \cellcolor{C0} & \footnotesize $.5n + 6.01$ \cellcolor{C1} & \footnotesize $.5n + 6.01$ \cellcolor{C1} & \footnotesize $.5n + 6.01$ \cellcolor{C2}
\\\hline
10 & \footnotesize \cellcolor{C0} $.0229n -3.66$  & \footnotesize \cellcolor{C0} $.19n -6.91$  & \footnotesize \cellcolor{C0} $.308n -8.83$  & \footnotesize \cellcolor{C1} $.375n -9.76$  & \footnotesize \cellcolor{C1} $.441n -10.6$  & \footnotesize \cellcolor{C2} $.494n -11.2$  & \footnotesize \cellcolor{C2} $.541n -11.9$  & \footnotesize \cellcolor{C2} $.53n -12.1$ \\ & \footnotesize $.231n + 12.5$ \cellcolor{C0} & \footnotesize $.29n + 11.7$ \cellcolor{C0} & \footnotesize $.327n + 11.3$ \cellcolor{C0} & \footnotesize $.348n + 11.0$ \cellcolor{C1} & \footnotesize $.37n + 10.8$ \cellcolor{C1} & \footnotesize $.387n + 10.6$ \cellcolor{C2} & \footnotesize $.396n + 10.6$ \cellcolor{C2} & \footnotesize $.405n + 10.5$ \cellcolor{C2}
\\ & \footnotesize $.303n + 16.8$ \cellcolor{C0} & \footnotesize $.363n + 16.0$ \cellcolor{C0} & \footnotesize $.417n + 15.3$ \cellcolor{C0} & \footnotesize $.433n + 15.2$ \cellcolor{C1} & \footnotesize $.455n + 14.9$ \cellcolor{C1} & \footnotesize $.481n + 14.6$ \cellcolor{C2} & \footnotesize $.498n + 14.4$ \cellcolor{C2} & \footnotesize $.486n + 14.6$ \cellcolor{C2}
\\ & \footnotesize $.5n + 6.01$ \cellcolor{C0} & \footnotesize $.5n + 6.01$ \cellcolor{C0} & \footnotesize $.5n + 6.01$ \cellcolor{C0} & \footnotesize $.5n + 6.01$ \cellcolor{C1} & \footnotesize $.5n + 6.01$ \cellcolor{C1} & \footnotesize $.5n + 6.01$ \cellcolor{C2} & \footnotesize $.5n + 6.01$ \cellcolor{C2} & \footnotesize $.5n + 6.01$ \cellcolor{C2}
\\\hline
11 & \footnotesize \cellcolor{C0} $.0822n -3.78$  & \footnotesize \cellcolor{C0} $.247n -6.51$  & \footnotesize \cellcolor{C1} $.368n -8.41$  & \footnotesize \cellcolor{C1} $.432n -9.23$  & \footnotesize \cellcolor{C2} $.504n -10.2$  & \footnotesize \cellcolor{C3} $.535n -10.4$  & \footnotesize \cellcolor{C3} $.578n -11.0$  & \footnotesize \cellcolor{C3} $.563n -11.7$ \\ & \footnotesize $.276n + 12.0$ \cellcolor{C0} & \footnotesize $.32n + 11.5$ \cellcolor{C0} & \footnotesize $.347n + 11.2$ \cellcolor{C1} & \footnotesize $.362n + 11.0$ \cellcolor{C1} & \footnotesize $.383n + 10.8$ \cellcolor{C2} & \footnotesize $.397n + 10.6$ \cellcolor{C3} & \footnotesize $.405n + 10.6$ \cellcolor{C3} & \footnotesize $.414n + 10.5$ \cellcolor{C3}
\\ & \footnotesize $.351n + 16.4$ \cellcolor{C0} & \footnotesize $.38n + 16.0$ \cellcolor{C0} & \footnotesize $.45n + 15.1$ \cellcolor{C1} & \footnotesize $.451n + 15.1$ \cellcolor{C1} & \footnotesize $.448n + 15.2$ \cellcolor{C2} & \footnotesize $.505n + 14.4$ \cellcolor{C3} & \footnotesize $.52n + 14.3$ \cellcolor{C3} & \footnotesize $.503n + 14.6$ \cellcolor{C3}
\\ & \footnotesize $.5n + 6.01$ \cellcolor{C0} & \footnotesize $.5n + 6.01$ \cellcolor{C0} & \footnotesize $.5n + 6.01$ \cellcolor{C1} & \footnotesize $.5n + 6.01$ \cellcolor{C1} & \footnotesize $.5n + 6.01$ \cellcolor{C2} & \footnotesize $.5n + 6.01$ \cellcolor{C3} & \footnotesize $.5n + 6.01$ \cellcolor{C3} & \footnotesize $.5n + 6.01$ \cellcolor{C3}
\\\hline
12 & \footnotesize \cellcolor{C0} $.0974n -2.22$  & \footnotesize \cellcolor{C0} $.252n -4.78$  & \footnotesize \cellcolor{C1} $.357n -6.44$  & \footnotesize \cellcolor{C1} $.419n -7.35$  & \footnotesize \cellcolor{C2} $.508n -8.74$  & \footnotesize \cellcolor{C3} $.546n -9.14$  & \footnotesize \cellcolor{C3} $.579n -9.55$  & \footnotesize \cellcolor{C2} $.592n -11.2$ \\ & \footnotesize $.271n + 12.4$ \cellcolor{C0} & \footnotesize $.321n + 11.7$ \cellcolor{C0} & \footnotesize $.343n + 11.5$ \cellcolor{C1} & \footnotesize $.375n + 11.0$ \cellcolor{C1} & \footnotesize $.386n + 10.9$ \cellcolor{C2} & \footnotesize $.404n + 10.7$ \cellcolor{C3} & \footnotesize $.411n + 10.6$ \cellcolor{C3} & \footnotesize $.419n + 10.6$ \cellcolor{C2}
\\ & \footnotesize $.279n + 17.9$ \cellcolor{C0} & \footnotesize $.483n + 14.7$ \cellcolor{C0} & \footnotesize $.483n + 14.7$ \cellcolor{C1} & \footnotesize $.433n + 15.6$ \cellcolor{C1} & \footnotesize $.433n + 15.6$ \cellcolor{C2} & \footnotesize $.638n + 12.5$ \cellcolor{C3} & \footnotesize $.638n + 12.5$ \cellcolor{C3} & \footnotesize $.491n + 14.9$ \cellcolor{C2}
\\ & \footnotesize $.5n + 6.01$ \cellcolor{C0} & \footnotesize $.5n + 6.01$ \cellcolor{C0} & \footnotesize $.5n + 6.01$ \cellcolor{C1} & \footnotesize $.5n + 6.01$ \cellcolor{C1} & \footnotesize $.5n + 6.01$ \cellcolor{C2} & \footnotesize $.5n + 6.01$ \cellcolor{C3} & \footnotesize $.5n + 6.01$ \cellcolor{C3} & \footnotesize $.5n + 6.01$ \cellcolor{C2}
\\\hline
\end{tabular}
\caption{Each grid in these tables represents a distribution of $k$-SAT instances with differing amounts of structure $\beta$, fixed to contain only \textit{unsatisfiable} instances (where lower $\beta$ means more structure). For each grid we sample SAT instances and fit the classical runtime and quantum \textit{query complexity} of the form $2^{a\cdot n + b}$. Each cell contains the exponent $a\cdot n+b$ in the order: classical runtime, quantum backtracking detection, quantum backtracking search, Grover. We color a cell blue if the classical algorithm scales best (i.e. has the lowest $a$ value), orange if the detection algorithm scales best, yellow if Grover scales best, and green if search scales better than classical and Grover. Query complexity of Grover depends only on $n$ (and the number of solutions, which is constant for unsatisfiable instances), not on $k$, $\beta$ and hence is constant.}
\end{table}

\end{document}